\documentclass{ar-1col}

\usepackage{url}

\usepackage[]{geometry}

\usepackage{natbib}
\usepackage{amssymb,amsmath}
\usepackage[backref,breaklinks,colorlinks,citecolor=blue]{hyperref}
\usepackage[export]{adjustbox}

\jname{Xxxx. Xxx. Xxx. Xxx.}
\jvol{59}
\jyear{2019}
\doi{10.1146/((please add article doi))}

\newcommand{\farcm}{\mbox{$.\mkern-4mu^\prime$}}
\newcommand{\farcs}{\mbox{$.\!\!^{\prime\prime}$}}
\newcommand\la{\lower.5ex\hbox{$\; \buildrel < \over \sim \;$}}
\newcommand\ga{\lower.5ex\hbox{$\; \buildrel > \over \sim \;$}}

\newcommand{\Gaia}{{\it Gaia}}
\newcommand{\msun}{M_{\odot}}

\newcommand{\tcr}{t_{\rm cr}}

\def\Usol {\ifmmode{U_\odot} \else {$U_\odot$}\fi} 
\def\Vsol {\ifmmode{V_\odot} \else {$V_\odot$}\fi} 
\def\Wsol {\ifmmode{W_\odot} \else {$W_\odot$}\fi} 
\def\Vsolar {\ifmmode{{\bf v}_\odot} \else {${\bf v}_\odot$}\fi} 
\def\vsolar {\ifmmode{v_\odot} \else {$v_\odot$}\fi} 
\def\VLSR {\ifmmode{{\bf \Theta}_{\rm 0}} \else {${\bf \Theta_{\rm 0}}$}\fi} 
\def\Vlsr {\ifmmode{{\bf v_{\rm LSR}}} \else {${\bf v_{\rm LSR}}$}\fi} 
\def\Rsolar{\ifmmode{R_0}\else {$R_0$}\fi}
\def\zsolar{\ifmmode{z_0}\else {$z_0$}\fi}
\def\phisolar{\ifmmode{\phi_0}\else {$\phi_0$}\fi}
\def\Omegasolar{\ifmmode{{\bf \Omega}_\odot}\else {${\bf \Omega}_\odot$}\fi}

\def\avir	{\alpha_{\rm vir}}
\def\epss	{{\epsilon_*}}
\def\aclump	{{\alpha_{M,\,\rm clump}}}

\newcommand{\gmc}		{{\rm GMC}}
\newcommand{\kb}		{k_{\rm B}}
\newcommand{\tff}  {t_{\rm ff}}

\newcommand{\epsff}{\epsilon_{\mathrm{ff}}}
\newcommand{\vectheta}{\boldsymbol{\theta}}

\newcommand{\beq}	{\begin{equation}}
\newcommand{\eeq}	{\end{equation}}
\newcommand{\beqa}{\begin{eqnarray}}
\newcommand{\eeqa}{\end{eqnarray}}
\newcommand{\e}	{$^{-1}$}
\newcommand{\ee}	{$^{-2}$}
\newcommand{\eee}	{$^{-3}$}
\def\simlt{\lower.5ex\hbox{$\; \buildrel < \over \sim \;$}}
\def\simgt{\lower.5ex\hbox{$\; \buildrel > \over \sim \;$}}
\def\la{\simlt}
\def\ga{\simgt}
\newcommand{\avg}[1]  {{\langle #1 \rangle}}

\begin{document}

\markboth{Krumholz, McKee, \& Bland-Hawthorn}{Star clusters}

\title{Star Clusters Across Cosmic Time}
\author{Mark R. Krumholz$^{1,2}$, Christopher F. McKee$^3$, and Joss Bland-Hawthorn$^{2,4,5}$
\affil{$^1$Research School of Astronomy and Astrophysics, Australian National University, Canberra, ACT 2611, Australia; email: mark.krumholz@anu.edu.au}
\affil{$^2$Centre of Excellence for All Sky Astrophysics in Three Dimensions (ASTRO-3D), Australia}
\affil{$^3$Departments of Astronomy and of Physics, University of California, Berkeley, CA 94720, USA}
\affil{$^4$Sydney Institute for Astronomy, School of Physics A28, University of Sydney, NSW 2006, Australia}
\affil{$^5$Miller Professor, Miller Institute, University of California, Berkeley, CA 94720, USA}
}

\hoffset = -5.0cm
\begin{abstract}
Star clusters stand at the intersection of much of modern astrophysics: the interstellar medium, gravitational dynamics, stellar evolution, and cosmology. Here we review observations and theoretical models for the formation, evolution, and eventual disruption of star clusters. Current literature suggests a picture of this life cycle with several phases:

\begin{minipage}{3.93in}
\vspace{0.1in}
\begin{itemize}
\item Clusters form in hierarchically-structured, accreting molecular clouds that convert gas into stars at a low rate per dynamical time until feedback disperses the gas.
\item The densest parts of the hierarchy resist gas removal long enough to reach high star formation efficiency, becoming dynamically-relaxed and well-mixed. These remain bound after gas removal.
\item In the first $\sim 100$ Myr after gas removal, clusters disperse moderately fast, through a combination of mass loss and tidal shocks by dense molecular structures in the star-forming environment.
\item After $\sim 100$ Myr, clusters lose mass via two-body relaxation and shocks by giant molecular clouds, processes that preferentially affect low-mass clusters and cause a turnover in the cluster mass function to appear on $\sim 1-10$ Gyr timescales.
\item Even after dispersal, some clusters remain coherent and thus detectable in chemical or action space for multiple galactic orbits. 
\end{itemize}
\vspace{0.1in}
\end{minipage}
In the next decade a new generation of space- and AO-assisted ground-based telescopes will enable us to test and refine this picture.
\end{abstract}

\maketitle

\hoffset = -1.75cm

\tableofcontents


\section{INTRODUCTION}

\smallskip
\begin{quote}
The galaxy is in fact nothing but a congeries of innumerable stars grouped together in clusters\ldots
\vskip 0.1cm
Siderius Nuncius (Sidereal Messenger), Galileo Galilei, 1610
\end{quote}

\subsection{Historical background and motivation}

\begin{figure}
\includegraphics[width=\textwidth]{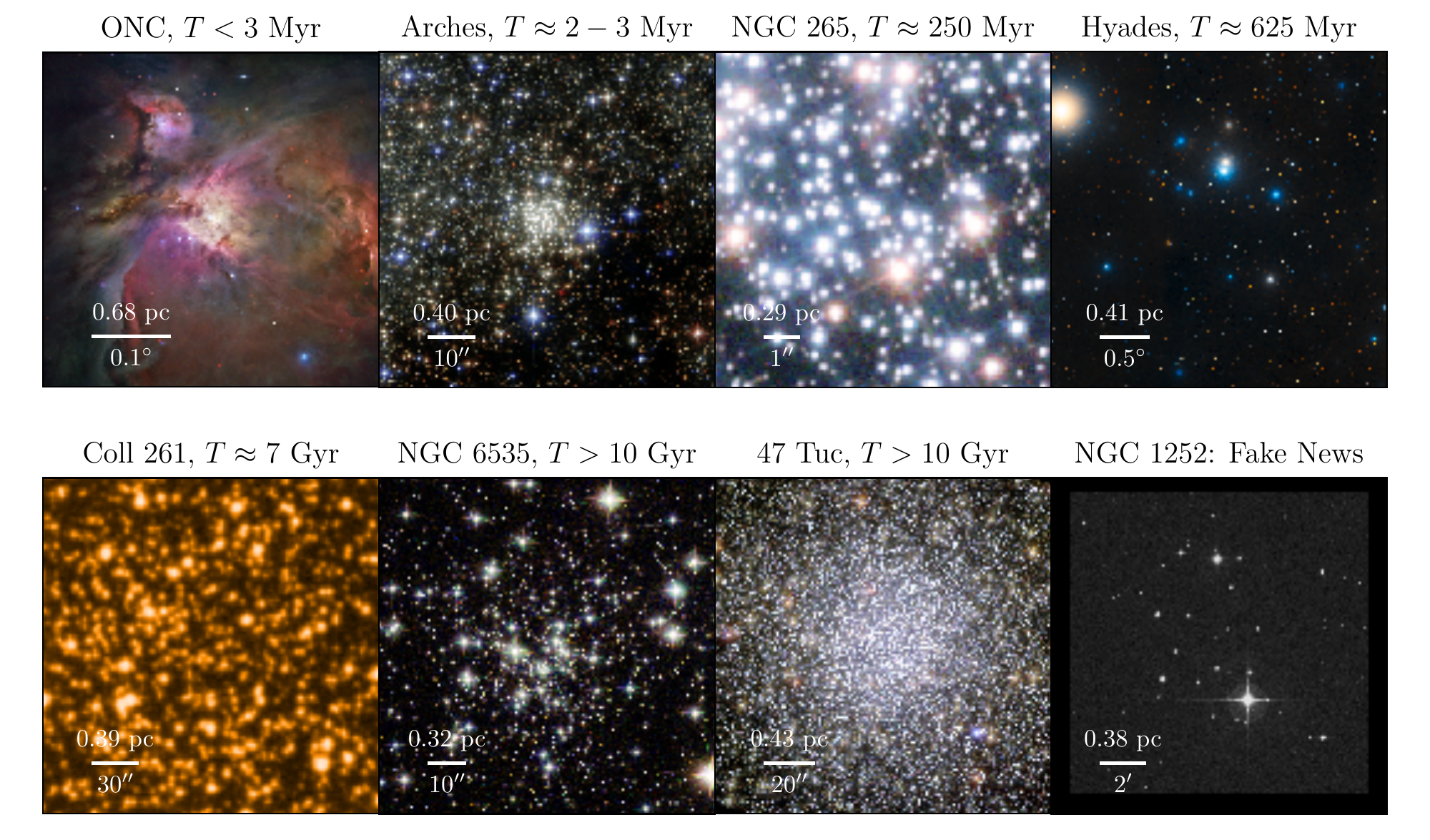}
\caption{
\label{fig:postage_stamps}
Images of a range of star clusters, along with NGC 1252, an object previously classified as a cluster but now known to be an asterism. The field of view in all frames is $3\mbox{ pc}\times 3\mbox{ pc}$, and North is up; angular sizes are indicated by scale bars. Image sources are: Orion Nebular Cluster (ONC) -- \citet{Robberto13a}; Arches cluster -- NASA \& ESA; NGC 265 -- NASA \& ESA; Davide De Martin (ESA/Hubble) and Edward W. Olszewski (University of Arizona, USA); Hyades -- NASA, ESA, \& STScI; Collinder 261 -- ESO/Digitized Sky Survey; NGC 6535 -- ESA/Hubble \& NASA, Gilles Chapdelaine; 47 Tuc -- NASA, ESA, and the Hubble Heritage (STScI/AURA)-ESA/Hubble Collaboration;  J. Mack (STScI) and G. Piotto (University of Padova, Italy); NGC 1252 -- WEBDA database, \url{https://www.univie.ac.at/webda/}.
}
\end{figure}

After four centuries since Galileo's first observations, star clusters remain mysterious. What are the unifying principles that govern the formation of globular and open clusters, groups, associations and super-associations? As illustrated in \autoref{fig:postage_stamps}\footnote{The source code and data tabulations used to produce all the figures in this review are available at \url{https://bitbucket.org/krumholz/cluster_review/}, and also are provided as Supplementary Materials. We make use of the following software packages: \texttt{scipy} \citep{Oliphant07a, Millman11a}, \texttt{matplotlib} \citep{Hunter07a}, \texttt{astropy} \citep{Astropy-Collaboration13a}, and \texttt{SLUG} \citep{da-Silva12a,Krumholz15b}.}, clusters cover a huge range of mass, size, and density scales. The clusters shown span ages from $\sim 1$ Myr to $>10$ Gyr, and masses from $\sim 10^2 - 10^6$ $M_\odot$. Some are so compact and rich that stars become lost in confusion in the 3 pc-frames shown, others are so sparse and extended that most cluster members are outside the frame. Some are classified as open clusters (Arches, NGC 265, Hyades, Coll 261), some as globular clusters (NGC 6535, 47 Tuc). The Orion Nebula Cluster (ONC) is still forming and, depending on the author, might not even be classified as a cluster at all. NGC 1252 had been classified as an open cluster since 1888, but in 2018 was shown to be merely an asterism \citep{Kos2018b}. Cluster formation is central to the star formation process. Conceivably, {\it all} stars formed in groups, clusters, or hierarchies, although, for this to be true, most clusters must have dissolved into the Galactic background soon after formation. However, our understanding of when, how, and why stars cluster remains primitive (see reviews by \citealt{Krumholz14b}, \citealt{Renaud18a}, and \citealt{Adamo18a}).

A review of clusters is timely for several reasons. One is the emerging overlap between the 
traditionally separate communities focused on globular clusters and modern-day star clusters and star formation. These have been separate in the past because star clusters in the disk of the Milky Way, generally classified as open clusters (OCs), and those in the halo, generally classified as globular clusters (GCs), appear to occupy distinct loci in the parameter space of mass, age, and metallicity. New data have begun to blur this distinction: even in the Milky Way there is substantial overlap in both metallicity and density between OCs and GCs, and in rapidly star-forming local galaxies such as the Antennae \citep{Zhang99b} and M82 \citep{McCrady07a} there is overlap in mass as well. While there is still a bimodality in the star cluster age and spatial distribution, this may simply reflect the combined effects of cluster dissolution and the history of galaxy assembly. In this view, toward which we shall mostly tend here (see also \citealt{Kruijssen14a} and \citealt{Forbes18a}), GCs are no different than any other type of star cluster in their formation and internal dynamics; only their cosmological history is different. Their origins are therefore best addressed in the context of general models for the formation and evolution of star clusters, hence the need for this review.\footnote{We do note, however that nuclear star clusters probably do represent a physically distinct class, in that their formation and evolution are inseparably linked with those of the central black hole around which they orbit; for this reason we will not include nuclear clusters within the scope of this review.}

An additional motivation for this review is to bring together the results of modern studies of stellar kinematics and abundances with the formation and evolution of star clusters. As with GCs and OCs, these have traditionally been somewhat separate fields, with the stellar kinematics and abundances community 
focused more on the long-term dynamical evolution of stars in and around the Galaxy, and the star formation community paying more attention to the dynamics of interstellar gas and the effects of star formation and stellar feedback on it. This separation is no longer viable. In the era of \textit{Gaia} and massive spectroscopic surveys, kinematic and abundance data are now becoming accurate enough that it should be possible to trace stars that are now part of the field back to their birth places, or at least to reconstruct {\it some} of the now-dissolved structures in which they were born. To guide such reconstructions, however, will require theoretical and observational input from the star formation community.

With these motivations in mind, we first review current observational constraints on star cluster populations, and then discuss our current understanding, or lack therefore, of the processes by which clusters form, evolve, and eventually disperse.

\begin{marginnote}[]
Surveys referenced in this review:
\entry{PHAT}{Panchromatic Hubble Andromeda Treasury \citep{Dalcanton12a}}
\entry{LEGUS}{Legacy Extragalactic UV Survey \citep{Calzetti15a}}
\entry{GOALS}{Great Observatories All-sky LIRG Survey \citep{Armus09a}}
\entry{GRS}{Galactic Ring Survey \citep{Jackson06a}}
\entry{ATLASGAL}{APEX Telescope Large Area Survey of the Galaxy \citep{Schuller09a}}
\entry{GALAH}{Galactic Archaeology with HERMES \citep{DeSilva2015}}
\entry{RAVE}{Radial Velocity Experiment \citep{Steinmetz06a}}
\end{marginnote}
\subsection{Prelude: what is a star cluster?}
\label{ssec:definition}

\subsubsection{The need for a definition}

The first question that any discussion of star clusters must face, going back at least to \citet{Trumpler30a}, is how to define a cluster and distinguish clusters from multiple systems. In their recent review, \citet{Portegies-Zwart10a} define a star cluster as a set of stars that are gravitationally bound to one another, while the earlier review by \citet{Lada03a} defines a cluster as a collection of stars with a mass density large enough ($\gtrsim 1$ $M_\odot$ pc$^{-3}$) to resist tidal disruption in Solar neighborhood conditions, and numerous enough to avoid $N$-body evaporation for at least 100 Myr. One might also consider defining clusters as concentrations of stars whose mass density significantly exceeds the mean in their galactic neighborhood ($\approx 0.1$ $M_\odot$ pc$^{-3}$ near the Sun -- \citealt{McKee15a}). For older stellar populations these definitions are in practice nearly identical. A group of stars of mean density $\rho_*$ and radius $r$ is unbound only if its velocity dispersion $\sigma \gtrsim r\sqrt{G\rho_*}$, so the time required for it to disperse is of order the crossing time $t_{\rm cr} = r/\sigma \lesssim 1/\sqrt{G \rho_*}$. At the density threshold proposed by \citet{Lada03a}, this is $\lesssim 10$ Myr, so overdensities older than this must be either held together by self-gravity or external forces, or be short-lived transients.

\begin{figure}
\includegraphics[width=\textwidth]{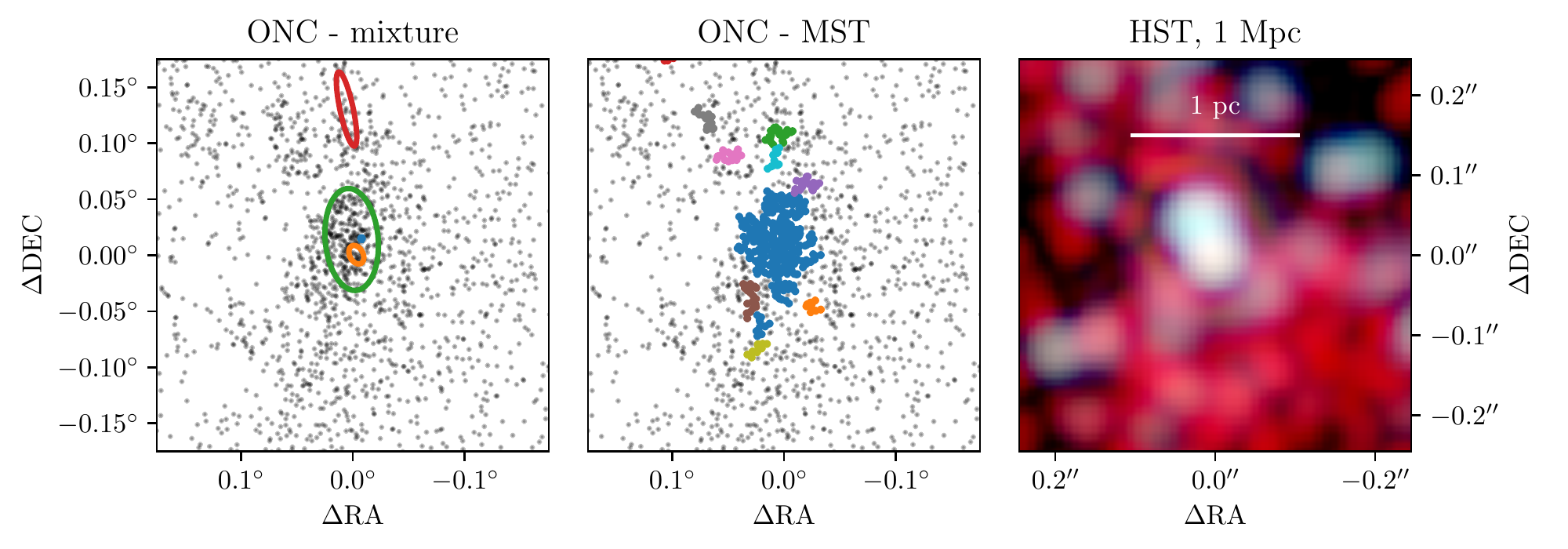}
\caption{Sample decompositions of the stars in and around the Orion Nebula Cluster (ONC). In the left two panels, gray dots show stars from the catalog of \citet{Da-Rio09a}; the axes indicate offset from the position of $\theta^1$c Ori, the most luminous star in the cluster. In the left panel, ellipses mark the clusters identified by \citet{Kuhn14a} based on a Bayesian decomposition into isothermal ellipsoids. The middle panel shows a minimum spanning tree (MST) decomposition, performed on the \citet{Da-Rio09a} catalog using a pruning length of $0\farcm 3$; different colours indicate the different sub-clusters identified by MST. The right panel is a synthetic UBV image produced by placing the ONC to a distance of 1 Mpc (assuming that the true distance is 389 pc -- \citealt{Kounkel18a}), and assuming diffraction-limited imaging with a 2.4 m-diameter telescope such as the \textit{Hubble Space Telescope}. The region shown in the right panel is the same as shown in the two other panels. Colors are scaled logarithmically to give a dynamic range of 100 in flux for all channels. 
}
\label{fig:orion_plot}
\end{figure}

In this review, however, we are interested in the formation of star clusters as well as their evolution, and thus we cannot limit ourselves to stellar populations that are older than their internal dynamical time \citep[c.f.~][]{Portegies-Zwart10a, Gieles11a}. However, younger populations often have complex internal structures such that there is no single way to decompose them into clusters. We illustrate this point with the example of the ONC in \autoref{fig:orion_plot}, where the left two panels show the results of applying two different cluster identification methods that show little agreement: a Bayesian decomposition into isothermal ellipsoids \citep[left]{Kuhn14a}, and a minimum spanning tree (middle). Numerous other decompositions are possible as well \citep{Schmeja11a}. It is not clear which of these, if any, correspond to physically meaningful concepts such as a distinction between structures with positive or negative total energy. Moreover, none of the clustering methods used on resolved stars in the Milky Way are likely to correspond to the clusters picked out by most extragalactic surveys. As the right panel of \autoref{fig:orion_plot} shows, the ONC would certainly be seen as a single cluster (were it detected at all) in a galaxy at a distance of $\approx 1$ Mpc, as targeted by the PHAT survey \citep[see Margin Note]{Dalcanton12a}, or $\sim 3-10$ Mpc, typical of the LEGUS survey \citep{Calzetti15a}. We return to the question of how to handle the ambiguity in defining clusters at young ages in \autoref{ssec:demo_limits}.

\subsubsection{Working definition}
\label{ssec:working}

Given the discussion above, we shall follow \citet{Trumpler30a} and define clusters very generally, so that they can be studied throughout their lives. For our purposes a star cluster is a group of at least 12 stars (to clearly distinguish it from a multiple star system -- \citealp{Trumpler30a}) with a mean density that (i) is at least a factor of a few times the background density (similar to the constraint adopted by \citealp{Lada03a}), with the factor being larger in larger galaxies so that the overdensity is statistically significant, and (ii) is much greater than the local density of dark matter. Such groups of stars are physically associated while they form, but there is no constraint on the formation timescale or lifetime; in contrast to \citet{Lada03a}, we do not require the cluster to be denser than the gas cloud out of which it forms, nor do we require that it have enough members to avoid evaporation for $100$ Myr.  Clusters generally dissipate in time (\autoref{sec:life_and_death}), and once they no longer satisfy our criteria we refer to them as ``dissolved clusters" (\autoref{sec:afterlife}).

Star clusters can be gravitationally bound or unbound, and young clusters can be a mixture. Unbound clusters are termed associations; bound clusters can be categorized as OCs or GCs. The most massive and young OCs are sometimes referred to as young massive clusters (YMCs) or super-star clusters (SSCs); they are sometimes also called young GCs, although it is unknown if they will in fact evolve into clusters like the GCs we observe today. YMCs have traditionally been defined by mass $>10^4\,M_\odot$ and age $<100$ Myr, but the lines are fuzzy and there is no particular physical significance to them. Similarly, OCs and GCs do not have precise definitions. In the Milky Way, most OCs have masses $\lesssim 5000\,M_\odot$ and ages $\lesssim 6$ Gyr, whereas most GCs have masses $\gtrsim 10^4\,M_\odot$ and ages $\gtrsim 6$ Gyr \citep{Kharchenko13a}, but in other galaxies these catagories have more overlap. The metallicity of OCs overlaps that of the thick-disk population of Galactic GCs, which has [Fe/H]~$>-0.8$ \citep{Zinn85a}. Clusters with ages $\lesssim$ few hundred Myr are often organized hierarchically, with OCs and YMCs being embedded in associations with sizes up to a few hundred pc, associations grouped together in complexes at $\sim$kpc scales, etc.; there is no physical division between these scales for the unbound structures \citep{Gouliermis18a}. Despite the wide range of physical properties of these various types of star clusters, there is no evidence that they do {\it not} form via the same basic mechanism (\citealp{Elmegreen97a}; \autoref{sec:birth}). The only clear intrinsic difference among these different types of clusters is that old GCs exhibit anti-correlations in some light element abundances and have multiple stellar populations, hallmarks that are {\it never} observed in OCs (\autoref{ssec:chemical_comp}). Since these phenomena, which are not understood, have been reviewed recently by \citet{Renzini2013} and \citet{Bastian2018}, we shall not discuss them further here.

\section{DEMOGRAPHICS: OBSERVATIONAL CONSTRAINTS}
\label{sec:demographics}

\begin{marginnote}[]
\entry{PDF}{probability distribution function}
\entry{Initial mass function (IMF)}{PDF of initial mass distribution of stars in a stellar ensemble}
\entry{Present day mass function (PDMF)}{PDF of observed mass distribution of stars in a stellar ensemble}
\entry{Initial cluster mass function (ICMF)}{PDF of initial mass distribution of star clusters}
\entry{Cluster mass function (CMF)}{PDF of observed mass distribution of star clusters}
\entry{Cluster age function (CAF)}{PDF of observed age distribution of star clusters}
\end{marginnote}

The physical processes that govern the formation and evolution of star clusters encode themselves in the distributions of star clusters' properties -- mass, age, size, etc. Any review of star clusters must therefore begin with the observational constraints and how they are obtained. In this section and the subsequent ones we will define a number of symbols, which we summarize for convenience in \autoref{tab:symbols}.

\begin{table}[h]
\tabcolsep7.5pt
\caption{Symbols used in this review}
\label{tab:symbols}
\begin{center}
\begin{tabular}{@{}l|l@{}}
\hline
Symbol & Meaning\\
\hline
$\alpha_M$ & Index of cluster mass distribution, $dN/dM \propto M^{\alpha_M}$ (\autoref{ssec:mass})\\
$M_c$ & Upper truncation in cluster mass function (\autoref{ssec:mass}) \\
$\alpha_T$ & Index of cluster age distribution, $dN/dT \propto T^{\alpha_T}$ (\autoref{ssec:age})\\
$\Gamma$ & Fraction of stellar mass in bound clusters (\autoref{ssec:boundfrac}) \\
$r_h$ & Cluster half-mass radius (\autoref{ssec:size}) \\
$\alpha_{\rm vir}$ & Virial ratio of a cloud or cluster (\autoref{ssec:initial_conditions}) \\
$M_J$ & Jeans mass (\autoref{ssec:initial_conditions}) \\
$t_{\rm cr}$ & Crossing time of a star cluster (\autoref{ssec:initial_conditions}) \\
$t_{\rm ff}$ & Free-fall time (\autoref{ssec:epsff}) \\
$\epsff$ & Fraction of gas converted to stars per free-fall time (\autoref{ssec:epsff}) \\
$\eta$ & Mass loading factor, i.e., ratio of mass ejected to mass converted to stars (\autoref{ssec:epsff}, \autoref{ssec:feedback}) \\
$\epsilon_*$ & Fraction of gas converted to stars at the end of cluster formation (\autoref{ssec:epsff}, \autoref{ssec:feedback}) \\
$t_{\rm sf}$ & Time over which star formation takes place (\autoref{ssec:epsff}, \autoref{ssec:tsf}) \\
$v_{\rm esc}$ & Escape speed from a star cluster (\autoref{sssec:outflows}) \\
$\Sigma_{\rm DR}$ & Surface density below which direct radiation pressure becomes important (\autoref{sssec:dr}) \\
$\Sigma_{\rm IR}$ & Surface density above which indirect radiation pressure becomes important (\autoref{sssec:ir}) \\
$\rho_{\rm SN}$ & Density above which cluster formation is complete before the first supernova (\autoref{sssec:sne}) \\
$\sigma$ & Cluster velocity dispersion (\autoref{ssec:cluster_struct}) \\
$r_c$ & \citet{King66a} or \citet{Elson87a} model cluster core radius (\autoref{ssec:cluster_struct}) \\
$r_{\rm tr}$ & \citet{King66a} model cluster truncation radius (\autoref{ssec:cluster_struct}) \\
$W_0$ & \citet{King66a} model cluster dimensionless central potential (\autoref{ssec:cluster_struct}) \\
$c$ & \citet{King66a} model cluster concentration parameter, $c = \log(r_{\rm tr}/r_c)$ (\autoref{ssec:cluster_struct}) \\
$r_{\rm ti}$ & Cluster tidal radius (\autoref{ssec:cluster_struct}) \\
$E_J$ & Jacobi energy (\autoref{ssec:cluster_struct}) \\
$t_{\rm rlx}$ & Cluster relaxation time at the half-mass radius (\autoref{sssec:relaxation}) \\
\hline
\end{tabular}
\end{center}
\begin{tabnote}
\end{tabnote}
\end{table}

\subsection{Methods}

Methods for determining the demographics of star clusters can be divided into those that operate on resolved stellar populations, and those that operate on unresolved populations. 

\subsubsection{Resolved Stellar Populations}

For star clusters in the Milky Way, the Magellanic Clouds, and M31 (using \textit{Hubble Space Telescope} resolution), it is possible to resolve individual main sequence stars down to masses of a few $M_\odot$ or less. This statement oversimplifies the situation somewhat; the luminosity or mass limit down to which it is possible to resolve stars is a complex function of the stellar density and luminosity distribution in the region being studied. It is possible to resolve massive main sequence stars in relatively sparse regions even beyond the Local Group \citep[e.g.,][]{Larsen11a}, while confusion is a problem for the most crowded regions even in the Milky Way \citep[e.g.,][]{Ascenso09a}. There exists a broad range of parameter space where stars are partially resolved, i.e., their separations are comparable to the observational point spread function (psf); the statistical and photometric techniques used to analyze such fields are beyond the scope of this review.

When individual stars can be resolved, placing them on a color-magnitude diagram (CMD) provides the most direct method of determining the properties of the parent star cluster. In principle, the CMD allows one to read off the masses of the individual stars almost directly and, if the main sequence turn-off is resolved, to infer the age of the population as well. By adding spectroscopy to photometry, one can also determine the stars' compositions. While this method is direct, it does encounter complications. In young clusters where the resolved stars are pre-main sequence, the choice of evolutionary tracks can induce significant uncertainties in the final properties \citep[e.g.,][]{Da-Rio12a}, and a non-negligible range of ages may be present, so that the system cannot be described by a single age \citep[e.g.,][]{Reggiani11a, Getman18a}. If the observations do not resolve down to the peak of the initial mass function (IMF), one may need to extrapolate using an assumed IMF to account for the mass of unresolved stars \citep[e.g.,][]{Lada03a, Beerman12a}. 

Despite these limitations, however, two other uncertainties are more important. One is the error introduced by the need to correct demographics for clusters that are too small or too old to make it into the catalog. For resolved observations, catalog completeness is invariably a function not just of the mass and age of the stellar population, but also its concentration, the exact masses and ages of its most luminous few stars, and the density of the background \citep{Silva-Villa11a, Johnson15a}. The other source of uncertainty is in assignment of stars to clusters. As already discussed in \autoref{ssec:definition}, for populations that are not old enough to have relaxed, there may be no unique way to decompose a collection of stars into clusters, and variations in how one defines a cluster can lead to order of magnitude differences in the inferred number of clusters and their properties \citep{Bressert10a}. However, even for somewhat older populations, catalogs constructed using different methods can easily differ by $\sim 10-20\%$ even when the underlying data are identical \citep{Johnson15a}. These errors are generally larger than the uncertainties on the properties of individual clusters. 


\subsubsection{Unresolved Populations}
\label{ssec:unresolved_pop}

While CMDs offer the most reliable method for determining star cluster properties, at distances $\gtrsim 1$ Mpc it is not possible to resolve any but the most luminous individual stars. Consequently, studies that probe a wide range of environments must use unresolved populations. In an unresolved observation, a star cluster is distinguished from a single star or other point source either by the fact that its light is more extended than the observational psf (for targets at distances below tens of Mpc) or that it is too luminous to be a single massive star (for more distant targets). In the earliest work with unresolved populations authors simply analyzed luminosity functions \citep[e.g.,][]{Whitmore93a, Whitmore99a}, but one can also infer physical properties using simple stellar population (SSP) models. In practice this procedure can be carried out in several different ways. The simplest is to generate theoretical luminosities as a function of mass and age by assuming a stellar population that fully samples the IMF (including models for nebular emission and dust extinction) and find the mass and age that best match the observations using $\chi^2$ or a similar goodness-of-fit statistic \citep[e.g.,][]{Zhang99b, Adamo10a}. More complex methods take into account the stochastic fluctuations in light output that occur as a result of partial sampling of the IMF in clusters smaller than $\sim 10^4$ $M_\odot$ \citep[e.g.,][]{Fouesneau10a,Popescu12a,de-Meulenaer13a,Krumholz15c} or add narrowband photometry targeting nebular lines (usually H$\alpha$) to provide better age discrimination \citep[e.g.,][]{Fouesneau12a, Bastian12a, Chandar16a}.

\begin{figure}
\includegraphics[width=\textwidth]{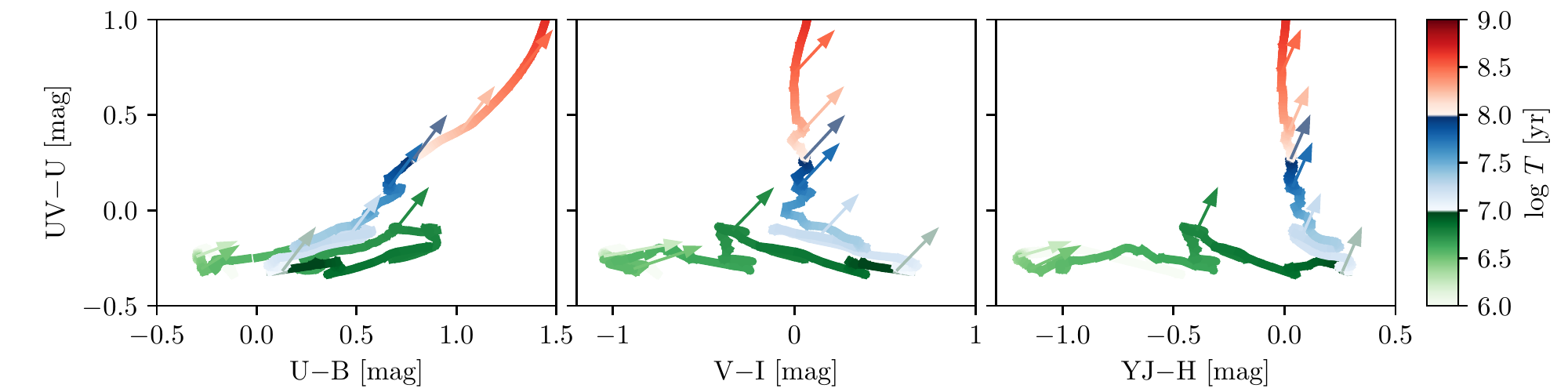}
\caption{
\label{fig:color_evol}
Color evolution of a simple stellar population with a fully sampled IMF, including nebular emission. In each panel the line shows the intrinsic color (in AB mag) as a function of population age. Arrows, placed at intervals of $0.25$ dex in age starting at $\log (T/\mbox{yr}) = 6.25$, show observed color for a visual extinction $A_V = 0.5$ mag for the stars, $A_V = 1.05$ mag for nebular light. This calculation uses \texttt{SLUG} stellar population synthesis code \citep{da-Silva12a, Krumholz15b}, run with a \citet{Chabrier05a} IMF, Solar metallicity, MIST rotating stellar evolution models \citep{Choi16a}, starburst99 atmosphere models \citep{Leitherer99a}, a \citet{Calzetti00a} starburst dust attenuation curve, and HST/WFC3 filters. However, other choices for these parameters yield qualitatively similar results.
}
\end{figure}

Fitting star cluster properties from photometry leads to substantially larger errors than doing so from resolved measurements of individual stars, with the magnitude of the error depending on the locus in color space. We illustrate this point in \autoref{fig:color_evol}, which shows the trajectory that an SSP traces out in color over its lifetime. The process of inferring cluster age and extinction essentially amounts to placing an observed point on a multidimensional color-color plot such as \autoref{fig:color_evol}, and finding the location on an extinction-adjusted SSP track such as the one shown that gets as close as possible to the observed point in every dimension. As is apparent from the figure, for ages below $\approx 5$ Myr or above $\approx 50$ Myr, in at least some color combinations, SSP tracks are relatively straight and not parallel to the extinction vector. Consequently, the matching process is straightforward and there is little ambiguity in the determination of ages; comparing photometric ages to those inferred from CMDs, \citet{Elson85a, Elson88a} estimate that photometric ages for massive clusters are accurate to $\approx 0.3$ dex. At ages of $\approx 5 - 50$ Myr, however, SSP colors oscillate rapidly and tracks in color space frequently cross themselves, making age determinations quite uncertain. For example, notice that SSPs with ages of $\approx 8$ Myr (dark green in the figure) have very similar colors to SSPs with ages of $\approx 30$ Myr (light blue in the figure), and that the separation between them is mostly along the extinction vector, making these two possibilities difficult to disentangle, particularly once stochastic color variation due to finite IMF sampling is taken into account (also see \citealt{Maiz-Apellaniz09a} and \citealt{Krumholz15c}). 

\begin{figure}
\includegraphics[width=\textwidth]{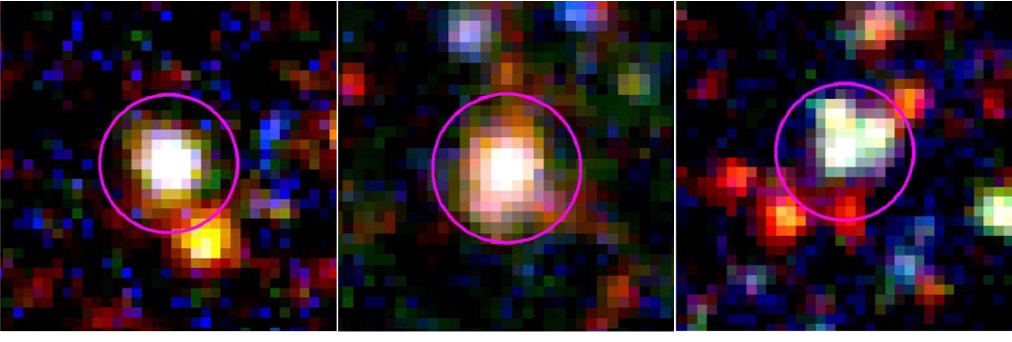}
\caption{
\label{fig:adamo_classes}
Examples of unresolved clusters of different morphological classes. Each panel shows a three-color UBV image of a star cluster in NGC 628, from \citet{Adamo17a}. The ring shows a radius of $0\farcs 28$, approximately $13.4$ pc at the distance of NGC 628. The morphologies are classified as, from left to right, compact and symmetric, compact and asymmetric, and multiply-peaked. An ``exclusive'' catalog, in the sense used in this review, would include the left two objects but exclude the one on the right, while an ``inclusive'' catalog would include all three. A comparison to \autoref{fig:orion_plot} suggests that the ONC might well be excluded from an ``exclusive'' catalog.
}
\end{figure}

\begin{marginnote}[]
\entry{Exclusive catalog}{a star cluster catalog that excludes objects that do not pass morphological tests for compactness and roundness}
\entry{Inclusive catalog}{a star cluster catalog without additional morphological requirements}
\end{marginnote}

On top of these errors that apply to individual clusters, catalog construction and completeness correction induce additional errors on measurements of population demographics. Completeness correction for unresolved observations faces essentially the same challenges as for resolved data. Recently \citet{Krumholz18a} proposed a method to carry out full forward-modelling of an entire photometric catalog, simultaneously accounting for completeness and stochasticity, but this method is thus far experimental. Catalog construction is the more serious issue, because it raises the question of how one should handle sources where the light distribution is multiply-peaked or contains significant color gradients, as illustrated in \autoref{fig:adamo_classes}. Some authors, who limit the definition of cluster to dynamically-relaxed structures (c.f.~\autoref{ssec:definition}), discard such sources on the grounds that they are not relaxed \citep[e.g.,][]{Bastian11a, Bastian12a}, while others adopt a broader definition and thus retain them \citep[e.g.,][]{Chandar10b, Chandar11a, Chandar14a}; some split the difference by including all objects but then reporting separate analyses for objects of differing morphology \citep[e.g.,][]{Adamo17a}. These different methods of catalog construction have historically led to extreme confusion in interpretation of observations, since inclusive catalogs tend to have $\approx 30-50\%$ more members in the youngest age bin than exclusive ones, with much smaller differences at older ages \citep{Chandar14a, Messa18a}. In the discussion that follows, we will minimize this confusion by identifying whether a particular catalog was constructed using criteria that are ``exclusive'' (i.e., the catalog excludes non-symmetric objects) or ``inclusive'', and pointing out when the choice of one method or the other leads to systematic differences. However, we caution that all existing cluster catalogs for galaxies at distances $\gtrsim 20$ Mpc are necessarily inclusive, because limited resolution makes morphological measurement impossible in such distant samples. 

Authors who build exclusive catalogs generally assume that morphology can be used as a proxy for boundedness. This likely holds in one direction: since star-forming regions are morphologically complex, if the observed stars have a round, compact morphology, it is likely that they have relaxed into it and thus are bound. However, the converse need not be true, i.e., there is no reason to assume that a population that is too young to have dispersed is unbound simply because it is not round. The implication is that, for young stellar populations, an exclusive approach probably omits everything that is not bound, but also discards some unknown number of bound systems. By contrast, an inclusive catalog captures all structures above some luminosity threshold without regard to their dynamical state.

\subsection{Mass distribution}
\label{ssec:mass}

\begin{figure}
\includegraphics[width=\textwidth]{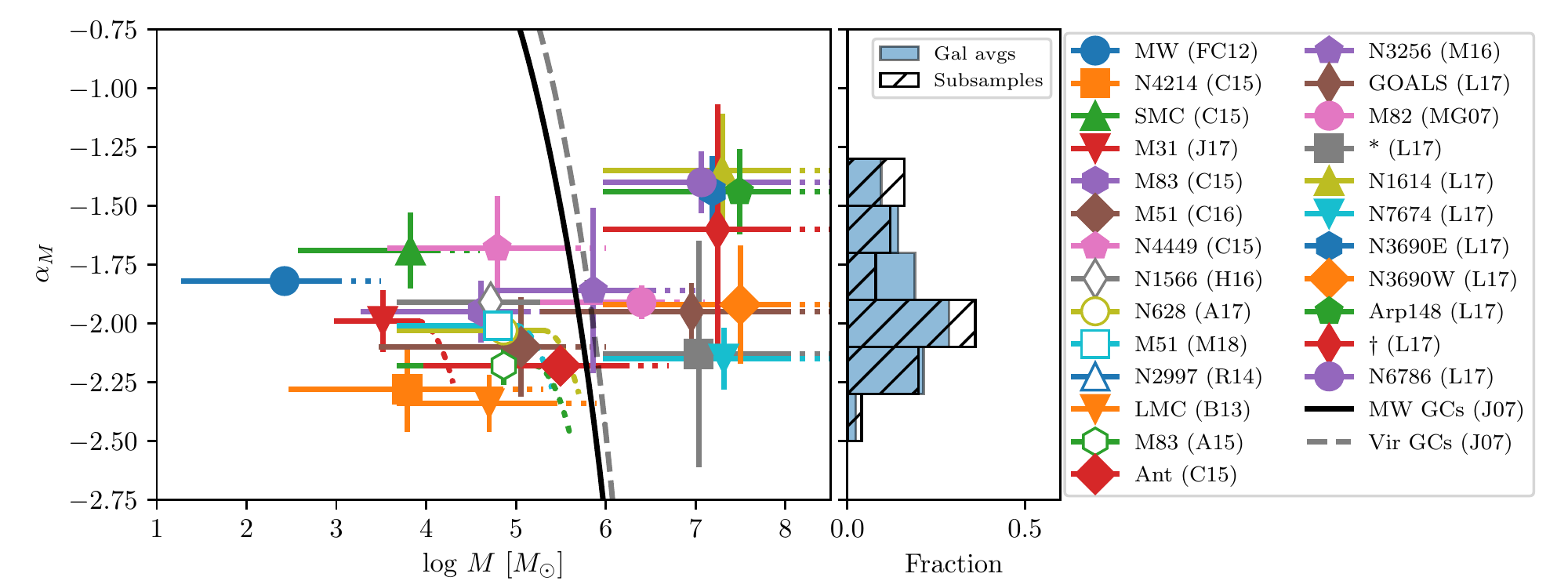}
\caption{
\label{fig:cmf}
Compilation of measured CMF slopes $\alpha_M$. In the main panel, colored points represent clusters in the disks of nearby galaxies, and solid and dashed black lines are evolved Schechter function fits to the Milky Way's GCs and to an average of GCs in Virgo, respectively. The horizontal error bars show the range in cluster mass over which the measurement was made, with open points indicating measurements from exclusive catalogs and filled points indicating inclusive catalogs. Points are displaced slightly from the centers of their corresponding mass ranges to minimize confusion. A straight dotted line on the high mass end of the horizontal error bar indicates that the observation did not detect a truncation to the CMF, while a downwards curve indicates an observation that reported a truncation. For the GC mass functions, $\alpha_M$ is not constant in mass, and the line shows $\alpha_M$ versus mass for the best-fitting evolved Schechter function. In the main panel only whole-galaxy average measurements are shown. The histogram on the right shows the distribution of measured CMF slopes for disk clusters; blue is the distribution of whole-galaxy average measurements, while hatched shows measurements in which galaxies are broken into independent sub-samples. In the legend, N is short for NGC, and the galaxies listed as $*$, $\dagger$, and GOALS are UGC09618NED02, IRAS20351+2521, and an average of 22 LIRGs from the GOALS sample, respectively. References listed in the legend are as follows: MG07 = \citet{McCrady07a}, J07 = \citet{Jordan07a}, FC12 = \citet{Fall12a}, B13 = \citet{Baumgardt13a}, R14 = \citet{Ryon14a}, C15 = \citet{Chandar15a}, A15 = \citet{Adamo15a}, C16 = \citet{Chandar16a}, H16 = \citet{Hollyhead16a}, M16 = \citet{Mulia16a}, J17 = \citet{Johnson17b}, A17 = \citet{Adamo17a}, L17 = \citet{Linden17a}, M18 = \citet{Messa18a}. The hatched histogram on the right also contains sub-samples from \citet{Messa18b}. 
}
\end{figure}

The most basic property of a star cluster is its mass, and thus the most basic distribution for star clusters within a galaxy is the observed cluster mass function (CMF).\footnote{Mass as used to define the CMF, and as we shall use the term throughout \autoref{sec:demographics}, refers to the mass that the stellar population would have had before mass loss due to stellar evolution.} \autoref{fig:cmf} summarises recent measurements for the CMFs of star clusters in the disks of Local Group galaxies. All observed disk CMFs appear to be reasonably well-described by a power-law $dN/dM \propto M^{\alpha_M}$, with values of $\alpha_M = -2 \pm 0.2$.\footnote{It is common in the cluster literature to use the letter $\beta$ rather than $\alpha_M$ to denote the CMF index. We use $\alpha_M$, and $\alpha_T$ for the analogous index of the cluster age function (\autoref{ssec:age}), because while there is general agreement on the usage of $\beta$, there is no equivalent agreement on the symbol used to represent the cluster age function index. On the contrary, some groups use $\gamma$ for this index, while others use $\gamma$ to denote a completely unrelated quantity.} A slope of $\alpha_M = -2$ corresponds to equal mass per logarithmic bin, and thus is the expected slope for a completely scale-free distribution. In normal star-forming galaxies, the mass function slope is generally measured over the range from $\approx 10^3 - 10^5$ $M_\odot$. The lower limit on this range is entirely a function of observational limitations -- only in a few cases do we have cluster samples that are complete enough below $\approx 10^3$ $M_\odot$ to enable measurement of a CMF, and in those cases the data are consistent with $\alpha_M \approx -2$ down to the completeness limit. Indeed, some authors have claimed that the power law continues to masses as small as $\approx 20$ $M_\odot$, corresponding to individual massive stars \citep[e.g.,][]{Lamb10a}.

The nature of the upper limit is less clear. In most galaxies, the total number of clusters is such that very few with masses $\gtrsim 10^5$ $M_\odot$ would be expected even for a pure power-law mass function, and in galaxies with cluster populations large enough that such clusters would be expected -- the Antennae, M82, and various luminous infrared galaxies (LIRGs) -- they are found. Consistent with this, the luminosity of the brightest cluster in a galaxy is well correlated with the luminosity of the galaxy and its star formation rate \citep{Larsen02a}. However, some authors report evidence that the observed CMF or luminosity function is better fit by power-law that is truncated at a mass $M_c$ (either a hard truncation or a Schechter function, $dN/dM \propto M^{\alpha_M} e^{-M/M_c}$) than a pure power-law \citep[e.g.,][]{Bastian08a, Larsen09a, Bastian12a, Adamo15a, Adamo17a}; others question the statistical robustness of this conclusion \citep{Mok18a}. Any truncation would need to vary systematically with some other galaxy property (e.g., the Jeans mass in the disk, \autoref{ssec:giant}) in order to explain the presence of more massive clusters in more luminous galaxies.

The question of whether there is a truncation has been difficult to settle due to the challenge of deriving realistic confidence intervals from observations where the dominant errors are catalog construction and translation from photometry to cluster masses; some authors using nearly identical data sets nonetheless reach differing conclusions about whether they provide evidence for a truncation (e.g., \citealt{Adamo15a} versus \citealt{Sun16a} on M83), and the results can be sensitive to the statistical power of the fitting method \citep{Messa18b}. The most convincing case for a deviation from a pure power-law mass function is in M31, where \citet{Johnson17b} robustly identify a truncation mass of $M_c \approx 8.5\times 10^3$ $M_\odot$ based on a sample of $\approx 1000$ clusters with masses determined from CMDs rather than unresolved photometry. However, this survey does not cover some of the more actively star-forming parts of M31, and it is possible these might host clusters larger than $M_c$. There is also indirect evidence for a truncated CMF in the Milky Way, in the form of an observed truncation in the luminosity function for radio recombination line emission from H~\textsc{ii} regions \citep{McKee97a}, although translating this into a truncation mass is not straightforward.

Values of $\alpha_M$ tend to be slightly larger for starbursting galaxies than for quiescent spirals. However, it is unclear if this represents a real physical difference or an observational artifact: the starburst galaxies are located at systematically larger distances, and thus observations of them have systematically lower spatial resolution, which will tend to flatten the mass function by blurring multiple small objects into a single large one. This matter will be ``resolved'' in the next decade by AO-assisted imagers and spectrographs on extremely large telescopes.

The mass function for the GCs found in the halos of galaxies is substantially different. It has a clear turnover mass below which the number of clusters is either flat or increasing with mass. The distribution can be reasonably well described as a lognormal with a characteristic mass of $\approx 10^{5.3}$ $M_\odot$ \citep{Brodie06a}. However, a more physically-motivated description is the evolved Schechter function \citep{Fall01a, Jordan07a}, which is simply a standard Schechter function modified by the assumption that all clusters have lost a fixed amount of mass $\Delta$; we discuss the physical motivation for this model in \autoref{sssec:relaxation}. This gives a distribution $dN/dM \propto \left(M+\Delta\right)^{\alpha_{M,i}} \exp\left[-\left(M+\Delta\right)/M_c\right]$, where $\alpha_{M,i}$ is the slope at formation. The corresponding present-day slope is
\begin{equation}
\alpha_M = \frac{d}{d\log M}\left(\log \frac{dN}{dM}\right)
= \frac{-M\left(M-\alpha_{M,i} M_c + \Delta\right)}{M_c\left(M+\Delta\right)},
\end{equation}
In \autoref{fig:cmf}, we show $\alpha_M$ for the Milky Way GC system and for the average of GC systems in the Virgo cluster from \citet{Jordan07a};\footnote{\citet{Jordan07a} give their fits in terms of the $z$-band luminosity. We have converted these values to masses using their favored $z$-band mass to light ratio. The line shown in \autoref{fig:cmf} corresponds to $M_c$ and $\Delta$ values equal to the average of the values given in their Table 3, where $\alpha_{M,i} = -2$ in their fits. Also note that, unlike elsewhere in this section, we have not corrected for mass loss due to stellar evolution. Doing so would shift the distribution to the right by a factor of $\approx 1.5-2$.} the best-fit $M_c$ and $\Delta$ values are both in the range $10^{5.3} - 10^{6.6}$ $M_\odot$, spanning the range of truncation masses inferred for disk clusters. However, the functional forms of the GC and disk cluster mass functions are clearly very different, and the relationship between the two is a topic to which we shall return below.

\subsection{Age distribution}
\label{ssec:age}

\begin{figure}
\includegraphics[width=\textwidth]{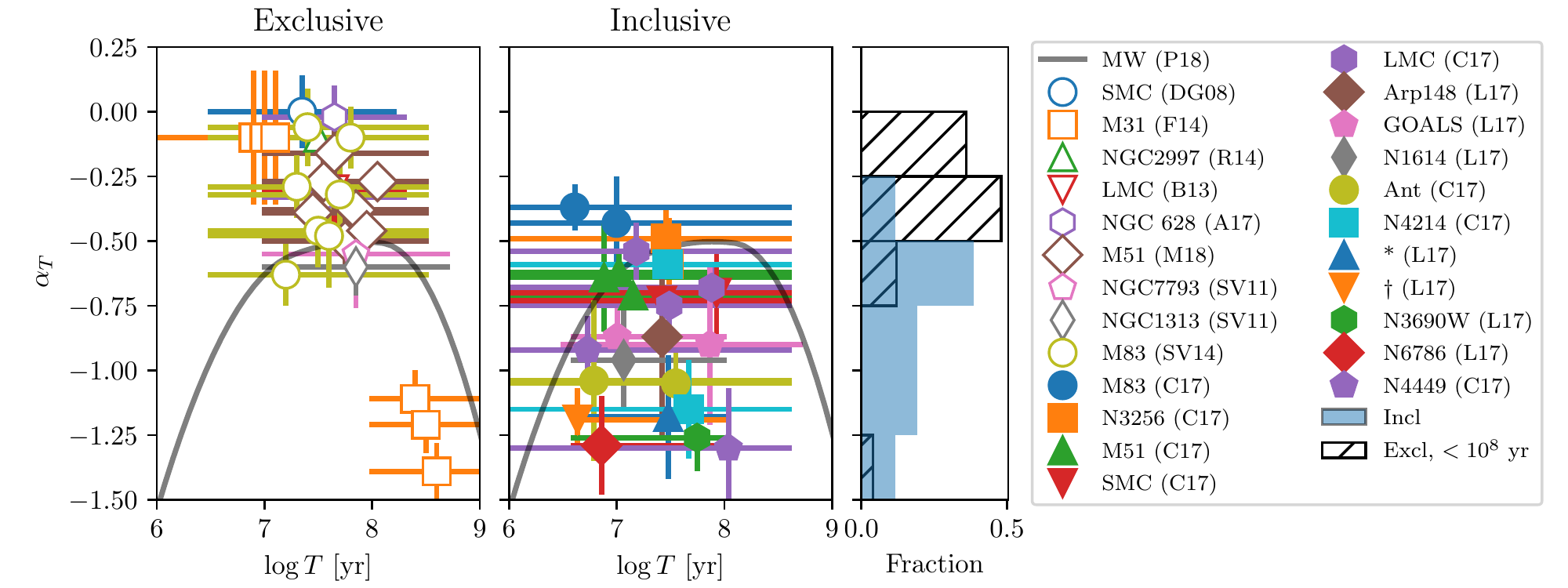}
\caption{
\label{fig:caf}
Observational determinations of the CAF in galaxy disks. The horizontal range indicates the age interval over which the measurement was made, while the value plotted in the vertical direction is the index $\alpha_T$ ($dN/dT \propto T^{\alpha_T}$) with error bars. For extragalactic data, open points correspond to exclusive catalogs (left panel), filled points to inclusive ones (middle panel). The galaxy and source are as indicated in the legend (using the same abbreviations as in \autoref{fig:cmf}), and multiple points for a single galaxy correspond to subsamples of clusters in that galaxy, separated either by mass or by location; we have added a small random displacement to the horizontal positions of points to prevent them from overlapping. The black line shown in both the left and middle panels is a smoothed basis spline fit to the catalog of \citet{Piskunov18a} for clusters within 2 kpc in the Sun. The right panel shows a histogram of $\alpha_T$ values for the extragalactic catalogs, with inclusive catalogs and exclusive catalogs for ages below $10^8$ yr shown separately. References are as follows: F14 = \citet{Fouesneau14a}, SV14 = \citet{Silva-Villa14a}, R14 = \citet{Ryon14a}, C17 = \citet{Chandar17a}, A17 = \citet{Adamo17a}, L17 = \citet{Linden17a}, M18 = \citet{Messa18b, Messa18a}, P18 = \citet{Piskunov18a}.
 }
\end{figure}

A second distribution of equal importance to the CMF is the age distribution of star clusters, the cluster age function (CAF). For Milky Way GCs this distribution is centred at ages $\sim 10$ Gyr with a width of a few Gyr \citep{Bastian2018}. For clusters in disks, observers generally fit the CAF as a power-law $dN/dT \propto T^{\alpha_T}$, where $T$ is the cluster age\footnote{Throughout this review we shall use $T$ to denote cluster ages, and $t$ for all other time quantities.}; some observers use a single $\alpha_T$ for all clusters, while others consider more complex forms where $\alpha_T$ can be different for different age or mass ranges. However, in practice most extragalactic samples cover a small enough dynamic range in mass and age that one can define at most two or three bins with potentially differing $\alpha_T$ values.

In a galaxy with a constant cluster formation rate, $\alpha_T$ is a measure of how long clusters survive. Consider a simple model, in the spirit of \citet{Boutloukos03a}, where the probability per unit time that a cluster of age $T$ is destroyed is $1/aT$ for some constant factor $a$. If we consider a cohort of clusters at a single age, our assumed destruction rate implies that their number will decrease with time as $dN/dT = - N/aT$. Integrating, we find that the number that survive to age $T$ is $N = N_0 (T/T_0)^{-1/a}$, where $N_0$ is the number of clusters at age $T_0$. This corresponds to a CAF with $\alpha_T = -1/a$. Thus $\alpha_T \rightarrow 0$ corresponds to the case $a\gg 1$, i.e., clusters survive for many times their current age, while $\alpha_T = -1$ corresponds to $a=1$, i.e., the expected time required to destroy a cluster is equal to its current age.

While the interpretation of $\alpha_T$ is straightforward, divining a consistent value of it from the published literature is much more difficult. In part this is because different groups tend to subdivide the sample to be fit in different ways, and to reach divergent conclusions both about the value of $\alpha_T$ and whether it is different at different ranges of cluster age or mass. \autoref{fig:caf} shows a summary of recent observational determinations for extragalactic systems, along with our fit to the completeness-corrected catalog of \citet{Piskunov18a} for Milky Way clusters within $\approx 2$ kpc of the Sun.\footnote{\citet{Piskunov18a} report cluster density in age bins. To derive the curve shown for $\alpha_T$ versus age, we computed a basis spline fit to their data and calculated the derivative from it. The source code for our calculation is provided in the Supplementary Material.} 

\begin{figure}
\includegraphics[width=\textwidth]{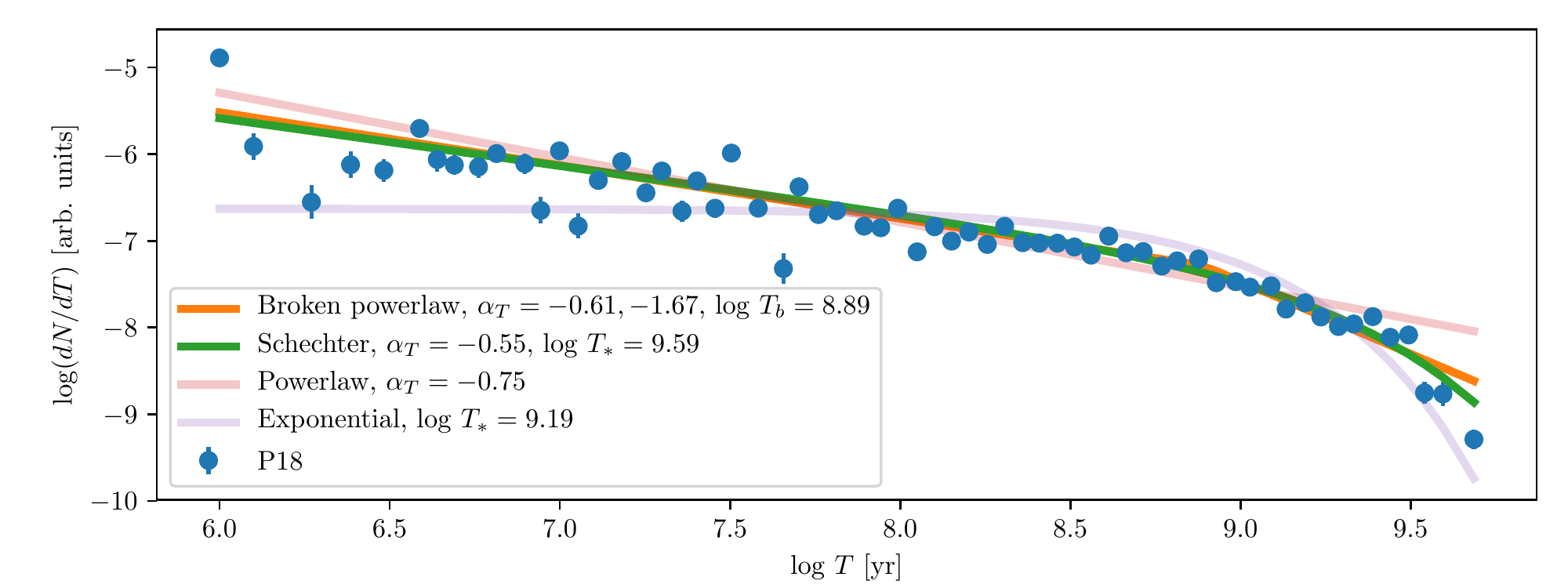}
\caption{
\label{fig:mw_age_dist}
CAF for star clusters within $\approx 2$ kpc of the Sun. Points with error bars show measurements from \citet[P18]{Piskunov18a}. Lines show $\chi^2$ fits to several functional forms, with parameters indicated in the legend: broken power-law (orange; $dN/dT \propto T^{\alpha_{T,1}}$ for $\log T < \log T_b$, and $dN/dT \propto T^{\alpha_{T,2}}$ for $\log T > \log T_b$), Schechter function (green; $dN/dT \propto T^{\alpha_T} e^{-T/T_*}$), single power-law (faded red; $dN/dT\propto T^{\alpha_T}$), and exponential (faded purple; $dN/dT\propto e^{-T/T_*}$). The broken power-law and Schechter forms both have reasonable $\chi^2$ values, while the other fits are poor.
}
\end{figure}

\autoref{fig:caf} shows that there is a systematic difference in slopes between inclusive and exclusive catalogs at young ages. The fits from exclusive catalogs at $T\lesssim 10^8$ yr have $\alpha_T \approx -0.3$ to $-0.2$, while the inclusive catalogs cluster around $\alpha_T \approx -1$ to $-0.7$. Thus exclusive catalogs suggest that clusters have long survival times compared to their current ages, while inclusive catalogs imply survival times comparable to cluster ages. In addition, although this is not shown in the figure, analysis of exclusive catalogs suggests that more massive clusters have larger values of $\alpha_T$ than less massive ones \citep[e.g.,][]{Silva-Villa14a, Adamo17a, Messa18a}, implying that cluster survival time is mass-dependent. A second, more subtle dependence worth noting is that starburst galaxies seem to have a slight bias toward slopes near $\alpha_T = -1$ compared to normal spirals and dwarfs -- for example, the GOALS galaxies and the Antennae all have $\alpha_T \lesssim -0.9$, while, even for inclusive catalogs, the mean slope for spirals is closer to $\alpha_T \approx -0.7$. A similar increase in $\alpha_T$ is seen in the centers of massive spirals compared to their outskirts \citep[e.g.,][]{Silva-Villa14a, Messa18b}. While such an environmental dependence is in accord with some theoretical expectations \citep[e.g.,][]{Kruijssen11a}, one should be cautious in putting too much weight on the observations. The regions with steeper $\alpha_T$ values are on average more crowded and distant than the systems with shallower $\alpha_T$, and both crowding and distance could create a systematic bias in measured color that would manifest as a difference in the inferred age distribution.

Milky Way clusters show a complex dependence of $\alpha_T$ on age, which we highlight in \autoref{fig:mw_age_dist}. In the Milky Way, $\alpha_T \approx -0.5$ at ages $\lesssim 10^9$ yr, indicating moderately strong cluster destruction. The age distribution steepens sharply above $\approx 10^9$ yr, indicating much more rapid disruption. This change in slope is not seen in the extragalactic samples, which are generally limited to clusters younger than $\approx 10^8$ yr for reasons of sensitivity. However, there are also two additional cautions to be mentioned in comparing the Milky Way and extragalactic samples. First, because the Milky Way sample is limited to $\lesssim 2$ kpc from the Sun, it consists entirely of low-mass clusters ($M \lesssim 10^3$ $M_\odot$), while the extragalactic sample is for much larger masses, $M \gtrsim 10^{3.5}$ $M_\odot$. Second, the extragalactic sample ages are derived from photometry, while the Milky Way ages are based on CMDs, which may lead to systematic differences. CMDs are more reliable in general, but this might not hold for the Milky Way sample because many of the clusters in it contain only a small number of stars bright enough to allow placement on the CMD, leading to large age uncertainties.

\subsection{Bound mass fraction}
\label{ssec:boundfrac}

A third basic statistic for clusters is the fraction of the total stellar mass in gravitationally-bound clusters, denoted $\Gamma$. Since this fraction changes with stellar age unless $\alpha_T = 0$, $\Gamma$ in general is a function of $T$. To measure $\Gamma$ one must determine the total mass in bound clusters and the total mass in all stars within the same age interval. Measuring the former invariably involves some degree of extrapolation to account for the mass of clusters too small to detect, but for $\alpha_M \approx -2$ the extrapolation is fairly modest. For the latter quantity, at ages above $\sim 10$ Myr the most reliable means of determining total stellar mass is from CMDs of field stars. However, these are only available for relatively nearby sources. A second-best option is to estimate the total stellar mass by multiplying the star formation rate by the length of the age interval, assuming the star formation rate has been constant. Before proceeding, we note that several authors describe $\Gamma$ as the ``cluster formation efficiency'' \citep[e.g.,][]{Larsen00a, Bastian08a, Goddard10a, Kruijssen12a}, meaning the fraction of stars formed in bound clusters. This definition implicitly assumes that one one can cleanly distinguish bound from unbound at all cluster ages, and that $\alpha_T \approx 0$ so that $\Gamma$ is the same when measured for any age range. We have argued neither of these assumptions is strictly correct, but if one makes them, our more general definition of $\Gamma$ reduces to theirs.

\begin{figure}
\includegraphics[width=\textwidth]{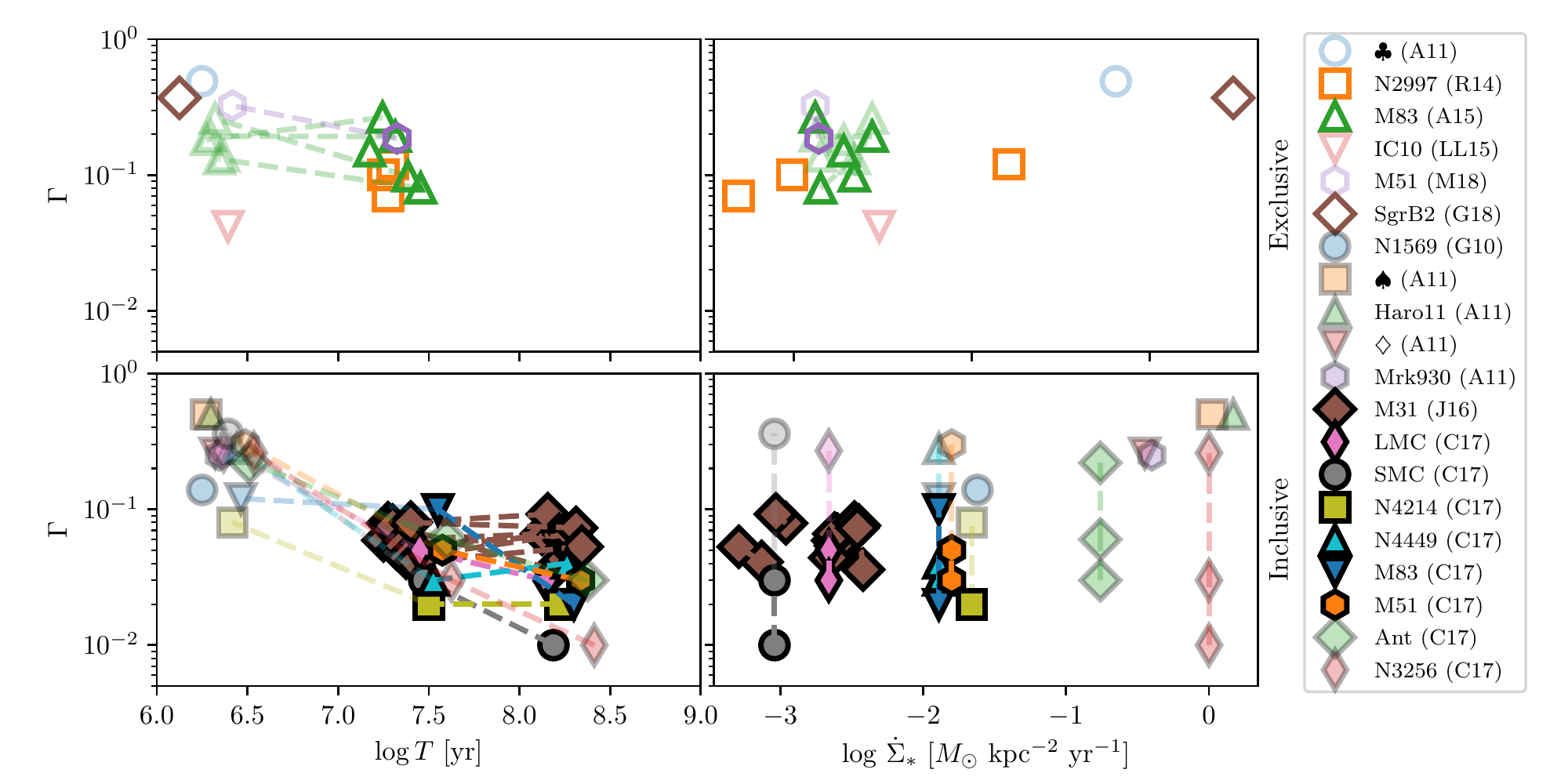}
\caption{
\label{fig:gamma}
Observational estimates of the bound mass fraction $\Gamma$. The top row shows values of $\Gamma$ taken from exclusive catalogs, while the bottom row shows inclusive catalogs. In the left column we show $\Gamma$ as a function of cluster age $T$, while in the right column we show it as a function of galaxy star formation rate $\dot{\Sigma}_*$. Measurements of $\Gamma$ in multiple age ranges for the same galaxy are connected by dashed lines. Faded points indicate measurements that are potentially dubious: either measurements of $\Gamma$ at ages $<10$ Myr where it is not feasible to separate bound from unbound systems, or measurements of $\Gamma$ for ages $>10$ Myr in starburst systems where the star formation rate may have fluctuated on this time scale. In the legend, $\clubsuit$, $\spadesuit$, and $\diamondsuit$ stand for SBS0335-052E, ESO338-IG04, and ESO185-IG13, respectively. References are as follows: G10 = \citet{Goddard10a}, A11 = \citet{Adamo11a}, R14 = \citet{Ryon14a}, A15 = \citet{Adamo15a}, LL15 = \citet{Lim15a}, J16 = \citet{Johnson16a}, C17 = \citet{Chandar17a}, M18 = \citet{Messa18a}, G18 = \citet{Ginsburg18a}.
}
\end{figure}

As with $\alpha_T$, measurements of $\Gamma$ in the literature span a wide range of values. We summarize the currently available set of measurements in \autoref{fig:gamma}. These data must be examined with extreme caution, even in comparison to compilations of $\alpha_M$ and $\alpha_T$. The most secure measurements, for steadily star-forming galaxies at ages above $\approx 10$ Myr, generally show $\Gamma \approx 0.01 - 0.1$, consistent with the observation that the majority of early B stars are found outside clusters, as is most of the UV flux \citep{Pellerin07a}. The low value of $\Gamma$ at ages $\gtrsim 10$ Myr is sometimes referred to as ``infant mortality'': essentially all stars are born in clusters (using our expansive definition of cluster that includes both bound and unbound systems), but a majority of the stars are not gravitationally bound, so by ages of a few tens of Myr, most stars are not members of clusters any more.

As shown in \autoref{fig:gamma}, a number of authors have reported measurements of $\Gamma$ at younger ages and in galaxies undergoing starbursts, and have attempted to deduce trends in $\Gamma$ with either age \citep[e.g.,][]{Chandar17a} or star formation activity \citep[e.g.,][]{Goddard10a, Adamo11a}. We regard such claims as dubious given the methodological uncertainties, and in \autoref{fig:gamma} we fade the points for which significant methodological concerns exist. With regard to age trends, as pointed out by \citet{Kruijssen16a}, estimates of $\Gamma$ from inclusive catalogs at ages below $\approx 10$ Myr are likely to be heavily contaminated by the presence of non-bound structures that have simply not yet had time to disperse. Conversely, however, exclusive catalogs may be missing a substantial population of bound but not yet relaxed clusters at similar ages. Thus $\Gamma$ estimates at ages below 10 Myr derived from inclusive and exclusive catalogs should be viewed as upper and lower limits, respectively. An exception is the point from \citet{Ginsburg18a} for the Sgr B2 region in the Milky Way's Central Molecular Zone, for which the authors check boundedness directly using stellar velocities estimates from the radio recombination lines of the ultracompact H~\textsc{ii} regions around each massive star.

Possible trends in $\Gamma$ with star formation activity are also potentially contaminated by bias. While $\Gamma$ measurements in low-star formation rate (SFR) galaxies tend to be made at ages $\gtrsim 10$ Myr to avoid the issues with inclusive versus exclusive catalogs, measurements for high-SFR galaxies are almost exclusively at younger ages. This is both because high SFR galaxies also tend to be merging systems, for which measuring $\Gamma$ at older ages is problematic because the total SFR may be variable, and because high SFR galaxies are rare and thus tend to be distant, resulting in a magnitude limit that precludes measurements of $\Gamma$ at ages $\gtrsim 10$ Myr \citep{Chandar17a}. 
This means that there is a degeneracy between two possibilities: high values of $\Gamma$ in starbursts could be because $\Gamma$ is in fact larger at higher SFR, but it could equally well be that $\Gamma$ is higher at younger ages. The Sgr B2 point from \citet{Ginsburg18a} does not suffer from this methodological problem, but also unfortunately does not help break the degeneracy either, because it has both high SFR and young age. It is unclear to which of these factors its high value of $\Gamma$ should be attributed. Consequently, we conclude that there is not yet convincing evidence that $\Gamma$ varies with either age or SFR.

\begin{textbox}[h]
\section{IMPROVING PHOTOMETRIC DETERMINATIONS OF CLUSTER PROPERTIES}
The preceding sections should make clear that one of the significant uncertainties for CMFs, CAFs, and $\Gamma$ are the difficulties of assigning masses and ages to clusters based on photometry. Progress will require an extensive effort ``ground-truthing'' photometric property determinations against measurements either CMDs or spectroscopy. Such a systematic comparison has not been performed since the work of \citet{Elson85a, Elson88a} for the Magellanic Clouds using ground-based data. Much more accurate work with space-based CMDs should now be possible in both the Clouds and M31 \citep{Johnson16a}. Similarly, spectroscopy can be used to break the degeneracy between ages of $\approx 5-10$ and $\approx 50$ Myr from photometry; spectroscopic measurements of a small sample of ambiguous clusters would at a minimum provide a useful prior for Bayesian analysis methods that generally return bimodal posterior PDFs when applied to clusters in the ambiguous parts of color space \citep{Fouesneau14a, Krumholz15c}. Unfortunately the improved infrared photometry provided by the \textit{James Webb Space Telescope} is likely to be of limited use, since clusters' colors in the IR are nearly constant at ages $\gtrsim 6$ Myr, when they become dominated by red giants and supergiants \citep{Gazak14a}.
\end{textbox}

\subsection{Size and density}
\label{ssec:size}

In addition to masses and ages, we can also study the physical sizes of star clusters, or equivalently their densities. The data on the distribution of cluster sizes are more limited than those available for mass or age, because measuring a size is obviously only possible if the object in question is at least marginally resolved. While the most physically-meaningful characterization of cluster size and density would be a half-mass radius or similar description of how mass is distributed, the limited resolution available for galaxies beyond the Milky Way and its closest neighbors generally precludes measuring anything other than the projected half-light radius, defined as the projected radius containing half the cluster light. Even this quantity is difficult to estimate for targets more distant than M31, because many clusters have relatively shallow light profiles where much of the light is in an extended halo (see \autoref{ssec:cluster_struct}) that tends to be lost against the background. For resolved stellar populations, on the other hand, it is usually possible to extract multiple structural parameters from fits of stellar number counts to theoretically-motivated profiles; see \autoref{ssec:cluster_struct} and Section 1.3.2 of \citet{Portegies-Zwart10a} for a summary of the various size parameters that can be defined for star clusters. Here we focus on projected half-mass radii (or half-number radii if masses are not available) from the resolved samples, as these are the closest to what is available from the photometric size measurements.

\begin{figure}
\includegraphics[width=\textwidth]{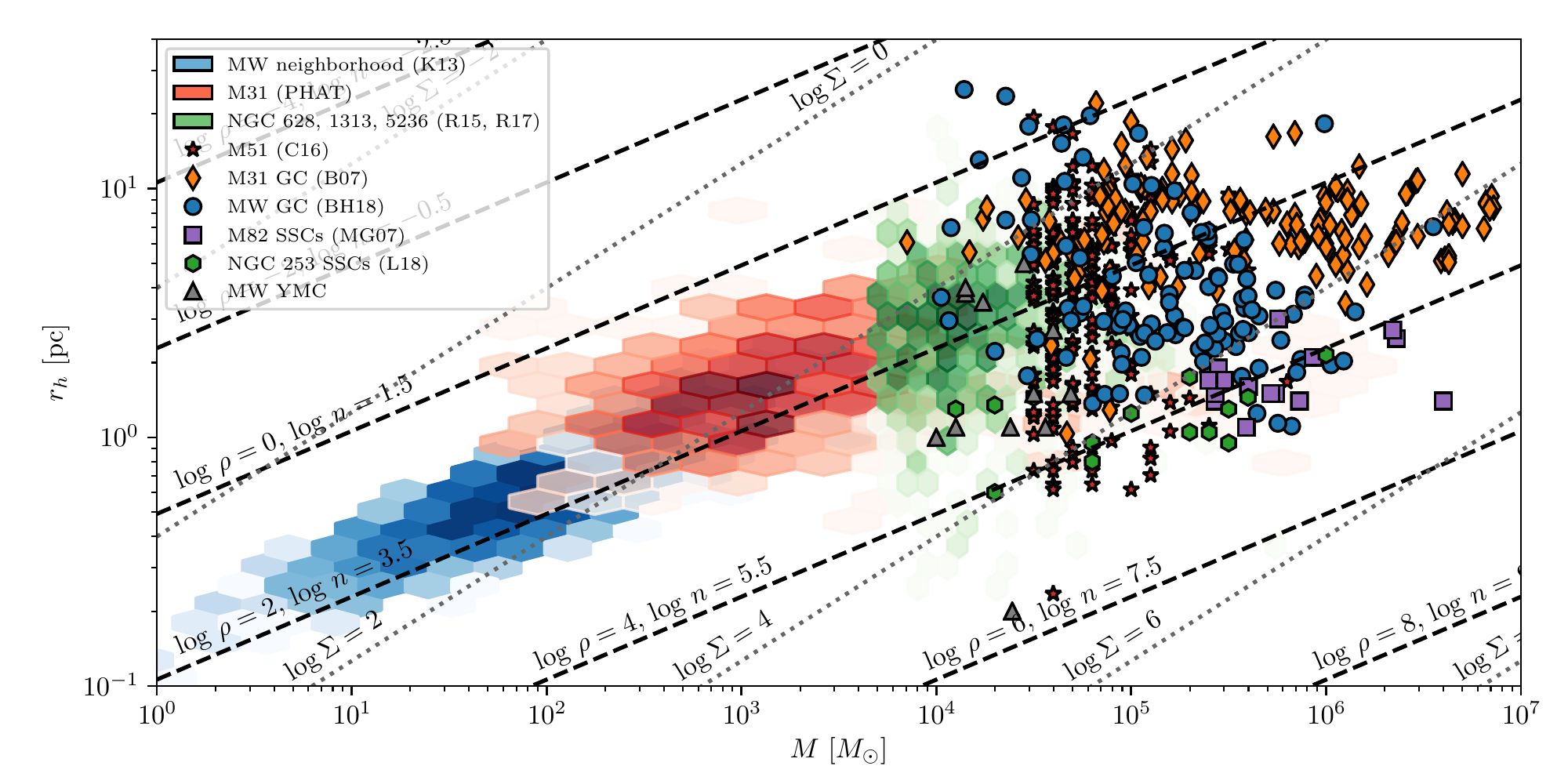}
\caption{
\label{fig:mass_radius}
Mass-radius relation for star clusters in nearby galaxies. For Milky Way and globular clusters, we plot half-mass radii where available, half-number radii otherwise; for all other data sets, we plot half-light radii. Dashed black lines indicate loci of constant density within $r_h$, with the mass density $\rho=3M/8\pi r_h^3$ (in units of $M_\odot$ pc$^{-3}$) and the corresponding number density $n$ (units of H nuclei cm$^{-3}$) as indicated; dotted lines show loci of constant surface density $\Sigma=M/2\pi r_h^2$ (units of $M_\odot$ pc$^{-2}$). Hexagonal density plots show the log of the density of clusters in the $(M,r_h)$ plane for Milky Way within 2 kpc of the Sun (data from \citealt[K13]{Kharchenko13a}, with masses estimated from tidal radii using the method of \citealt{Piskunov07a}), clusters in M31 from PHAT (radii from \citealt{Johnson12a} and masses from \citealt{Fouesneau14a}), and clusters in NGC 628, NGC 1313, and NGC 5236 \citep[galaxies grouped together;][R15, R17]{Ryon15a, Ryon17a}. The low-mass edges in the extragalactic data are a result of observational limitations, not a physical truncation. Points show individual clusters from smaller samples: clusters in the disk of M51 \citep[C16]{Chandar16a}, GCs in M31 \citep[B07]{Barmby07a} and the Milky Way \citep[BH18]{Baumgardt18a}, SSCs in M82 \citep[MG07]{McCrady07a} and NGC 253 \citep[L18]{Leroy18a}, and YMCs from our own compilation (\autoref{tab:ymcs}).
}
\end{figure}

\begin{table}[h]
\tabcolsep7.5pt
\caption{Summary of masses and radii in current cluster samples}
\label{tab:cluster_samples}
\begin{center}
\begin{tabular}{@{}l|c|c|c|c@{}}
\hline
Sample   &   $\log M$ [$M_\odot$]$^{\rm a}$   &   $r_h$ [pc]$^{\rm a}$   &   $\log\Sigma$ [$M_\odot$ pc$^{-2}$]$^{\rm a,b}$    &   Reference$^{\rm c}$ \\
\hline
Milky Way ($d<2$ kpc)   &   $1.90_{-0.55}^{+0.51}$   &   $0.58_{-0.20}^{+0.32}$    &   $1.55_{-0.30}^{+0.31}$   &    K13   \\
M31 (PHAT)   &   $3.18_{-0.55}^{+0.55}$   &   $1.56_{-0.46}^{+0.99}$    &   $1.92_{-0.45}^{+0.59}$   &    J12, F14   \\
NGC 5236 (M83)   &   $4.20_{-0.15}^{+0.34}$   &   $2.51_{-0.97}^{+1.20}$    &   $2.62_{-0.34}^{+0.56}$   &    R15   \\
NGC 628 (LEGUS)   &   $4.00_{-0.21}^{+0.40}$   &   $3.09_{-1.18}^{+2.16}$    &   $2.25_{-0.45}^{+0.56}$   &    R17   \\
NGC 1313 (LEGUS)   &   $4.10_{-0.26}^{+0.39}$   &   $2.69_{-1.67}^{+1.78}$    &   $2.51_{-0.47}^{+0.82}$   &    R17   \\
M51   &   $4.70_{-0.10}^{+0.30}$   &   $3.89_{-2.63}^{+2.99}$    &   $2.84_{-0.64}^{+0.84}$   &    C16   \\
M31 GC   &   $5.57_{-0.70}^{+0.84}$   &   $7.41_{-2.11}^{+2.54}$    &   $3.22_{-0.96}^{+0.77}$   &    B07   \\
Milky Way GC   &   $5.22_{-0.56}^{+0.42}$   &   $3.20_{-1.10}^{+3.88}$    &   $3.39_{-1.06}^{+0.60}$   &    BH18   \\
M82 SSC   &   $5.72_{-0.28}^{+0.56}$   &   $1.60_{-0.20}^{+0.80}$    &   $4.49_{-0.33}^{+0.26}$   &    MG07   \\
NGC 253 SSC   &   $5.18_{-0.86}^{+0.32}$   &   $1.27_{-0.32}^{+0.18}$    &   $4.13_{-0.66}^{+0.41}$   &    L18   \\
Milky Way YMC ($>10^4$ $M_\odot$)   &   $4.39_{-0.25}^{+0.18}$   &   $1.50_{-0.42}^{+2.35}$    &   $3.21_{-0.97}^{+0.37}$   &    See \autoref{tab:ymcs}   \\\hline
\end{tabular}
\end{center}
\begin{tabnote}
$^{\rm a}$The first number gives the sample median, while plus and minus values indicate the 16th to 84th percentile range; $^{\rm b}$$\Sigma_h$ is the surface density at the half-mass radius, $\Sigma_h = M/2 \pi r_h^2$; $^{\rm c}$References are as follows: K13 = \citet{Kharchenko13a}, J12 = \citet{Johnson12a}, F14 = \citet{Fouesneau14a}, R15 = \citet{Ryon15a}, R17 = \citet{Ryon17a}, C16 = \citet{Chandar16a}, B07 = \citet{Barmby07a}, BH18 = \citet{Baumgardt18a}, MG07 = \citet{McCrady07a}, L18 = \citet{Leroy18a}.
\end{tabnote}
\end{table}

\begin{table}[h]
\tabcolsep7.5pt
\caption{Young massive ($M>10^4$ $M_\odot$) clusters (YMCs) in the Milky Way}
\label{tab:ymcs}
\begin{center}
\begin{tabular}{@{}l|c|c|c@{}}
\hline
& $\log M^{\rm a}$ & $r_h^{\rm a}$ & \\
Name & ($M_\odot$) & (pc) & Reference$^{\rm b}$ \\
\hline
Westerlund 1 & $4.69\pm 0.045$	& 1.5 &	G11 \\
Westerlund 2 & $4.56\pm 0.035$	& 1.1 &	Z17	\\
Arches & $4.38\pm 0.17$	& $1.1\pm 0.61$ & C12 \\
NGC 3603	& $4.1\pm 0.10$ & $1.1\pm 0.40$ & HEM08,R10 \\
Quintuplet & 4.0 &	1.0	& FMM99 \\
DBS2003 179 & 4.39 & 0.2 & B08,B12 \\
RSGC01 & 4.5 &	1.5	& F06 \\
RSGC02 (Stephenson 2) & 4.6	& 2.7 &	D07	\\
RSGC03 & $4.45\pm 0.15$	 & 5.0 & C09 \\
RSGC04$^{\rm c}$ (Alicante 8) &   $4.15\pm 0.15$	& 3.8 &	N10 \\
RSGC05$^{\rm c}$ (Alicante 7) &   $4.24\pm 0.24$	& 3.5 &	N11 \\
Alicante 10$^{\rm c}$ & $4.15\pm 0.15$ & 4.0 & GFN12 \\
\hline
\end{tabular}
\end{center}
\begin{tabnote}
$^{\rm a}$Error estimates are given when provided by the referenced source; $^{\rm b}$References are as follows: FMM99 = \citet{Figer99a}, F06 = \citet{Figer06a}, D07 = \citet{Davies07a}, HEM08 = \citet{Harayama08a}, B08 = \citet{Borissova08a}, C09 = \citet{Clark09a}, R10 = \citet{Rochau10a}, N10 = \citet{Negueruela10a}, N11 = \citet{Negueruela11a}, G11 = \citet{Gennaro11a}, GFN12 = \citet{Gonzalez-Fernandez12a}, B12 = \citet{Borissova12a}, C12 = \citet{Clarkson12a}, Z17 = \citet{Zeidler17a}; $^{\rm c}$Cited reference does not list a radius; the value we quote is a by-eye fit based on their published images, and should only be considered accurate to the factor of $\sim 2$ level.
\end{tabnote}
\end{table}

In \autoref{fig:mass_radius} we show the mass-radius relation for cluster samples in nearby galaxies, and we summarize the properties of the cluster samples shown in the Figure in \autoref{tab:cluster_samples}; for the Milky Way, we have updated \citet{Portegies-Zwart10a}'s compilation of YMCs in \autoref{tab:ymcs}\footnote{Our cluster list is slightly different than \citet{Portegies-Zwart10a}'s compilation. Several from their list are omitted from our table because more recent observations have reduced their estimated masses below the threshold of $10^4$ $M_\odot$ commonly used to delineate YMCs, while our list includes several clusters discovered more recently than their compilation.}. The plot reveals a few notable features. First, there is a mass-radius relation, but it is weak in terms of both slope and tightness. Cluster radius increases with mass slightly more slowly than $r_h \propto M^{1/3}$, so that more massive clusters have mildly higher average density. We refrain from giving a formal fit to the $\log r_h-\log M$ relation, because the results would be highly sensitive to how we weighted the various data sets. Regardless of the index of the mass-radius relationship, it is clearly very broad, in the sense that, at any given mass, one can find clusters whose radii vary by a factor of 5; at least some of the spread is undoubtedly the result of measurement errors. There is no obvious difference in either the breadth or slope of the mass-radius relation between the inclusive and exclusive catalogs, although a comparison is not straightforward since, unlike for measurements of $\alpha_T$ and $\Gamma$, there is little overlap in \autoref{fig:mass_radius} between measurements carried out using exclusive versus inclusive catalogs. Second, the distribution is bounded at both large and small radii. Cluster densities are bounded below by $\rho \sim 1$ $M_\odot$ pc$^{-3}$ (or equivalently $n\sim 30$ H cm$^{-3}$); GCs in galaxy halos go to only slightly lower densities. On the small radius side, the data appear to be limited to the region $\Sigma \lesssim 10^5$ $M_\odot$ pc$^{-2}$, as noted previously by \citet{Hopkins10a}. Third, there is no separation between the loci occupied by clusters in galaxy disks and GCs; the two distributions blend continuously into one another.

\subsection{Elemental abundances}
\label{ssec:chemical_comp}

In systems close enough to permit spectroscopy of individual low-mass stars, we can study the composition of clusters. The stars in all OCs in the Galaxy are observed to be highly homogeneous ($\Delta$[Fe/H] $\lesssim$ 0.05 dex) in their abundances of all elements, while no old GCs are fully homogeneous, at least not in their light elements (Li, C, N, O, F, Na, Mg, and/or Al). Essentially all massive GCs also show an anti-correlation in certain light elements \citep[e.g., Na-O, Mg-Al;][]{Carretta2010}. Like multiple stellar populations observed in CMDs, this has become a distinguishing feature of globulars, including for globulars in the Magellanic Clouds, that is never observed in OCs \citep[e.g.,][]{Bragaglia2017}. This difference has led to the view that GC history is more complex than that of OCs \citep{Gratton2012}. However, there is a remarkable exception to the rule: the 12-Gyr old, $10^{4.8}\msun$ globular Ruprecht 106 appears to be a true ``single stellar population'' with homogeneous abundances and no light element anti-correlations \citep{Villanova2013}; unusually, it is enriched in r-process and s-process elements but does not show the enhanced [$\alpha$/Fe] signature of the oldest stars, a characteristic shared by all other Galactic GCs. The globulars Terzan 7 and Palomar 12 also do not have multiple populations and, intriguingly, may be extragalactic through their association with the disrupting Sgr dwarf. \citet[][Table 9]{Bragaglia2017} catalog all modern attempts to detect multiple populations in open and globular clusters.

Typically, the heavy elements (Fe and beyond) in globulars exhibit much less scatter than for the light elements \citep[e.g.][]{Gratton2012}, except in a few systems (e.g., $\omega$ Cen, Terzan 5, M22, NGC 5824) that have unusual properties compared to the general family of clusters. Unusually, NGC 5824 has $\Delta$[Fe/H] variations of up to 0.3 dex but it also has a remarkable stellar halo with no tidal truncation detected out to 400 pc in radius, far beyond other clusters \citep{DaCosta14a}. There are numerous other oddities \citep[e.g.,][]{Marino2018}: NGC 2419 has a large scatter in K-Mg; some globulars show s-process variations (e.g., M22, $\omega$ Cen), others (e.g. M15) exhibit r-process variations. With globulars, there are exceptions to every rule. In one particularly deep study of the globular NGC 6752, \citet{Yong2013} demonstrate weak inhomogeneities in {\it all} elements, even the Fe peak. The existence of inhomogeneities in globulars may be related to the presence of multiple populations 
in these systems but there is no clear association at the present time \citep[][]{Bragaglia2014}. Multiple stellar populations in globulars are simply not understood \citep{Gratton2012,Renzini2013,Bastian2018}.

Abundance studies in OCs are more limited. The young OC Hyades is nominally homogeneous \citep{DeSilva2006,DeSilva2011} although a deep study reveals weak variations at the level of 0.02$-$0.03 dex \citep{Liu2016a}. The older OC M67 exhibits slightly larger variations (still $\lesssim 0.05$ dex) from a differential analysis of two fortuitous solar twins \citep{Liu2016b}. To date, all OCs have been shown to be single stellar populations with a few rare counter claims \citep[e.g.,][]{Geiser2012} refuted in follow-up analysis \citep[e.g.,][]{Bragaglia2014}. The Hyades and M67 clusters have been subject to the most sensitive searches for abundance variations, so it is plausible that the majority of OCs will be shown to have low-level inhomogeneities in future studies. Weak metallicity variations may reflect atomic diffusion in stellar atmospheres \citep{Dotter2017,Souto2018}, planetary infall \citep{Melendez2017}, or simply intrinsic variations that existed at the onset of star formation \citep{Feng2014}. We return to this point in \autoref{ssec:chem_form}.

\subsection{The limits of demography at young ages}
\label{ssec:demo_limits}

Thus far we have discussed the demographics of star clusters without too much worry about the issues raised in \autoref{ssec:definition} about how one defines a cluster, though our discussion about the differences between inclusive and exclusive catalogs should already hint at potential problems. We now turn directly to the question: at what stellar population age does it make sense to treat as meaningful concepts such as the CMF, CAF, and all the other demographic distributions discussed above?

To answer this question, it is helpful to conduct a thought experiment: consider an annulus within the Milky Way centered on the Solar Circle, and for every star of age $T \pm \Delta T$, find its $N$th nearest neighbor (either in 3D or in projection) in the same age range, and plot the distribution of nearest neighbor distances. Qualitatively, what does this distribution look like? For stars at an age $T \sim 1$ Gyr, the answer is fairly straightforward: almost all stars at this age are field stars, and after 1 Gyr the orbits of these stars through the Galaxy are thoroughly phase-mixed. Separations therefore follow a Poisson distribution, perhaps with some overall structure at $\sim$kpc scales as a result of spiral arms. A small fraction of 1 Gyr-old stars are members of open clusters, and for these the neighbor distances are much smaller. Consequently, the overall separation distribution is a Poissonian with an excess at very small separation. Indeed, one sees exactly this sort of distribution if one plots the distribution of distances between molecular clouds and young star clusters in external galaxies \citep[e.g.,][]{Kawamura09a}. This makes it straightforward to define old clusters: they are the structures made up of the stars in the small-separation excess.

The results are very different for pre-main sequence stars (selected based on optical colours, infrared spectral energy distributions, or other methods) with ages $T \lesssim 10$ Myr. The angular correlation function for these stars is well described as a power-law $\xi(\theta) \propto \theta^{-p}$ with $p \approx 0.5 - 1.5$ on scales from $\sim 0.01$ pc to $\sim 30$ pc \citep[e.g.,][]{Hennekemper08a, Kraus08a, Schmeja09a, Bressert10a}. The cluster-cluster angular correlation function is a power-law up to $\sim$kpc scales \citep[e.g.,][]{Elmegreen01a, Bastian07a, Gouliermis17a, Grasha17a, Grasha17b} -- see \citet{Gouliermis18a} for a comprehensive review. For such distributions, there is no obvious break or scale that could be used to delineate between cluster and field populations. This presents a challenge for describing the demographics of star clusters, since it is not obvious how to go about defining the clusters to begin with.

This situation leads to two main options. One is a Press-Schechter-like approach, whereby one considers the stellar field smoothed with some characteristic window that defines a size scale. However, there is an important difference between our situation and the cosmological one: in cosmology one can compute from first principles the critical overdensity of $\approx 200$ that separates bound and unbound structures, while no such calculation is available for star clusters (though there are attempts -- see \autoref{ssec:feedback}). Alternately one can simply appeal to the empirical result that older clusters have a characteristic size $\sim 1-10$ pc, and adopt a similar averaging scale at younger ages. This is what many authors do implicitly by using clustering algorithms such as MST that use a free parameter to set the scale. A second approach is to abandon discussion of cluster demographics entirely for stellar populations $\lesssim 10$ Myr old. One can instead discuss \textit{clustering} -- i.e., the non-Poissonian distribution of stellar separations, and the correlations between separation in space and separation in other quantities (age, velocity, elemental abundance, etc.) -- rather than discuss the properties of discrete clusters, an approach previously suggested by \citet{Zinnecker10a}. This removes the need to impose a size scale on a distribution where none is immediately apparent.

As stars age, their separation distribution must evolve from the power-law form seen at young ages to the Poisson plus small-separation excess form that prevails at older ages. Observations suggest that much of this evolution happens relatively quickly, over about a crossing time of the relevant size scale \citep[e.g.,][]{Elmegreen00a, Grasha17b}. Thus the type of demographic analysis we have discussed to this point is sensible as applied to structures more than about a crossing time old. However, from a theoretical standpoint the question that any successful theory of cluster formation must answer is how the power-law, scale-free distribution that prevails among young clusters turns into the bimodal distribution that prevails at older ages, and what determines the demographics of the clustered portion of that bimodal distribution. In other words, how does stellar \textit{clustering} turn into star \textit{clusters}? 

\section{BIRTH}
\label{sec:birth}

In this section we examine clusters' early life, when gas dominates the mass; \autoref{sec:life_and_death} covers the gas-free phase. Here we first review the gaseous initial conditions for star cluster formation (\autoref{ssec:initial_conditions}). We then discuss observational constraints and theoretical models for the rates at which these structures form stars (\autoref{ssec:epsff}) the stellar feedback processes that inhibit this transformation (\autoref{ssec:feedback}), and the star formation histories that result (\autoref{ssec:tsf}). We conclude with an attempt to synthesise this material into a coherent model for the physical properties (\autoref{ssec:emergence}) and elemental abundances (\autoref{ssec:chem_form}) of the final gas-free cluster population.

\subsection{Initial conditions for star cluster formation}
\label{ssec:initial_conditions}

\subsubsection{Giant molecular clouds}
\label{ssec:giant}

Clusters are born in molecular clouds, primarily giant molecular clouds (GMCs). There is considerable evidence that GMCs obey several relations discovered by \citet{Larson81}: (1) The 1D velocity dispersion, $\sigma$, is supersonic and varies with size $L$ as $\sigma\propto L^p$, where $p\approx 0.5$ in the Galaxy \citep{Solomon87,Falgarone09}; (2) GMCs have roughly equal kinetic and gravitational potential energies, indicating that they are bound or nearly so; (3) the column density $N\approx nL$ (or, equivalently, the mass surface density $\Sigma=M/\pi R^2$) is approximately constant within a given galaxy. As Larson pointed out, only two of these are independent. To see this, note that the degree of gravitational binding can be measured by the virial parameter, $\avir\equiv 5\sigma^2R/GM$, which is unity for a sphere of constant density in virial equilibrium with no surface pressure; such a sphere is bound if it has $\avir<2$ (\citealp{Bertoldi92a}, who also considered non-spherical clouds). This definition can be rewritten as $\sigma=\sqrt{(\pi/5) G\avir\Sigma R}$. Thus bound clouds ($\avir\sim 1$) with the same surface density, $\Sigma$, automatically satisfy $\sigma\propto R^{1/2}$. In terms of the virial parameter, the ratio of the crossing time $t_{\rm cr}=R/\sigma$ to the free-fall time is $t_{\rm cr}/\tff\approx 2/\avir^{1/2}$.

Observations of molecular gas since \citet{Larson81}'s work have generally confirmed his findings. For the linewidth-size relation, \citet{Solomon87} found $\sigma=0.72 R_0^{0.5}$~km~s\e, where $R_0=R/(1\mbox{ pc})$, in their survey of GMCs in the first Galactic quadrant. \citet{Falgarone09} compiled observations of molecular gas in the Milky Way and found $\sigma\propto R^{0.5}$ within a factor 3 over the size range 0.1 to $>100$ pc. Subsequent observations of Galactic \citep{Rice16,Miville17} and extragalactic \citep{Bolatto08,Faesi18} molecular gas are generally consistent with the \citet{Solomon87} result. \citet{Faesi18} attribute the absence of a clear linewidth-size relation in some extragalactic observations to low spatial resolution.  Finally, we note that the value $p\approx 0.5$ is expected for supersonic turbulence on theoretical grounds and has been seen in many simulations (e.g., \citealp{Padoan14a}).

One can determine if a cloud is gravitationally bound by comparing the ``turbulence parameter" $C\equiv\sigma/R^{1/2}$ to the surface density $\Sigma$ \citep{Keto86}:
\beq
\avir=\frac{3.70C^2}{\Sigma_2},
\label{eq:avir}
\eeq
where $C$ is in units of km~s\e~pc$^{-1/2}$ and $\Sigma_2=\Sigma/(100\,M_\odot$~pc\ee) (c.f.~\citealp{Bolatto08}). The results of \citet{Solomon87} for GMCs in the inner Galaxy ($C=0.72$~km~s\e\ and $\Sigma_2=1.7$) imply $\avir\approx 1.1$, and they concluded that these GMCs are bound. \citet{Heyer09} and \citet{Roman10} confirmed this, the former by plotting $C$ vs.~$\Sigma$ from $^{13}$CO observations of the inner Galaxy. \citet{Sun18} were able to determine the virial parameter for the clouds in the different galaxies by measuring surface density and line width at a fixed linear scale. They found $\avir$ on the scale $R=60$~pc between 1.3 and 3.2 in all galaxies but M31 and M33 (where they were larger), showing that the clouds are bound or nearly bound. Visual inspection of their results shows that $\avir$ is almost always close to unity for the clouds with the highest surface densities. For a cloud to be bound, it must also be able to avoid tidal disruption. For a flat rotation curve, this requires that the mean cloud density exceed twice the mean density of the galaxy inside the orbit of the cloud \citep{Chernoff90a}, or $\bar\rho>5.4 \left(v_{c,220}/R_{\rm kpc}\right)^2$ $M_\odot$ pc$^{-3}$, where $v_{c,220}$ is the circular velocity in units of 220 km s$^{-1}$, and $R_{\rm kpc}$ is the distance to the center of the galaxy in kpc.

Larson's third relation, the constancy of the surface density of GMCs in the inner Galaxy (excluding the Galactic Center), was confirmed by \citet{Solomon87}, who found $\avg{\Sigma}=170\,M_\odot$~pc\ee\ and by \citet{Roman10}, who concluded that $\avg{\Sigma} = 140\,M_\odot$~pc$^{-3}$. (However, clouds with $M\la 10^4\,M_\odot$ can have lower $\Sigma$ -- \citealp{Heyer01a} -- and there can be systematic large-scale variations of the surface density within GMCs -- \citealp{Schneider15a}.) Different galaxies have different surface densities: in a study of 15 disk galaxies, \citet{Sun18} found that median surface density at 120 pc resolution ranged from about 10 to $200\,M_\odot$~pc\ee\ excluding the Antennae, for which it is $2300\,M_\odot$~pc\ee.  The variation in $\Sigma$ for the normal galaxies could reflect variations in the beam filling factor of GMCs as well as variations in the intrinsic properties of GMCs \citep{Lada13a}. The dispersion in these values within individual galaxies is about $\pm 0.4$ dex.

The surface density of a GMC is directly related to both its internal pressure and the pressure exerted on its surface, and insofar as $\Sigma$ is constant within a galaxy, these pressures tend to be also. The turbulent pressure within a self-gravitating cloud is
\beq
\bar P_{\rm turb}\equiv\bar\rho\sigma^2\equiv\frac{3\pi}{20}\avir G \Sigma_\gmc^2,
\label{eq:pturb}
\eeq
for spherical clouds (e.g., \citealp{McKee03}), so that $\bar P_{\rm turb}/\kb=1.0\times 10^5\avir\Sigma_2^2$ K cm\eee. This should exceed the external pressure, $P_{\rm ext}$, and indeed \citet{Hughes13a} found that $\bar P_{\rm turb}\sim (1-8)P_{\rm ext}$ in the eight galaxies they studied. For a typical GMC in the Milky Way, with $\Sigma_{\rm CO}\sim 140\,M_\odot$~pc\ee, one finds that the surface pressure on the CO cloud is comparable to the mean turbulent pressure inside the cloud. On the other hand, for GMCs  in high pressure environments, such as galactic nuclei or starbursts, the surface density must be large in order for the cloud to be self-gravitating: the virial theorem implies that $\avir\sim (1-P_{\rm ext}/\bar P)^{-1}$ in the absence of strong magnetic fields \citep{Bertoldi92a}, so the requirement that $\avir\sim 1$ implies $\Sigma\ga 2(P_{\rm ext}/G)^{1/2}$. High pressures also imply high volume densities: the mean density in a cloud is 
\beq
\bar\rho_\gmc=\frac 34 \left(\frac{\pi\Sigma_2^3}{M_6}\right)^{1/2}~~M_\odot\mbox{ ~pc\eee} = \frac{179}{\avir^{3/4} M_6^{1/2}}\left(\frac{\bar P_{\rm turb}/\kb}{10^8\mbox{ K cm\eee}}\right)^{3/4} ~~M_\odot\mbox{ pc\eee}
\label{eq:rhobar1}
\eeq
where the final step follows because for a non-spherical cloud $\Sigma\approx M/V^{2/3}$ has the value as for a spherical cloud of the same volume, $V$, and $P_{\rm turb}\ga 2 P_{\rm ext}$ for bound clouds.

\subsubsection{Clumps and clusters}
\label{ssec:clumps}

Clusters are born in density concentrations (clumps) within GMCs. Both the GMCs and the clumps within them are supersonically turbulent. In the absence of self-gravity and for an approximately isothermal equation of state, this turbulence leads to a log-normal density distribution (e.g., \citealp{McKee07a}) and a corresponding log-normal distribution of surface densities \citep{Brunt10a}. Self-gravity leads to the formation of a power-law tail in the PDF for dense gas \citep{Collins12}. This has been observed in the PDF of surface densities in a number of GMCs, $dP/d\log\Sigma\propto\Sigma^{-(2\pm 0.7)}$ for clouds with active star formation; this is often associated with an overall power-law density distribution in the cloud \citep{Lombardi15,Schneider15a}. The clumps in GMCs are formed by the combined action of supersonic turbulence and self-gravity. \autoref{eq:avir} implies that the typical surface density of a bound clump within a bound GMC (each of which therefore has $\avir\approx 1$) is about the same as that of the GMC, or $\Sigma_{\rm clump}\sim 100\,M_\odot$~pc\ee\ in the Galaxy. Observational studies of clumps have been carried out only for the Galaxy. Analyzing the results of the APEX 870 $\mu$m survey of the Galactic plane (excluding the central $\pm 5^\circ$ in longitude), which was complete for $\Sigma>700\,M_\odot$~pc\ee\ and $M>1000\,M_\odot$ within 20 kpc, \citet{Urquhart18a} found about $10^7\,M_\odot$ in 8000 dense clumps. A by-eye fit to their data on the surface densities of the clumps gives $dN_{\rm clump}/d\log \Sigma\propto\Sigma^{-1.6}$ over the range $\Sigma=600-6000\, M_\odot$~pc\ee; the median surface density is about $700\,M_\odot$~pc\ee\ and the maximum is about $3\times 10^4\,M_\odot$~pc\ee. The clumps are gravitationally bound with $\avir\sim 0.1-1$; \citet{Kauffmann13} and \citet{Tan13} have suggested that magnetic fields could provide support for sub-virial clumps. At high masses, the clump mass distribution has $\alpha_{M,\,\rm clump}\approx -2$ (indeed, \citealp{Heithausen98a} found $\alpha_{M,\, \rm clump}=-1.84$ over 5 decades in mass) and has an upper limit of about $10^5\,M_\odot$. In the Galaxy, the GMCs out of which the clumps form have a power-law mass with $\alpha_{M,\, \rm GMC}\approx -1.6$, so that most of the mass of GMCs is in massive clouds. This distribution is truncated at about $10^7\,M_\odot$ (\citealp{Rice16} and references therein, but see \citealp{Miville17} for a different view), as expected theoretically from the Jeans mass of the galactic disk (i.e., the Jeans mass expressed in terms of the gas surface density, $\Sigma_{\rm gal}$; \citealt{Kim01}),
\beq
M_J=\frac{\sigma^4}{G^2\Sigma_{\rm gal}}=10^7\left(\frac{\sigma}{7\mbox{ km s\e}}\right)^4\left(\frac{13\,M_\odot\mbox{ pc\ee}}{\Sigma_{\rm gal}}\right)~~~M_\odot.
\label{eq:mj}
\eeq
If the Toomre $Q$-parameter, $Q=\kappa\sigma/(\pi G\Sigma)\sim 1$, where $\kappa$ is the epicyclic frequency, this is about the maximum mass expected for a bound cloud, so in the Galaxy the most massive dense clump is about 1\% of the mass of the most massive GMC.

Observations of GMCs in external galaxies show that in some, but not all, cases the mass distribution can be described by 
a truncated power law with most of the mass in the most massive clouds, as observed in the Galaxy: this describes the GMCs in NGC 300 \citep{Faesi18} and those in the inner region and in the spiral density waves of M51, but not the GMCs in the rest of the galaxy, where 
$\alpha_{M,\, \rm GMC}<-2$, so that low-mass clouds contain most of the mass \citep{Colombo14a}.
In the regions of M51 in which 
$\alpha_{M,\, \rm GMC}>-2$, the upper limit on the mass distribution is similar to the Galactic value, $\sim 10^7\,M_\odot$.

The conditions required to produce massive clusters, such as globular clusters, occur in interacting galaxies such as the Antennae (e.g., \citealp{Whitmore14}), dwarf irregular galaxies, some of which are starbursts and many of which show signs of interaction  \citep{Billett02}, and nuclear starbursts 
in spiral galaxies such as NGC 253 (e.g., \citealp{Leroy18a}). Such conditions are rare now, but were common at redshifts $z\ga 2$ \citep{Kruijssen14a}. A detailed survey of the GMCs in the Antennae galaxies has yet to be made, but \citet{Wilson00} carried out a low-resolution ($\sim 300\times 500$ pc) survey of this system and identified five large molecular cloud complexes with masses in excess of $10^8\,M_\odot$; the largest one not associated with one of the galactic nuclei had a mass of $6\times 10^8\,M_\odot$. By contrast, \citet{Wilson03} pointed out that the most massive molecular cloud complex in M51 had a mass about 10 times less. They suggested that the interaction of the galaxies produced regions of less shear, which enables the existence of more massive GMCs; \citet{Billett02} emphasized the importance of low shear in producing more massive clusters in dwarf irregular galaxies. For disk galaxies, we can express
the Jeans mass (\autoref{eq:mj}) in terms of the Toomre $Q$-parameter, so that
\beq
M_J=\frac{\pi^4 G^2\Sigma^3 Q^4}{\kappa^4}=2500\frac{\Sigma_2^3Q_{1.5}^4}{\Omega_0^4}~M_\odot, 
\label{eq:mjq}
\eeq
where in the second expression a flat rotation curve has been assumed, $Q$ has been normalized to a typical critical value, and $\Omega_0$ is measured in units of Myr$^{-1}$ \citep{Krumholz05a}. Hence, if $Q$ self-regulates to be of order unity, a reduction in the epicyclic frequency, $\kappa$, by a factor 2 due to the effect of the interaction can increase the maximum GMC mass by more than an order of magnitude. More massive GMCs enable the production of more massive clumps \citep{Harris94,Reina-Campos17a}; \citet{Johnson15} have observed a clump in the Antennae with a mass of at least $5\times10^6\,M_\odot$ and a radius less than 24 pc, corresponding to $\Sigma\ga 3000\,M_\odot$~pc\ee\
and $\bar P_{\rm turb}/\kb\ga 10^8$~K~cm\eee.  

The maximum possible mass of a cloud that forms by gravitational instability in a disk is $M_{\rm GMC,\,max}\simeq (\lambda_{\rm T}/2)^2\Sigma$, where $\lambda_{\rm T}=4\pi^2 G\Sigma/\kappa^2$ is the Toomre length \citep{Escala08a}. Disks are stable for $Q\ga 1.5$ \citep{Kim01}, and the Jeans mass (eq. \ref{eq:mjq}) at $Q=1.5$ is close to $M_{\rm GMC,\,max}$. If the disk has a total mass (including stars and dark matter) inside a radius $R$ of $M_{\rm tot}(<R)$, a gas mass inside $R$ of $M_g(<R)\simeq \pi R^2\Sigma$ and it is rotationally supported, then 
\beq
\frac{M_{\rm GMC,\,max}}{M_g(<R)}\simeq\frac{4\pi}{(\kappa/\Omega)^4}\left[\frac{M_g(<R)}{M_{\rm tot}(<R)} \right]^2~~:
\eeq
the maximum fraction of the gas that can go into a single GMC scales as the square of the gas fraction. As a result, more massive GMCs, and therefore more massive cluster-forming clumps, are expected in gas-rich regions such as galaxies in the process of formation and in starbursts \citep{Escala08a}.

Galaxy interactions also produce shocks that can compress clouds. A collision at a relative velocity $v_{\rm rel}$ produces radiative shocks with a velocity $v_{\rm rel}/2$ and a pressure $P=\rho_0(v_{\rm rel}/2)^2$, corresponding to $P/\kb =3.8\times 10^6n_{\rm H,0}(v_{\rm rel}/300\,$km~s\e$)^2$. 
This leads to
localized regions of very high pressure when clouds collide, $P/\kb\sim 10^8$~K~cm\eee\ \citep{Jog92}, but only
$\sim 10^6$~K~cm\eee\ for shocks in the intercloud medium, which is not that much larger than the typical pressure in Galactic GMCs.

Based on this summary of the observed properties of molecular gas in galaxies, we have the following expectations for the gaseous precursors of star clusters currently forming in the Galaxy: clusters form from clumps with masses ranging from very small values (depending on the star formation efficiency) up to about $10^5\,M_\odot$ and surface densities from $\sim 100\,M_\odot$~pc\ee\ up to about $3\times 10^4\,M_\odot$~pc\ee. A substantial fraction of the clouds are bound ($\avir\la 2$). The typical clump has a column density comparable to that of the cloud in which it is embedded. For a clump mass of $10^3\,M_\odot$ and a typical GMC pressure corresponding to $\Sigma_\gmc=140\,M_\odot$~pc\ee, namely $P_{\rm turb}/\kb\sim 2\times 10^5$~K cm\eee, the density is about $50\,M_\odot$~pc\eee. Clumps have a distribution of surface densities extending to higher values, and the mean density of a clump at a given mass scales as $\Sigma_{\rm clump}^{3/2}$ (\autoref{eq:rhobar1}). Conditions in nearby spiral galaxies do not differ substantially from those in the Milky Way. These conditions are consistent with the production of the OCs observed in the local universe, but not of GCs, which are far more massive than the clumps observed in the Galaxy. Molecular clouds that can produce globulars are seen in interacting galaxies like the Antennae, dwarf starbursts such as He 2-10 \citep{Johnson18}, and nuclear starbursts \citep{Leroy18a}, and are expected in any disk with a high gas fraction (as observed in galaxies at high redshift -- \citealt{Genzel11a}) or a small epicyclic frequency (as observed in interacting and dwarf irregular galaxies).

\subsection{Conversion of gas to stars}
\label{ssec:epsff}

\begin{marginnote}[]
\entry{$t_{\rm ff}$}{free-fall time, the natural evolutionary timescale for a self-gravitating system}
\entry{$\epsilon_{\rm ff}$}{the fraction of a cloud's mass that is transformed into stars per cloud free-fall time}
\entry{$\epsilon_*$}{the fraction of a cloud's initial mass that is transformed into stars by the time all the initial gas has been been consumed or ejected; sometimes called the star formation efficiency, though we will mostly eschew this term to minimize confusion with $\epsff$}
\entry{$t_{\rm sf}$}{timescale over which conversion to stars occurs: $\epsilon_* \approx \epsff (t_{\rm sf}/t_{\rm ff})$}
\entry{$\eta$}{mass loading factor: ratio of mass ejected by star formation feedback to mass converted to stars}
\end{marginnote}

The collapse of molecular clouds eventually produces stars. Here we review only those aspects of this process most relevant to star clusters; for a more general treatment see the reviews by \citet{McKee07a} and \citet{Krumholz14a}. The CMF and the production of bound clusters depend strongly on the efficiency with which gas is converted into stars. Efficiency can be defined in multiple ways, which we can illustrate in an idealized example: a cloud of initial mass $M$ and free-fall time $t_{\rm ff}$ forms stars at a rate $\dot{M}_*$. At any given time, the remaining gas mass is $M_g$ and the mass of stars formed is $M_*$, and we can parametrize the relationship between the star formation rate, current gas mass, and free-fall time as
\begin{equation}
\dot{M}_* = \epsff \frac{M_g}{t_{\rm ff}},
\end{equation}
where $\epsff$ is one of the efficiencies we will define. A region in free-fall collapse with nothing inhibiting star formation has $\epsff \approx 1$. In addition to gas consumption by star formation, feedback from stars launches a wind at a rate $\dot{M}_w$, which we normalize to the star formation rate by defining the mass loading factor $\eta = \dot{M}_w/\dot{M}_*$. With these definitions, we have
\begin{equation}
\dot{M}_g = -\left(1+\eta\right)\epsff \frac{M_g}{t_{\rm ff}}.
\end{equation}

All the factors that appear in this equation -- $\eta$, $\epsff$, or $t_{\rm ff}$ -- can depend on $M_g$, $M_*$, time, or any number of other variables. Moreover, real clouds are not closed boxes with a fixed initial mass; there is certainly mass inflow occurring simultaneously with star formation (see ``Conveyor-Belt Model for Cluster Formation" in the Supplementary Materials). Indeed, nothing in our formulation even requires that clouds be bound -- $M_g$ may can include both bound and unbound material. For the purposes of illustration, however, we can ignore these complications and treat $\eta$, $\epsff$, and $t_{\rm ff}$ as constant, in which case it is trivial to write down expressions for the instantaneous gas and stellar masses:
\begin{equation}
\label{eq:definitions}
\frac{M_g}{M} = e^{-t/t_{\rm sf}}
\qquad
\frac{M_*}{M} = \frac{1-e^{-t/t_{\rm sf}}}{1+\eta} \equiv \epsilon_* \left(1-e^{-t/t_{\rm sf}}\right)
\qquad
t_{\rm sf} \equiv \frac{\epsilon_*}{\epsff} t_{\rm ff}.
\end{equation}
These equations provide another definition of the star formation efficiency: $\epsilon_*$, the fraction of the initial cloud mass that has been transformed into stars once all the gas is consumed. In this idealized problem, $\epsilon_* = 1/(1+\eta)$. The term $t_{\rm sf}$ we have introduced defines the characteristic timescale for star formation, i.e., $M_*/M$ reaches its final value of $\epsilon_*$ on a timescale $t_{\rm sf}$. In this section we discuss $\epsilon_{\rm ff}$, and in the next two sections we examine $\epsilon_*$ and $t_{\rm sf}$. 

\subsubsection{Observational constraints on $\epsff$}
\label{ssec:obsepsff}

The value of $\epsilon_{\rm ff}$ is well constrained by observations. Formally, for any specified region of volume $V$ containing a gas mass $M_g$, we have $\epsff = \dot{M}_*/[M_g/\tff(\bar{\rho})]$, where $\dot{M}_*$ is the instantaneous star formation rate within $V$, $\tff(\bar\rho) = \sqrt{3\pi/32 G\bar{\rho}}$ is the free-fall time as a function of density, and $\bar{\rho} = M_g/V$ is the mean density of the region in question. There are several methods available to estimate these quantities. The most direct is to define a column density or extinction threshold, estimate the mass within that threshold, calculate the density and thus the free-fall time from the mass and projected area (assuming the unseen third dimension is comparable in size to the two observed dimensions), and estimate the star formation rate by counting young stellar objects (YSOs) and estimating their masses and the duration of the YSO phase (\citealt{Krumholz12a, Federrath13a, Evans14a, Salim15a, Heyer16a, Ochsendorf17a}; also see \citealt{Heiderman10a, Lada13a}). All published studies using this method rely on data from \textit{Spitzer}, which was only sensitive enough to detect $\sim 1$ $M_\odot$ YSOs within a few kpc of the Sun, and massive YSOs out to the distance of the Magellanic Clouds. Consequently, this method is only available out to the Magellanic Clouds, and studies that go beyond the Solar neighborhood must make a correction for the unseen part of the IMF.

A second approach, usable throughout the Milky Way and its satellites, is to match catalogs of star-forming regions identified by tracers such as infrared or free-free emission with catalogs of molecular clouds identified by CO or dust emission, matching them up if they are sufficiently close in position or velocity, and then using the mass, free-fall time, and star formation rates of the matched clouds and star-forming regions to estimate $\epsff$ \citep{Vutisalchavakul16a, Lee16a, Ochsendorf17a}. A third approach, available for extragalactic systems with extensive molecular gas and star formation tracer maps, is simply to pixelize the entire galaxy, and estimate masses, densities, and free-fall times in each pixel \citep{Krumholz12a, Leroy17a, Utomo18a}. A fourth method is to observe tracers of dense gas, most commonly HCN, and correlate these with tracers of star formation \citep[e.g.,][]{Krumholz07a, Garcia-Burillo12a, Hopkins13a, Usero15a, Gallagher18a}. For this method there is no need to spatially resolve the emission, because the molecule itself determines the density and free-fall time; e.g., \citet{Onus18a} show that the luminosity-weighted mean density of HCN line-emitting gas is $\approx 10^4$ cm$^{-3}$.\footnote{A fifth and final method is available for the special case of the Milky Way's Central Molecular Zone (CMZ), where one can use the positions of clouds and star clusters around their orbit through the CMZ as an absolute clock to time the transformation of gas into stars \citep{Barnes17a}. Since this method is applicable only to the CMZ we will not discuss it further.}

\begin{figure}
\includegraphics[width=\textwidth]{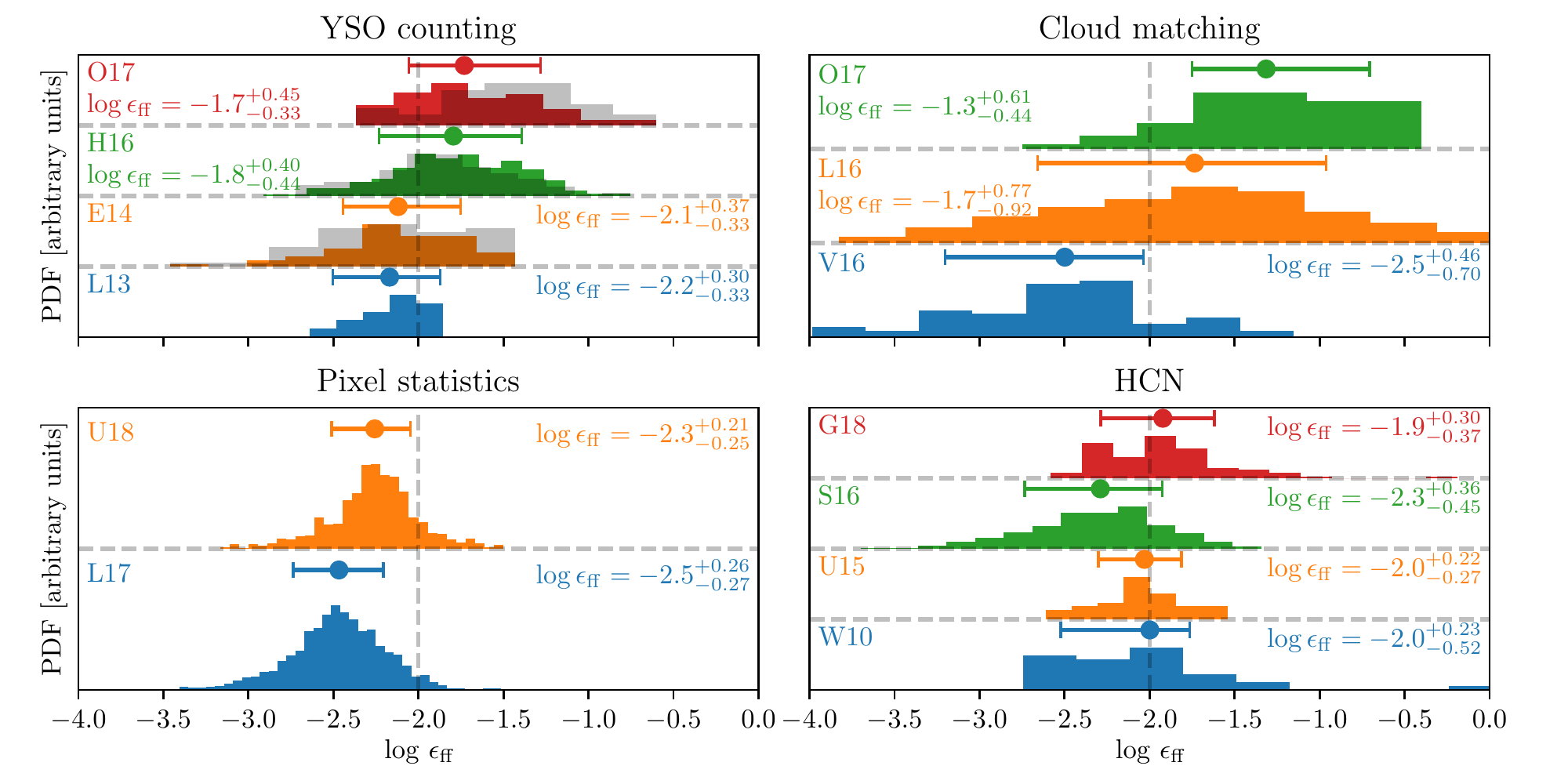}
\caption{
\label{fig:epsff}
Compilation of $\epsff$ measurements, grouped by method. Each histogram shows the distribution of $\epsff$ values measured in a single study. Points and error bars show the median and the 16th to 84th percentile; numerical values are printed in each panel. In panels with two histograms, the colored one is the distribution including non-detections plotted at their $2\sigma$ upper limits, while the gray histogram shows the distribution omitting non-detections; median and percentile values are for the colored distribution. The vertical dashed lines at $\log\epsff=-2$ are to help guide the eye. References are as follows: W10 = \citet{Wu10a}, L13 = \citet{Lada13a}, E14 = \citet{Evans14a}, U15 = \citet{Usero15a}, H16 = \citet{Heyer16a}, L16 = \citet{Lee16a}, V16 = \citet{Vutisalchavakul16a}, S16 = \citet{Stephens16a}, L17 = \citet{Leroy17a}, O17 = \citet[appears twice, because this study used two independent methods]{Ochsendorf17a}, G18 = \citet{Gallagher18a}, U18 = \citet{Utomo18a}. All values of $\epsff$ are as reported by the authors of the study, except that we have derived $\epsff$ for the \citet{Lada13a} sample following \citet{Krumholz14a}, and we have homogenized $\epsff$ from the HCN data sets to the calibration of \citet{Onus18a}.
}
\end{figure}

We summarize recent observational constraints on $\epsff$ in \autoref{fig:epsff}. Comparing the various methods, we see that essentially all studies based on YSO counting or HCN give $\epsff \approx 0.01$, with a study-to-study dispersion of $\approx 0.3$ dex, and a dispersion of about $0.3 - 0.5$ dex within any single study. This dispersion is probably an upper limit, since it includes both the true physical dispersion and the effects of measurement errors. Pixel statistics produce a similar dispersion, but with a median $0.3 - 0.5$ dex lower. However, these results are still consistent because there are systematic uncertainties in all the methods at the $\sim 0.5$ dex level. For pixel statistics one must assume a value for the poorly-constrained line of sight depth through the target galaxy. For YSO counting, one must estimate the duration of the phase during which newly-formed stars would be classified as YSOs, the volume density of the gas seen only in projection, and the IMF correction. For HCN, there is uncertainty in the mean density of the emitting gas and the emissivity per unit mass, which is needed to estimate the total mass.\footnote{\autoref{fig:epsff} uses the \citet{Onus18a} calibration, which is intermediate between the alternative \citet{Kauffmann17a} and \citet{Gao04a} estimates.}

\begin{figure}
\includegraphics[width=\textwidth]{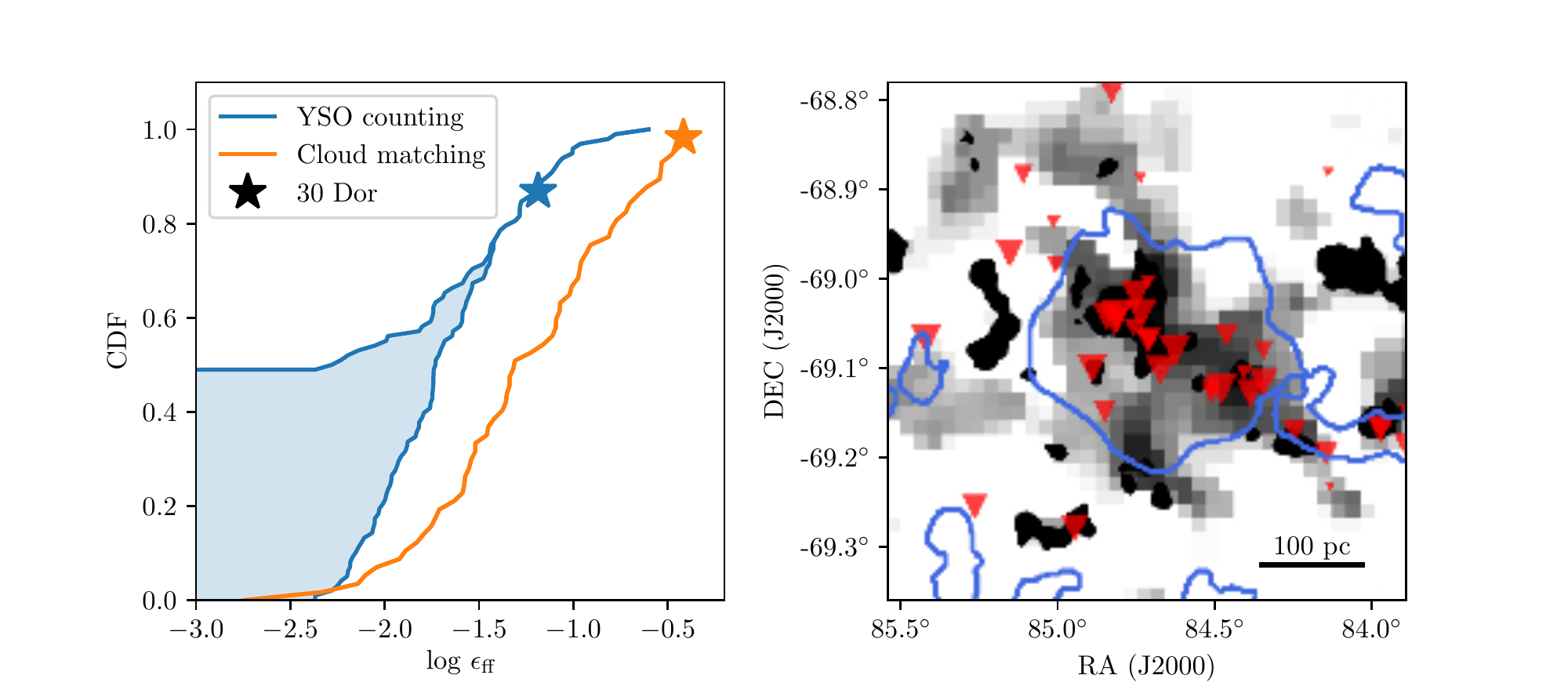}
\caption{
\label{fig:epsff_meth}
\textit{Left:} cumulative distribution of $\epsff$ values measured in the study of \citet{Ochsendorf17a} by YSO counting versus cloud matching, both using the \citet{Jameson16a} H$_2$ map. The star shows the 30 Doradus region. For the YSO counting method, the blue band indicates the uncertainty range associated with non-detections, with the lower value corresponding to assuming that clouds without YSO detections have star formation rates at the sensitivity limit, and the upper value corresponding to assuming that such regions have star formation rates of zero. \textit{Right:} zoom-in on the 30 Doradus region, reproduced from \citet{Ochsendorf17a} by permission of the AAS. Grayscale shows the H$_2$ column density, black contours show CO clouds from the catalog of \citet{Wong11a}, red triangles show the positions of massive YSOs, with size indicating inferred mass \citep{Ochsendorf16a}, and blue contours show H~\textsc{ii} regions.
}
\end{figure}

The cloud matching studies disagree with the other three methods, yielding much larger dispersions and mostly higher medians. The main difference between cloud matching and the other methods is that cloud matching does not require that the star formation and gas tracers be co-spatial, which makes the results sensitive how one constructs and matches catalogs of clouds and star-forming regions; the $0.8$ dex difference in median $\epsff$ values derived by \citet{Vutisalchavakul16a} and \citet{Lee16a} is almost entirely due to this sensitivity, since the regions they target and the underlying data sets they use are nearly identical. To gain more insight into why cloud matching differs from other methods, it is helpful to examine the study of \citet{Ochsendorf17a}, who use both YSO counting and cloud matching to derive $\epsff$ values for the same molecular gas map of the LMC.\footnote{\citet{Ochsendorf17a} analyze both a dust-based molecular cloud map from \citet{Jameson16a} and a CO-based map from \citet{Wong11a}. We focus on the dust-based map because, due to the LMC's low metallicity, is contains substantial amounts of CO-dark molecular mass.} For YSO counting, they adopt the standard approach of counting YSOs within the same contours used to define clouds, while for cloud matching they decompose an H$\alpha$ map into H~\textsc{ii} regions and match clouds and H~\textsc{ii} regions if there is any spatial overlap. We show the distribution of $\epsff$ values produced by the two methods in the left panel of \autoref{fig:epsff_meth}.

Clearly cloud matching yields a broader distribution with a higher median. To understand why, we can look at the example of the 30 Doradus region (indicated by the star in the left panel of \autoref{fig:epsff_meth}) for which YSO counting gives $\epsff = 0.065$, while cloud matching gives $\epsff = 0.38$, a factor of 6 larger. Since the same GMC properties are used in both estimates, the difference arises solely from the imputed star formation rate. Examining the right panel of \autoref{fig:epsff_meth}, it is clear how a substantial difference could arise: while the great majority of the YSOs are within the footprint of the cloud, there is significant H$\alpha$ emission that is not, but that partly overlaps it. Treating all this emission as if it comes from the cloud causes the cloud matching method to infer a much larger star formation rate. This emission could indeed be associated with the existing molecular cloud, but it could also be associated with molecular gas that has been dispersed or displaced by stellar feedback.

Similar effects are also apparent in other galaxies. For example, in a survey of NGC 628 at $\sim 50$ pc resolution, \citet{Kreckel18a} identify $\approx 1500$ H~\textsc{ii} regions and $\approx 750$ GMCs, but find $<100$ overlaps between the two. Naive application of a cloud matching method to this data set would lead one to conclude that most GMCs are inert and have $\epsff = 0$, while most H~\textsc{ii} regions formed with $\epsff \gg 1$, an obviously unphysical result. The data are better explained by the hypothesis that bright H~\textsc{ii} regions rapidly disperse or displace the molecular clouds in which they form. In principle it should be possible to account for this effect and extract corrected estimates of $\epsff$ from the statistics of the GMC-H~\textsc{ii} region correlation \citep{Kruijssen18a}, but this method has not yet been widely applied.

This analysis suggests that cloud matching-based estimates of $\epsff$ should be treated with caution. Given the reasonable consistency in results produced by the other methods, we conclude that the preponderance of the current observational evidence favors $\epsff \approx 0.01$ for regions $\gtrsim 1$ pc in size, with a dispersion of $\lesssim 0.5$ dex and a comparable level of systematic uncertainty.

\begin{textbox}[ht!]
\section{THE FUTURE OF YSO COUNTING}
YSO counting is the most reliable method of estimating $\epsff$, but at present it is limited by the sensitivity and angular resolution of \textit{Spitzer}, the telescope used for all existing YSO counting studies. With \textit{Spitzer}, the only YSOs we can reliably detect beyond the Milky Way have masses $\gtrsim 8$ $M_\odot$ \citep{Ochsendorf16a}. This sensitivity limit is responsible for the large uncertainty in the shape of the low end of $\epsff$ distribution shown in \autoref{fig:epsff_meth}. However, the \textit{James Webb Space Telescope} will be far more sensitive and have significantly better angular resolution. With \textit{JWST} it will possible to detect $\approx 1$ $M_\odot$ YSOs in the Magellanic Clouds, and to obtain YSO catalogs comparable to those currently available for the LMC and SMC out to M31 at least, possibly farther. This will greatly improve cloud-scale measurements of $\epsff$ beyond the Milky Way.
\end{textbox}

\subsubsection{Theory of $\epsff$}

The observed value of $\epsff$ is surprisingly low, and a number of authors have proposed theoretical models aimed at explaining it. The earliest models relied on magnetic regulation of star formation. Interstellar gas is magnetized, and if the field strong enough it can prevent gas from collapsing until non-ideal magnetohydrodynamic effects reduce the magnetic flux threading the gas. However, more recent measurements of magnetic fields have generally shown that they are too weak to support the gas -- see \citet{Crutcher12a} for a review. Consequently, attention has focused on three possibilities: $\epsff$ may be low because clouds are not bound, because they are turbulent, or due to feedback.

Unbound cloud models propose that $\epsff$ is low because most of the material observationally-defined as molecular clouds is not in fact self-gravitating \citep[e.g.,][]{Dobbs11a, Meidt18a}. We have argued in \autoref{ssec:giant} that the bulk of the evidence does not support this view, but if it were the case that only $\sim 1\%$ of the mass in a molecular cloud were bound, this would naturally explain why $\epsff\approx 1\%$. While this is a potentially-viable explanation for the low $\epsff$ values of GMC as traced by CO, most of the observations compiled in \autoref{ssec:obsepsff} are for the much denser gas traced by HCN, or by the YSO counting studies, which is almost certainly bound \citep[e.g.,][]{Kauffmann13}. The unbound cloud hypothesis does not explain why this gas also shows low $\epsff$.

Turbulent regulation models propose that $\epsff$ is low because star-forming gas is turbulent enough to render most sub-regions of a molecular cloud unbound, even if that cloud is bound on its largest scales. Qualitatively the argument stems from the linewidth-size relation discussed in \autoref{ssec:initial_conditions}: since velocity dispersion varies with size as $\sigma \propto R^{1/2}$, then the kinetic energy per unit mass contained within a region of size $R$ scales as $e_K\propto R$, while the binding energy per unit mass of a region of mass $M$ scales as $e_G \propto GM/R$. If one chooses a region close to the mean density then $M \propto R^3$ and $e_G \propto R^2$. Consequently, for a randomly-chosen sub-region of a cloud, the virial ratio obeys $\alpha_{\rm vir} \propto 1/R$, and thus most regions smaller than an entire cloud are unbound. This argument can be made quantitative by integrating over the density PDF. The earliest version of such an argument appeared in \citet{Krumholz05a}, but this has been superseded by numerous later works that model the density PDF and the evolution of its self-gravitating parts with increasing accuracy \citep{Padoan11a, Hennebelle11a, Padoan12a, Federrath12a, Hopkins13b, Burkhart18a, Burkhart18b}.

The consensus of recent models is that turbulence does substantially reduce $\epsff$, but not all the way to $\epsff\approx 0.01$ as required by the observations. As time goes by the density PDF of a self-gravitating cloud deviates develops an increasingly-prominent power-law tail on its high-density end that causes $\epsff$ to rise with time \citep{Murray15a, Burkhart18a}. This density build-up is most likely counteracted by localized feedback processes that break up high-density regions and suppress the growth of the power-law tail; the most likely mechanism for this is protostellar outflows, since stars begin to launch outflows as soon as they form, and even low-mass stars can produce significant outflow feedback. Modern simulations that include both protostellar outflows and magnetic fields\footnote{Non-MHD simulations find that outflows are much less effective at lowering $\epsff$, since without magnetic fields outflows couple poorly to the bulk of the material \citep[e.g.,][]{Krumholz12b, Murray18a}.} that help couple them to gas are able to achieve values of $\epsff$ that, while still a factor of $2-3$ too high compared to the consensus observational value, are approaching agreement with the data \citep{Wang10a, Myers14a, Federrath15a, Cunningham18a}. An alternative possibility is that cloud disruption by feedback is so fast and efficient that typical clouds never have time to develop power-law tails extensive enough to drive $\epsff$ to large values \citep{Grudic18a}.

A final caveat on theoretical explanations for the low value of $\epsff$ is that turbulent regulation is effective only if the gas is actually turbulent. There is significant debate in the literature about whether the observed large linewidths of molecular clouds should be interpreted as turbulence, or whether they might represent coherent gravitational collapse \citep[e.g.,][]{Heitsch09a, Ballesteros-Paredes11a, Zamora-Aviles12a, Traficante18a}. While there is significant evidence from both analytic theory \citep{Klessen10a, Goldbaum11a} and numerical simulations \citep{Robertson12a, Lee16b, Lee16c, Birnboim18a} that accretion flows invariably drive turbulence, and recent kinematic measurements with \textit{Gaia} also appear to disfavor the idea that GMCs are in global collapse (\autoref{ssec:proposed}), the question is not settled.

\subsection{Feedback and the termination of star formation}
\label{ssec:feedback}

\begin{figure}
\includegraphics[width=\textwidth]{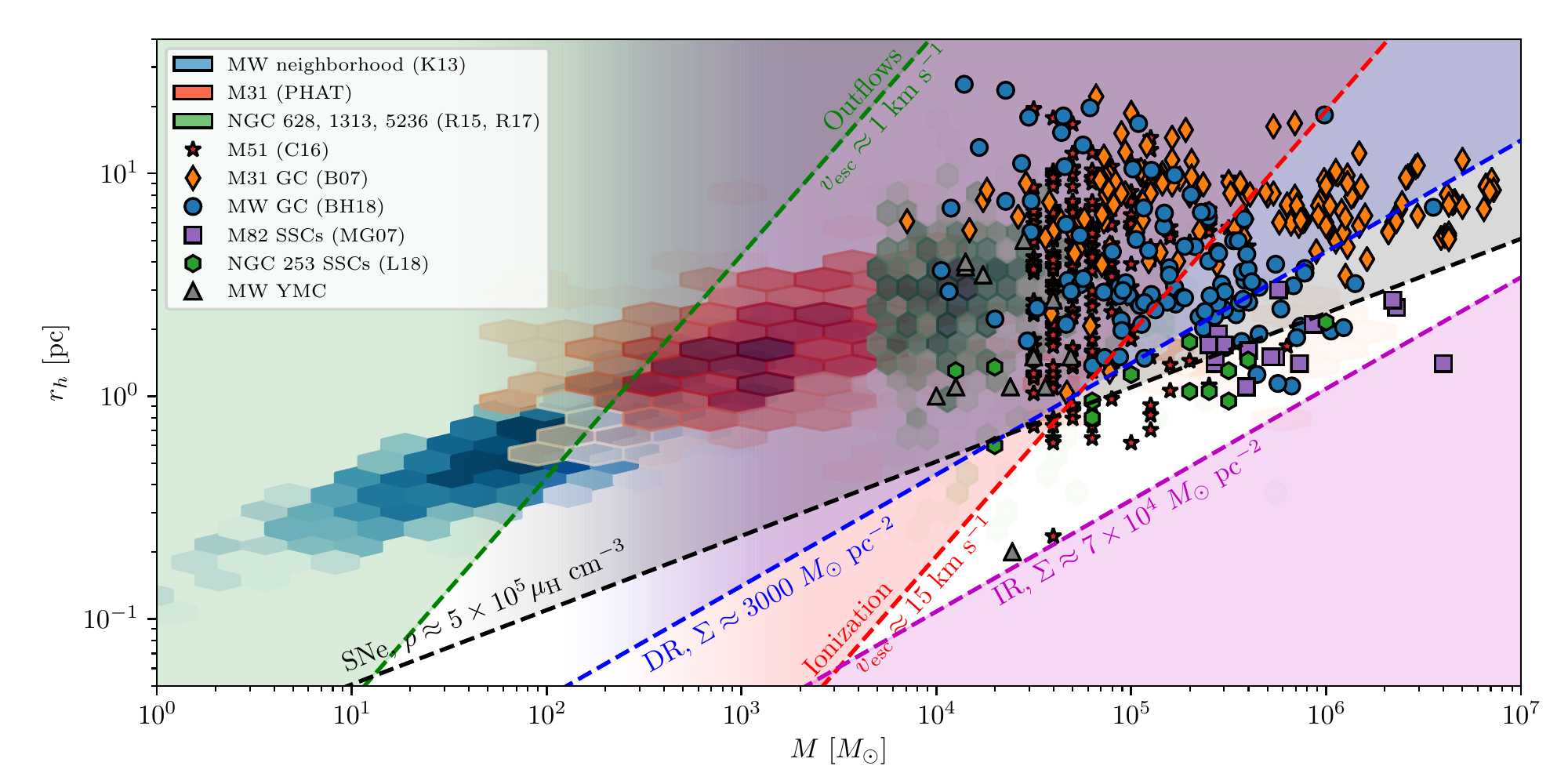}
\caption{
\label{fig:mass_radius_feedback}
Star cluster mass-radius relation, showing the same data as \autoref{fig:mass_radius}, with shaded regions indicating where various feedback mechanisms are potentially significant. In constructing this diagram, we have for simplicity considered the case of star-forming clouds where the gaseous and stellar components have comparable masses and radii, but we emphasize that in reality this need not be the case. The feedback mechanisms shown are outflows (\autoref{sssec:outflows}), supernovae (SNe; \autoref{sssec:sne}), direct radiation pressure (DR; \autoref{sssec:dr}), ionization (\autoref{sssec:ionization}), and indirect radiation pressure (IR; \autoref{sssec:ir}). Shaded regions fade at masses below which stochastic sampling of the IMF makes them unlikely to be active (\autoref{sssec:outflows}). For all feedback types except IR, the region where the mechanism is effective is to the left of and above the line.
}
\end{figure}

The final star formation efficiency, $\epsilon_*$, depends both on $\epsff$ and on the ability of stellar feedback to eject gas from nascent star clusters. Unfortunately we cannot measure $\epsilon_*$ for individual clouds directly, because for a single region we cannot measure both the gas mass at the onset of star formation and the stellar mass at its conclusion. One can attempt to measure the mass-loading factor $\eta$, but this is technically very challenging (see \citealt{Yang18a} for a recent attempt). The best prospect for measuring $\epsilon_*$ is likely the use of statistical methods to analyze populations of clouds \citep{Kruijssen18a}, but thus far measurements using this technique are not widely available. For these reasons, estimates of $\epsilon_*$ come primarily from theory. In the remainder of this section we follow the approach taken by several previous authors \citep[e.g.,][]{Matzner02a, Fall10a, Matzner15a, Rahner17a} by considering a variety of feedback mechanisms and attempting to determine under what circumstances they become relevant. We summarize these findings in \autoref{fig:mass_radius_feedback}.

\subsubsection{Protostellar outflows}
\label{sssec:outflows}

\citet{Bally16a} provides a comprehensive recent review of protostellar outflows, so we focus only on the details most relevant for star cluster formation. Outflows are critical to breaking up dense regions and thus keeping $\epsff$ small because they eject $\sim 2/3$ of the mass from individual protostellar cores \citep[e.g.,][]{Matzner00a, Offner17a}. However, this mass is ejected at relatively low velocities, and thus may not escape the larger-scale protocluster. For this reason, theoretical models \citep[e.g.,][]{Matzner00a, Matzner15a}, numerical simulations \citep[e.g.,][]{Wang10a, Krumholz12b, Murray18a}, and observations \citep[e.g.,][]{Nakamura14a, Li15a, Plunkett15a} all suggest that outflows have a limited role in setting final star formation efficiencies in most clusters. There has yet to be a comprehensive numerical survey of the parameter space, but \citet{Matzner15a} estimate analytically that outflows produce $\epsilon_*\lesssim 0.5$ only in clusters with escape speeds $v_{\rm esc} \lesssim 1$ km s$^{-1}$. We show this limit in \autoref{fig:mass_radius_feedback}.

Despite this limitation, outflows play a unique role, because they do not depend on the presence of massive stars. The momentum per unit mass of stars formed delivered by protostellar outflows to their surroundings is of order the escape speed from a protostellar surface \citep[e.g.,][]{Matzner00a}. Low-mass and high-mass protostars have similar surface escape speeds, so low-mass stars are as effective at providing outflow feedback as massive ones. By contrast, for all the other mechanisms we discuss below, feedback is dominated by massive stars unlikely to be found in small clusters. To quantify this point, we use the \texttt{SLUG} stochastic stellar populations code \citep{da-Silva12a, Krumholz15b} to measure the PDFs for the number of supernovae, and bolometric and ionizing luminosity, in clusters at a range of masses for a \citet{Chabrier05a} IMF. We find that the number of expected supernovae exceeds unity only for clusters with masses $\gtrsim 100$ $M_\odot$, and that the median (bolometric, ionizing) luminosity of clusters is $<50\%$ of the value for a fully-sampled IMF in clusters with mass $\lesssim (400, 700)$ $M_\odot$. Consequently, protostellar outflows are likely to be the only feedback mechanism that limits $\epsilon_*$ for stellar populations smaller than a few hundred $M_\odot$. This limit is very blurry due to stochasticity; some clusters with masses of only a few hundred $M_\odot$ nevertheless have substantial ionizing luminosities \citep{Andrews14a}.

\subsubsection{Photoionization feedback}
\label{sssec:ionization}

A population of stars that fully samples the IMF produces ionizing photons at a rate of $\Psi_i \approx 6\times 10^{46}$ s$^{-1}$ $M_\odot^{-1}$ \citep{Murray10b}. These can heat gas in star-forming clouds to $\approx 10^4$ K, giving it a sound speed of $\approx 10$ km s$^{-1}$. If not trapped either by the surrounding material or by gravity, this gas will flow outward in a wind known as a champagne flow, directly removing mass from the cloud. The back-reaction created when the ionized gas pushes off the surrounding neutral material, known as the rocket effect, may eject even more mass and exert significant forces, ultimately unbinding clouds entirely. The mass removed in this process can be considerable -- the analytic calculations of \citet{Williams97a} and \citet{Matzner02a} suggest that a $10^4$ $M_\odot$ population of stars can eject $\approx 10$ times its own mass over a few Myr, giving $\epsilon_* \sim 0.1$. Numerical simulations generally confirm this result, though they show that the value of $\epsilon_*$ produced by ionization also depends strongly on the initial boundedness of clouds, and on whether or not they are magnetized \citep[e.g.,][]{Dale12a, Dale13a,  Howard17a, Geen15b, Geen16a, Gavagnin17a, Dale17a, Kim18a}.

The main limitation on photoionization as a feedback mechanism is that the ionized gas must be able to escape, but there is some tension in the simulation literature about how much of a limitation this represents. \citet{Dale12a, Dale13a} find that photoionization is unable to unbind more than a few percent of the mass for clouds with $v_{\rm esc} > 10$ km s$^{-1}$, while \citet{Kim18a} find $\epsilon_* \lesssim 0.5$ at all escape speeds $\lesssim 20$ km s$^{-1}$, though it is possible that this is because they include direct radiation pressure as well as photoionization (see \autoref{sssec:dr}). None of the other published studies explore a range of $v_{\rm esc}$ values. It is plausible that even regions with $v_{\rm esc} > 10$ km s$^{-1}$ could lose mass due to photoionization, since freshly-ionized gas rocketing off a dense neutral surface will generally accelerate to small multiples of the ionized gas sound speed. In the analogous problem of photoionized gas driving winds off accretion disks, simulations show that even regions with escape speeds $\approx 30$ km s$^{-1}$ launch significant winds \citep{Woods96a}. Thus escape speed required to suppress the effects of photoionization is of order 10 km s$^{-1}$, but its exact value is uncertain; in \autoref{fig:mass_radius_feedback} we adopt 15 km s$^{-1}$.

\subsubsection{Direct radiation pressure} 
\label{sssec:dr}

A zero-age population of stars that fully samples the IMF has a light to mass ratio $\Psi \approx 1100$ $L_\odot$ $M_\odot^{-1}$ \citep[e.g.,][]{Fall10a}. The light is emitted mostly at UV wavelengths where the ISM is highly opaque, so that even a modest dust column is sufficient to ensure that essentially all the radiation momentum is deposited in the gas. Whether this affects $\epsilon_*$ depends on how the associated radiation force compares to pressure and gravitational forces. H~\textsc{ii} region gas pressures vary with radius as $r^{-3/2}$ due to ionization balance, while direct radiation pressure varies as $r^{-2}$, so gas pressure dominates once H~\textsc{ii} regions reach a characteristic size $r_{\rm ch} \approx 0.03 S_{49}$ pc, 
where $S_{49}$ is the ionizing luminosity in units of $10^{49}$ s$^{-1}$ \citep{Krumholz09a}. For a stellar population of mass $M_*$ with ionizing luminosity $S = \Psi_i M_*$, one can show that this radius is larger than the size of the cluster, and thus radiation pressure rather than photoionization dominates mass removal, only if the escape speed $v_{\rm esc} \gtrsim 10$ km s$^{-1}$. Thus direct radiation pressure becomes dominant roughly where photoionized gas pressure ceases to be effective.

The importance of direct radiation pressure relative to gravity depends on the cloud column density \citep{Fall10a, Murray10a}. Consider a spherical gas cloud of total (gas plus stellar) mass $M$ and radius $R$, with a stellar population of mass $\epsilon_* M$ forming at its center. The column-averaged outward force per unit gas mass supplied by radiation is $\Psi \epsilon_* M/4\pi R^2 (1-\epsilon_*) \Sigma c$, where $\Sigma=M/\pi R^2$ is the mass per unit area. The corresponding column-averaged inward force per unit mass from gravity is $\approx G M/R^2$. Thus the Eddington ratio, which defines the ratio of radiative to gravitational force, is $f_{\rm Edd} = \Sigma_{\rm DR} \epsilon_*/(1-\epsilon_*)$, where
\begin{equation}
\Sigma_{\rm DR} = \frac{\Psi}{4\pi G c} \approx 340\,M_\odot\mbox{ pc}^{-2}
\end{equation}
is the surface density below which radiation direct pressure becomes important. Naively we would expect direct radiation pressure to expel gas and thus limit $\epsilon_*$ only for $f_{\rm Edd} \gtrsim 1$, requiring clouds with $\Sigma \lesssim \Sigma_{\rm DR}$. However, \citet{Thompson16} point out that in a turbulent medium there will be sightlines with $\Sigma < \Sigma_{\rm DR}$ even if $\Sigma > \Sigma_{\rm DR}$ on average. Gas can be ejected on these low-column sightlines, leading to a wind. Their models suggest that, in a cloud with $\epsff \approx 0.01$, this wind will remove mass fast enough to produce $\epsilon_* \approx 0.5$ at $\Sigma \approx 10 \Sigma_{\rm DR}$; we show this line in \autoref{fig:mass_radius_feedback}.

Simulations of the effects of direct radiation pressure have produced mixed results. All simulators find that direct radiation pressure is ineffective at reducing $\epsilon_*$ at surface densities $\gtrsim 10\Sigma_{\rm DR}$ \citep[e.g.,][]{Grudic18a}, but at the surface densities of $\Sigma \approx 100$ $M_\odot$ pc$^{-2}$ characteristic of Milky Way GMCs some simulators find that direct radiation pressure yields $\epsilon_* \lesssim 0.1$ \citep{Grudic18a}, while others find substantially larger values of $\approx 0.3-0.5$ \citep{Raskutti16a, Kim18a, Howard18a}. These disagreements may be due to differences in the initial conditions, or to resolution problems in the simulations that can lead to a significant overestimate in the effectiveness of direct radiation pressure in some numerical methods \citep{Krumholz18b}.

\subsubsection{Indirect radiation pressure}
\label{sssec:ir}

When interstellar dust absorbs radiation, it re-emits the energy at IR wavelengths. IR dust opacities are a factor of $\sim 100$ smaller than UV ones, but regions of sufficiently high column density may nonetheless be opaque to the re-radiated IR. In this case the IR photons will be absorbed again, and can deposit additional momentum; in clouds opaque enough to require many cycles of absorption and re-emission before the energy escapes, the resulting force will greatly exceed the direct radiation force \citep{Thompson05a, Murray10a, Thompson15a}.

Indirect (i.e., dust-reprocessed) radiation pressure feedback is governed by two processes: frequency diffusion and radiation Rayleigh-Taylor (RRT) instability. Frequency diffusion is important because IR wavelengths are much larger than interstellar grain radii, so dust opacity varies with frequency as $\kappa \propto \nu^2$. Thus as stellar radiation diffuses outward through a dusty envelope and mean photon frequency decreases, dust-radiation coupling weakens. Eventually photons shift to frequencies low enough to escape. This effect greatly limits the importance of dust-reprocessed radiation \citep{Wolfire87a, Reissl18a}, and means that the mass-weighted mean opacity of a cloud is a function of the total gas column. Many subgrid models of radiation pressure ignore this effect and simply assume a constant IR opacity \citep[e.g.][]{Rosdahl15a, Tsang18a, Hopkins18a}; given the importance of diffusion, their results should be treated with caution.

RRT instability \citep{Jacquet11a} causes gas that is being accelerated out of a gravity well by indirect radiation to become clumpy, reducing the effectiveness of its coupling to radiation. Early numerical work suggested this would prevent gas ejection entirely \citep{Krumholz12c}, but this result proved to be sensitive to the details of the radiation transfer method. Simulations based on flux-limited diffusion \citep{Krumholz12c} or M1 closures \citep{Rosdahl15a, Skinner15a} find that RRT prevents gas ejection unless the dust opacity is greatly boosted compared to observed Milky Way values, while those based on variable Eddington tensor \citep{Davis14a} or implicit Monte Carlo \citep{Tsang15a} find that it does not, though it does greatly reduce the ejection velocity. The latter two methods are likely more reliable, so we expect that indirect radiation pressure can limit $\epsilon_*$.

Under what conditions will this happen? \citet{Krumholz12c} show, and simulations by \citet{Davis14a} confirm, that for any column of gas confined by gravity, there is a maximum radiation flux that can pass through it without triggering gas ejection. Calculations properly accounting for frequency diffusion show that a gas cloud forming stars at its center will exceed the critical flux before the star formation efficiency $\epsilon_*$ reaches 0.5 only if its surface density exceeds \citep{Crocker18b, Crocker18a} 
\begin{equation}
\label{eq:Sigma_IR}
\Sigma_* \gtrsim \Sigma_{*,\rm IR} = \frac{16 \sqrt{\pi G \sigma_{\rm SB}/c}}{\Psi}\left(\frac{\kappa_0}{T_0^2}\right)^{-1} \approx 7\times 10^4 \left(\frac{\kappa_0}{0.03\;\mathrm{cm}^2\;\mathrm{g}^{-1}}\right)^{-1}\,M_\odot\,\mathrm{pc}^{-2},
\end{equation}
where $T_0 = 10$ K and $\kappa_0$ is the opacity per unit gas mass to radiation with a color temperature $T_0$; the normalization used above is typical of Milky Way dust. We show this condition in \autoref{fig:mass_radius_feedback}. Note that indirect radiation pressure has a \textit{minimum} surface density at which it becomes effective, while all the feedback mechanisms discussed previously have a \textit{maximum} surface density (for direct radiation) or escape velocity (for ionization) below which they are effective. The reason for this difference is the frequency diffusion effect: for a fixed radiation flux, a larger gas column traps heat more effectively and thus the amount of indirect radiation force delivered rises superlinearly with the gas column. Consequently, indirect radiation force becomes more effective relative to gravity as the total column density increases.

\subsubsection{Hot star winds}

Stars with surface temperature $\gtrsim$ 25,000 K and metallicity $\gtrsim 0.5Z_\odot$ drive winds with speeds of several thousand km s$^{-1}$ \citep[e.g.,][]{Vink01a}, producing temperatures $\gtrsim 10^7$ K when the wind shocks against the ISM. At this temperature the radiative cooling time is long compared to the dynamical time (unlike for protostellar outflows -- \autoref{sssec:outflows} -- or winds from intermediate-mass stars, \citealt{Offner15a}), leading to the formation of hot bubbles that can push on surrounding colder gas and potentially eject it from a cloud. There are a number of analytic models describing this process in one dimension without cooling (see \citealt{Bisnovatyi-Kogan95a} for a review) and with cooling \citep{Koo92a}, and several authors have published 1D numerical results including cooling \citep[e.g.,][]{Krause16a, Fierlinger16a, Rahner17a, Silich18a, Naiman18a}. The general conclusion from these models is that winds will be the dominant feedback mechanism unless their effects are reduced by two phenomena that cannot be properly captured in 1D: hydrodynamic leakage of hot gas through low-density channels, and sapping of hot gas thermal energy via turbulent mixing at the hot-cold interface. 
The general conclusion from these models is that winds will be the dominant feedback mechanism unless their effects are reduced by two phenomena that cannot be properly captured in 1D: hydrodynamic leakage of hot gas through low-density channels, and sapping of hot gas thermal energy via turbulent mixing at the hot-cold interface. 

To constrain the importance of leakage and mixing, and thus the effectiveness of hot star winds at reducing $\epsilon_*$, there are two approaches available: multi-dimensional simulations and observations. Published simulations find that, as a result of efficient leakage \citep{Rogers13a, Calura15a, Wareing17a} or mixing-induced losses \citep{Mackey15a}, wind feedback lowers $\epsilon_*$ by at most tens of percent. However, there has yet to be a comprehensive parameter study. In addition, only \citet{Wareing17a} consider magnetized clouds, and the fields they assume are weak. It is possible that more realistic, stronger magnetic fields might reduce mixing losses \citep{Gentry18a}. Moreover, winds may be important not so much because they remove mass, but because they halt ongoing accretion of low-density gas onto molecular clouds \citep[e.g.,][]{Gatto17a, Haid18a}.

X-ray observations probe wind feedback because the pressure in hot gas is directly related to its X-ray luminosity. The observed luminosities of diffuse gas around YMCs is rule out the possibility that the wind is trapped \citep{Dunne03a, Townsley03a, Townsley06a, Townsley11a}, and require significant leakage or mixing-induced radiative loss \citep{Harper-Clark09a, Rosen14a, Rogers14a, Toala18a}. Indirect diagnostics based on IR line ratios are consistent with this conclusion \citep{Yeh12a}. Comparing the hot gas pressure in H~\textsc{ii} regions to the pressure exerted by warm ionized gas and radiation, \citet{Lopez11a, Lopez14a} find that hot gas is subdominant in the massive 30 Doradus region, and in a wide range of smaller H~\textsc{ii} regions in the LMC and SMC. \citet{Pellegrini11a} reach the opposite conclusion for 30 Doradus, but only under the assumption that the hot gas has a small volume filling factor, which \citet{Lopez14a} point out would make it locally significant but unimportant as an agent of gas clearing on large scales. Thus there is currently no observational evidence that hot star winds are important in setting $\epsilon_*$. However, as with the simulations, observations cover only a narrow range of parameter space, and it is possible that future work might identify regimes where winds are important.

\subsubsection{Supernovae}
\label{sssec:sne}

Supernovae (SNe) deliver about the same energy as hot star winds, but all at once rather than gradually over several Myr. Whether this makes them more effective is uncertain. Some simulators find that SNe effectively destroy molecular clouds \citep{Rogers13a, Calura15a, Wareing17a}, while others find that their effects are modest compared to other feedback mechanisms \citep{Kortgen16a, Geen16a, Rey-Raposo17a}. Simulation results appear to be sensitive to both resolution and numerical method \citep{Gentry17a, Gentry18a}, and there has yet to be a systematic survey of parameter space. Moreover, analytic models suggest that photoionization feedback prior to the first SN strongly affects the final outcome \citep{McKee84a, Matzner02a}, but of the published studies, only \citet{Geen16a} include this effect.

One important limitation on the importance of SN feedback is time delay: even the most massive stars require $t_{\rm SN} \approx 3$ Myr to explode, and sufficiently dense regions may convert a significant fraction of their gas to stars before the first SN occurs \citep{Fall10a, Kruijssen12a}. A region with free-fall time $t_{\rm ff}$ and star formation rate efficiency $\epsff$ and no feedback mechanisms except SNe will convert a fraction $\epsilon_* = \epsilon_{\rm ff} t_{\rm SN}/t_{\rm ff}$ of its mass to stars before the first SN. This yields $\epsilon_* <0.5$ only if the density obeys
\begin{equation}
\label{eq:rho_SN}
\rho < \rho_{\rm SN} \approx \frac{3\pi}{128 G \epsff^2 t_{\rm SN}^2} \approx 5\times 10^5 \mu_{\rm H} \left(\frac{\epsff}{0.01}\times\frac{ t_{\rm SN}}{4\mbox{ Myr}}\right)^{-2}\mbox{ cm}^{-3},
\end{equation}
where $\mu_{\rm H} = 2.3\times 10^{-24}$~g is the mean mass per H nucleus for the usual cosmic composition of 27\% He by mass. We show this criterion in \autoref{fig:mass_radius_feedback}. Moreover, SNe may be unimportant even in regions that obey \autoref{eq:rho_SN} because other mechanisms might remove gas on a timescale below $t_{\rm SN}$. There is substantial observational evidence that this is the case in at least some regions: clusters with ages $\approx t_{\rm SN}$ are observed to be gas free both in the Milky Way \citep{Longmore14a} and in M83 \citep{Hollyhead15a}, and the spectra of clusters in the process of gas clearing frequently show Wolf-Rayet features, indicating an evolutionary phase that precedes the first SN \citep{Sokal16a}. These observations strongly disfavor SNe as an important mechanism for gas removal. However, much like winds, the main effect of SNe may not be to clear dense gas, but rather to remove low-density gas and thereby prevent star formation from restarting after gas clearing.

\subsubsection{Summary of feedback mechanisms and expectations on $\epsilon_*$}

\autoref{fig:mass_radius_feedback} provides an overall picture of stellar feedback. For the smallest clusters, $M \lesssim 100$ $M_\odot$, protostellar outflows are the only mechanism of interest because the massive stars responsible for other feedback mechanisms are likely to be absent. Moving to slightly higher masses, in the great majority of the parameter space occupied by observed clusters in the disks of modern galaxies, and even for a substantial portion of the globular cluster population, SNe, photoionization, and direct radiation pressure are all at least potentially effective. When all three are capable of operating, both simulations and observations suggest that photoionization will dominate because it has no time delay, and photoionized gas pressure exceeds radiation pressure except in very compact star clusters with large escape speeds. Most star formation in this region should be characterized by efficiencies $\epsilon_* \lesssim 0.3$, since simulations suggest that photoionization feedback is quite effective when it operates. This is also in accord with \citet{Lada03a}'s observation that, for embedded clusters with masses $\gtrsim 10^3$ $M_\odot$, the stellar mass fraction seems to increase with age, but only to a maximum of $\approx 30\%$.

As we move to yet higher masses and densities, characteristic of globular clusters, YMCs, and SSCs, these three feedback mechanisms begin to become ineffective \citep[see also][]{Grudic18a}. Ionization likely becomes ineffective first, followed by direct radiation pressure and SNe; this ordering is dictated simply by the mechanisms' dependence on mass and radius, as shown in \autoref{fig:mass_radius_feedback}. Presumably $\epsilon_*$ increases in this region. Indirect radiation, unlike all the other feedback mechanisms, only becomes effective once the surface density rises above a threshold value. The observed cluster mass-radius distribution shows a maximum surface density (first pointed out by \citealt{Hopkins10a}) consistent with this limit, suggesting that indirect radiation pressure either lowers $\epsilon_*$ at surface densities above the threshold, or causes cloud radii to expand during star formation to drive them below it \citep{Crocker18b}.

Interestingly there is a locus, represented by the unshaded triangle in \autoref{fig:mass_radius_feedback}, where no known feedback mechanism is expected to be effective. This locus is occupied by a small population of massive, dense star clusters, generally classified as SSCs. Our analysis suggests that these clusters may form with very high $\epsilon_*$ values, simply because no feedback mechanism is capable of ejecting much gas from them. However, we caution that it is probably not realistic to assume that protoclusters spend their entire formation history on this locus; as we discuss below, real protoclusters almost certainly grow by gas accretion and form stars simultaneously, and thus will move in \autoref{fig:mass_radius_feedback}.

\subsection{Star formation histories and timescales}
\label{ssec:tsf}

\subsubsection{Observational constraints}

The final ingredient in our simple model of star cluster formation is its duration $t_{\rm sf}$, or, more generally, the star formation history. Age-dating stars is a fraught problem \citep{Soderblom14a}, so the conclusions we draw must be regarded as somewhat tentative, though the situation has improved considerably in the last few years.

\begin{figure}
\includegraphics[width=0.85\textwidth]{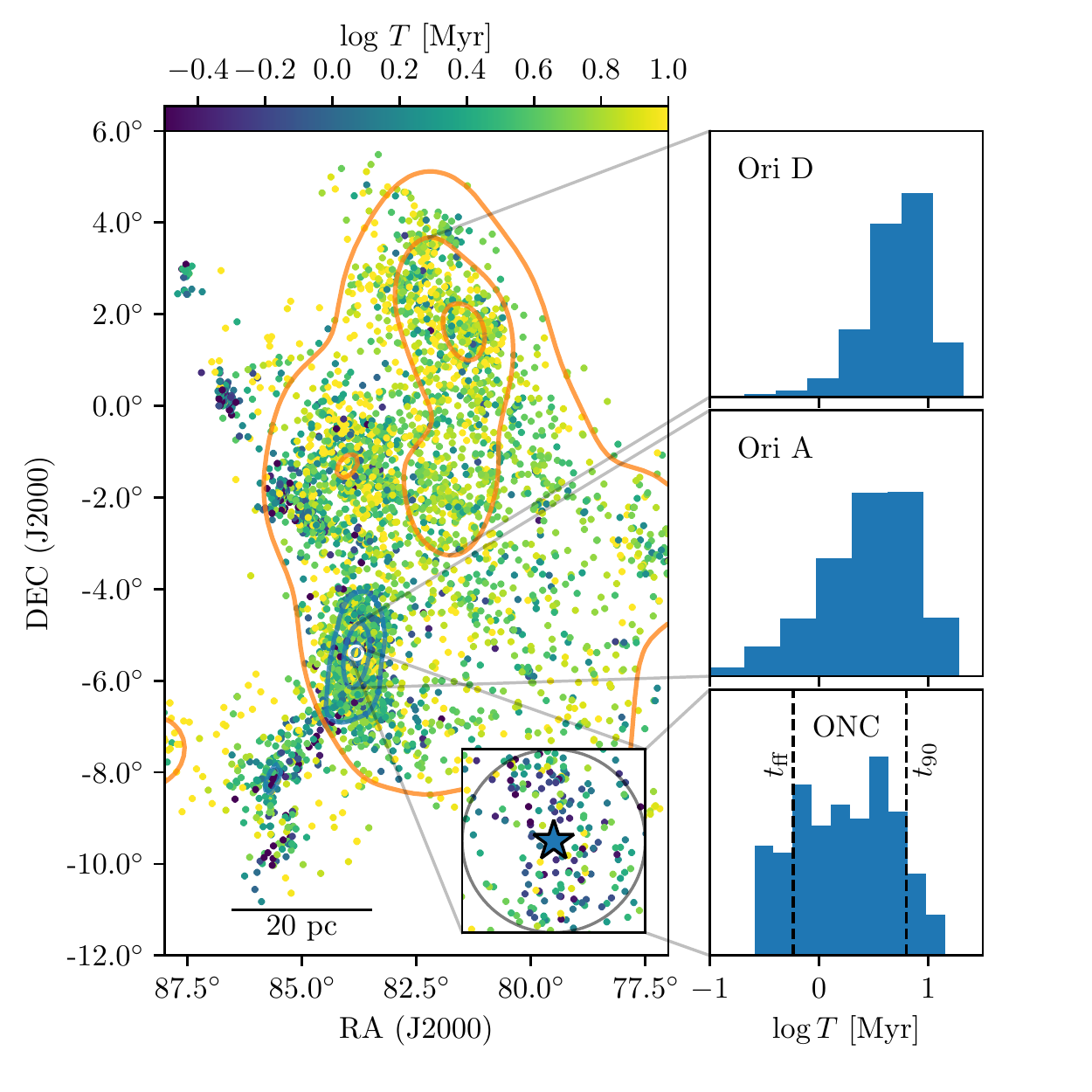}
\caption{
\label{fig:orion_age}
Distribution of stellar ages in the Orion star-forming complex. In the main panel, points are stars color-coded by age as estimated by \citet{Kounkel18a}. The orange and blue contours, respectively, show the density of stars on the sky associated with the components Orion D and Orion A identified in \citeauthor{Kounkel18a}'s phase-space analysis; contours are placed at densities of 10\%, 50\%, and 90\% of maximum density. The inset shows a zoom-in on the ONC, defined here as a 1 pc-radius region centered on $\theta^1$ C (shown by the star). The flanking histograms show the age distributions for Orion D, Orion A, and the ONC, as indicated. In the ONC panel, we also show $t_{\rm ff}$, the free-fall time in the central pc of the ONC as estimated by \citet{Da-Rio14a}, and $t_{\rm 90} \approx 10 t_{\rm ff}$, the time over which 90\% of the stellar population formed. The ages we show are those estimated based on spectroscopy (\citeauthor{Kounkel18a}'s age$_{\rm HR}$) where available, and based on color (\citeauthor{Kounkel18a}'s age$_{\rm CMD}$) elsewhere. However, in the ONC inset and histogram we show only spectroscopic ages, since the color-based ones are unreliable in regions of high extinction.
}
\end{figure}

By far the most comprehensively-studied region is the Orion complex. The young stars in Orion are spread over several tens of pc, but molecular gas and ongoing star formation are limited to a smaller $\sim 10$ pc region around the ONC, which is $\approx 1$ pc in size (\citealt{Hillenbrand98a} find $r_h \approx 0.8$ pc) and is centered on $\theta^1$ Ori C, the most massive star in the complex. The ONC is the densest region, and, despite its small volume, represents a significant fraction of the total number of stars in the complex. We illustrate one recent measurement of the age distribution in Orion in \autoref{fig:orion_age}, which suggests three important conclusions that appear to apply to other star-forming regions as well \citep[e.g.,][]{Tan06a, Azimlu15a, Getman18a, Caldwell18a}. First, the densest regions show age spreads that are substantially larger than their dynamical times. Using the ages shown in \autoref{fig:orion_age}, the time over which 90\% of the ONC stars formed is $\approx 6$ Myr \citep[other studies give similar results:][]{Reggiani11a, Da-Rio16a, Beccari17a}, while the free-fall time is $\approx 0.6$ Myr \citep{Tan06a, Da-Rio14a}. Consequently, star formation has been ongoing for $\approx 10$ free-fall times. While the absolute ages and the magnitude of the age spread depend on the choice of age indicator, the conclusion that there is a substantial age spread is robust against this choice, and against possible contamination due to photometric variability \citep{Messina17a} or variable accretion histories (see Sidebar). This conclusion is also bolstered by kinematic observations showing that the ONC is very close to virial equilibrium \citep{Kim19a}, implying that it must have had time to relax dynamically. 

Second, while star formation has an extended history, the star formation rate has not been constant over this period; star formation has accelerated \citep{Palla00a, Caldwell18a}. (In \autoref{fig:orion_age}, recall that the time axis is logarithmic, so constant star formation rate corresponds to a line of slope unity rather than zero.) Using the ages shown in \autoref{fig:orion_age}, we find that the time over which 90\% of the stars formed is $\approx 10t_{\rm ff}$, but roughly 50\% of stars formed in the last $\approx 3t_{\rm ff}$.

\begin{textbox}[ht!]
\section{FALSE AGE SPREADS FOR ACCRETING YSOS?}
Some authors have suggested that the dispersion in YSO luminosity at fixed effective temperature, usually interpreted as an age spread, might instead be due to variation in YSO accretion histories \citep[e.g.,][]{Baraffe09a, Baraffe12a}. If some YSOs were to undergo exclusively ``cold'' accretion, meaning that accreting material radiates away all its entropy at the accretion shock, they would descend the Hayashi track more rapidly than YSOs with less efficient shocks, leading to a luminosity spread that could be mistaken for an age spread. However, the current consensus is that observed stellar age spreads are \textit{not} due to this effect. Although both cold accretion and Kelvin-Helmholtz contraction drive stars down the Hayashi track, they do so at very different rates as a function of stellar mass. As a result, rather than generating a mass-independent age spread, cold accretion would make it appear that stars with effective temperatures $\gtrsim 3500$ K are systematically older than cooler stars \citep{Hosokawa11a}. For these higher $T_{\rm eff}$ stars, cold accretion would also produce a small population that would be extremely luminous and thus appear young \citep{Vorobyov17a, Jensen18a}, rendering the overall apparent age distribution bimodal. Neither a bimodal inferred age distribution for hotter stars nor a systematic difference in inferred age as a function of effective temperature are observed, which puts strong limits on how much of the apparent age spread is due to cold accretion.
\end{textbox}

Third, mean age varies with position. The stars in the ONC are significantly younger than the mean of the larger Orion A region, which is in turn younger than the still-larger Orion D region. Consequently, there is a gradient of increasing mean age and age spread as one moves from denser to less dense regions \citep{Getman14a, Getman18a}. However, age does not increase as quickly as free-fall time, so while the ONC is $\approx 10$ free-fall times old, the lower density regions around it in Orion A are only $\sim 1-3$ free-fall times old \citep{Da-Rio14a, Jaehnig15a}. Thus the outskirts of clusters are older than their centers in an absolute sense, but are dynamically younger. This statement likely applies to Orion D as well, but we cannot be sure because the gas has already been cleared, so the stellar density only provides an upper limit on the free-fall time.\footnote{This observation reinforces the problem with cloud matching identified in \autoref{ssec:epsff}: cloud matching would likely attribute the stars in Orion D to the present-day molecular gas cloud that is confined to the Orion A region. However, kinematic measurements demonstrate that the stars in Orion D did not form in the Orion A cloud, but from a separate, now-dispersed reservoir of molecular gas.}

\subsubsection{Proposed scenarios}
\label{ssec:proposed}

For an isolated star-forming cloud, the natural evolutionary timescale is $\tff$, and simulations of clouds that do not include an external environment (or effectively mock one up by using periodic boundary conditions) generally find that all gas is either converted to stars or expelled on such timescales \citep[e.g.,][though for an exception see \citealt{Wang10a}]{Krumholz12b, Kim18a, Grudic18a}. Thus the observed spread of $\approx 10\tff$ in the ONC is inconsistent with expectations from isolated cloud models.\footnote{An analogous problem occurs on the scales of entire GMCs, where simulations of isolated GMCs with properties similar to observed ones generally yield lifetimes of $\lesssim 5$ Myr \citep[e.g.,][]{Grudic18a}, while observationally-estimated GMC lifetimes are closer to 20-30 Myr \citep{Dobbs14a}.} Observed star formation histories also present a second challenge: even with $t_{\rm sf} \approx 10 t_{\rm ff}$ as in the ONC, since $\epsff \approx 0.01$ (c.f.~\autoref{ssec:epsff}) the ONC should reach a maximum conversion efficiency $\epsilon_* \approx 0.1$, probably too low to yield a gravitationally-bound cluster (c.f.~\autoref{ssec:emergence}). It is possible that the ONC will in fact not remain bound \citep{Kroupa01a}, though recent \textit{Gaia} observations suggest that it will \citep{Kuhn18a, Kim19a}. Moreover, the general point holds regardless of the fate of the ONC: the observed low median value of $\epsff$ means that bound clusters are likely to arise only from regions where $\epsff$ is in the upper $\sim 20\%$ of the distribution (plausible, since only $\sim 10\%$ of star formation results in bound clusters), where the free-fall time is short enough for the star formation history to extend over $>10$ free-fall times \citep{Kruijssen12a}, or some combination of the two.

One possible explanation for the ONC's star formation history is that it did form in a single free-fall time, but in an extended region of lower density than that the stars currently occupy. This region was in a free-fall collapse that halted when massive stars dispersed the gas. In this scenario, $t_{\rm sf} \gg t_{\rm ff}$ not because star formation was extended in time, but because $t_{\rm ff}$ decreased after the stars formed \citep[e.g.,][]{Zamora-Aviles14a, Kuznetsova15a, Kuznetsova18a, Vazquez-Semadeni17a}. This would naturally give rise to an age gradient such as that observed, because stars that formed earlier in the process would retain a higher proportion of their kinetic energy, and thus end up further from the cluster center in the final configuration \citep{Getman18a}.

\begin{figure}
\includegraphics[width=\textwidth]{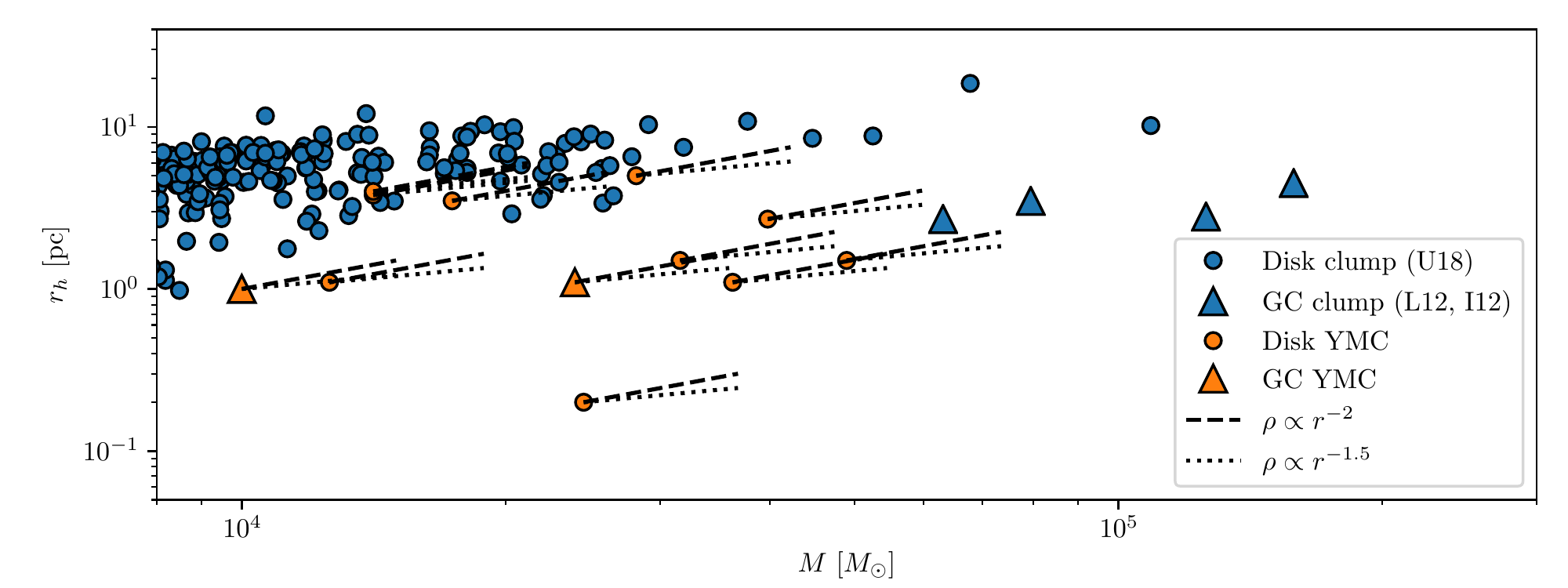}
\caption{
\label{fig:clump_cluster}
Comparison of masses and radii of young massive clusters (YMCs) in the disk of the Milky Way and near the Galactic Center, as indicated, and for clumps of molecular gas. The YMCs plotted are those listed in \autoref{tab:ymcs}. For gas clouds in the Galactic disk, we show the sample of \citet[U18]{Urquhart18a} from ATLASGAL, which is complete except near the Galactic Center. We supplement this with the Galactic Center clouds reported by \citet[L12]{Longmore12a} and \citet[I12]{Immer12a}. The dashed and dot-dashed lines marked for each cluster show the mass as a function of radius for clouds with density as a function of radius $\rho\propto r^{-2}$ and $r^{-1.5}$, as indicated. These lines indicate the loci below which progenitors clouds for these clusters would have to lie if their densities varied as the indicated power-law, i.e., a gas cloud with density versus radius $\rho \propto r^{-2}$ would only be dense enough to form the indicated star cluster if it were to lie below the dashed line emerging from that star cluster. For several disk clusters, there are no gas clouds that satisfy this condition.
}
\end{figure}

However, this scenario encounters serious observational difficulties. First, it would produce a distinct kinematic signature that is not observed. During collapse stellar motions should be radially inward, and once collapse is complete and gas has been expelled, the $\sim 90\%$ of the stars that do not remain bound (since $\Gamma \sim 10\%$; \autoref{ssec:boundfrac}) should be moving radially outward from the dense center; in either case stellar velocities should point radially toward or away from the densest point (in the example of \autoref{fig:orion_age}, the ONC). However, \textit{Gaia} shows that this is not the case in Orion or in other complexes \citep{Ward18a, Kuhn18a, Kounkel18a}. Note that this radial expansion is not seen in the simulations of \citet{Kuznetsova15a, Kuznetsova18a}, but this is because the simulations do not include any form of feedback, and thus nearly all the stars they produce remain bound. If the stars in their final configuration were to become unbound somehow, the velocities would become largely radial as soon as the stars expanded significantly compared to their original volume. A second problem is that this scenario would require even more extreme efficiencies, $\epsff \gtrsim 0.3$, to produce bound systems; excluding the discrepant cloud matching results, observed $\epsff$ values do not go this high (c.f.~\autoref{ssec:epsff}). A third problem is that this scenario requires star clusters to pass through a dense phase wherein the mass is still largely in the form of gas, so that for any given star cluster, there should be a comparably dense gas cloud. However, surveys have consistently failed to find gas clouds as dense as the densest YMCs \citep{Ginsburg12a, Longmore14a, Walker16a, Urquhart18a}. We illustrate this with a current compilation of gas clouds and clusters in \autoref{fig:clump_cluster}. Notice that, for the densest clusters in the disk of the Galaxy (though interestingly not for the Galactic Center), there are no clouds dense and massive enough to be their progenitors if the entire cloud had to be assembled at once. Given these problems, it is unlikely that the extended formation times are illusory.

How then are the long formation times to be explained? One possibility is that existing simulations do not yet have enough physics or enough resolution to produce realistic lifetimes. There would be some precedent for this: for decades simulations struggled to reproduce the small observed values of $\epsff$, but, as noted in \autoref{ssec:epsff}, simulations that include multiple feedback processes, initial turbulence, and magnetic fields are now approaching this goal. It is possible that the same will happen with the duration of star formation. More likely, however, the problem with current simulations is lack of the larger scale-environment that provides an ongoing mass supply. Observed protoclusters are generally located at the confluence of gaseous filaments, suggesting a ``conveyor belt'' picture in which gas accretion onto a protocluster and star formation occur simultaneously (\citealp{Longmore14a, Motte18a}; see Supplementary Materials). Inflow supplies energy to maintain turbulence and prevent contraction \citep{Klessen10a, Goldbaum11a, Matzner15a, Lee16b, Lee16c}, which together with local feedback processes keeps $\epsff$ low. This state persists as long as the accretion lasts, which can be many times the local value of $t_{\rm ff}$ because the accretion flow is dictated by processes occurring on much larger scales with much longer dynamical times. Because $\epsff$ is small and gas consumption slow, the star-forming gas mass grows with time, leading to an accelerating star formation rate. Once accretion ceases, and the turbulent energy supply and confining ram pressure of the accretion are removed, any remaining gas is converted to stars or expelled by feedback in a time of order $t_{\rm ff}$, as happens in the isolated cloud simulations. The final star formation efficiency is ultimately be dictated by the efficiency of feedback at ejecting gas, and if the ratio of mass ejected by feedback to mass converted into stars is $\lesssim 1$, then the final value of $\epsilon_*$ is large even if $\epsff$ is always fairly small.

This scenario is consistent with observed values of $\epsff$, naturally produces an accelerating rate of star formation, and also yields a cluster mass-size relation in good agreement with observations \citep{Matzner15a, Lee16c}. Because the final mass that goes into a cluster was never assembled in a single cloud all at once, there is no problem with finding clouds dense enough to be the progenitors of the densest clusters. It is not clear if this mechanism is capable of producing the core-halo age gradient observed by \citet{Getman14a} and \citet{Getman18a}, and visible in \autoref{fig:orion_age} -- for the ONC this gradient is $\sim 0.5$ Myr pc$^{-1}$. Such a gradient could arise as a result of earlier-forming stars having longer to undergo dynamical heating in the core of the forming cluster, but this possibility has not been examined quantitatively. A second possible challenge for this picture is that preliminary work suggests that it leads to clusters with significant rotational support \citet{Lee16b, Lee16c}. While some clusters do rotate \citep[e.g.][]{Davies11b, Henault-Brunet12b, Kamann18b}, others do not \citep{Kuhn18a}, and it is not clear if the predicted rotation rate distribution is consistent with observations. However, current theoretical predictions for the rotation rate distribution do not yet take into account the possibility of braking by dynamically-significant magnetic fields.

\subsection{Emergence of the gas-free cluster population}
\label{ssec:emergence}

\subsubsection{The initial cluster mass function}

The initial cluster mass function (ICMF), defined as the CMF immediately after gas removal, is the starting point for understanding the evolution of clusters. The ICMF is less well defined than its stellar counterpart because the stellar content of a cluster is constantly evolving, particularly for unbound clusters. \citet{Adams01a} showed that a cluster must contain $\gtrsim 100$ stars in order for the dynamical time to exceed the formation time, which they took to be $\sim 1-2$ Myr. Indeed, the ICMF for embedded clusters in the Solar neighborhood turns down below $N\sim 100$ stars, corresponding to a cluster mass $\sim 50\,M_\odot$ \citep{Lada03a}. As discussed in \autoref{ssec:mass}, the CMF for observed clusters usually can be fit with a Schechter function with a power-law index $\alpha_M\approx -2$ and a truncation mass $M_c$ that is related to the environment. \citet{Elmegreen06a} has shown that a slope $\alpha_M=-2$ naturally accounts for the fact that the stellar IMF averaged over a galaxy is very close to the average IMF of individual star clusters.

\paragraph{The clump mass function}
Theoretically, the ICMF is the product of the clump mass function (\autoref{ssec:clumps}) and the mass-dependent star formation efficiency $\epsilon_*(M)$. This is approximate: in a turbulent medium the clump mass evolves with time, and we have seen in \autoref{ssec:tsf} that bound clusters form over multiple dynamical times with continuous accretion, so that the star-forming gas is never assembled into a single monolithic clump. Nonetheless, clumps still provide a useful approximation for modeling the complex process of cluster formation, just as protostellar cores provide a useful basis for modeling the stellar IMF (e.g., \citealp{Hennebelle08a}). An important factor determining the clump mass function, $dN_{\rm clump}/dM\propto M^{\aclump}$, is that molecular clouds are supersonically turbulent and appear to be self-similar from the outer scale of the turbulence (generally encompassing the entire cloud) to the sonic scale ($\sim 0.1$~pc in Galactic GMCs). \citet{Fleck96a} was the first to argue that $\aclump=-2$ for a self-similar medium in which clumps form hierarchically. \citet{Elmegreen96a} came to a similar conclusion by using the observed fractal dimension of molecular clouds, and for the first time suggested that this would lead to a similar slope for the ICMF. \citet{Guszejnov18a} have elaborated the argument for $\aclump=-2$, which we paraphrase: assume that the clouds form hierarchically, either by fragmentation or by mergers, and that each level of the hierarchy, $n$, contains $N_n$ clumps that have a range of masses $\Delta M_n\propto M_n$. Then the total mass at level $n$ is
\beq
M_{\rm tot}(M_n)=M_n N_n=M_n\left(\frac{N_n}{\Delta M_n}\right)\left(\frac{\Delta M_n}{M_n}\right)M_n \propto M_n^2 \frac{dN}{dM},
\eeq
where the last step follows for large $n$. If the hierarchy is populated throughout, so that the mass at each level of the hierarchy, $M_{\rm tot}(M_n)$, is about constant, then $\aclump\approx -2$. This argument is clearly approximate; for example, GMCs have most of their mass at high masses in the Milky Way and at low masses in some other galaxies (\autoref{ssec:clumps}). In a dynamical system such as a GMC, the clumps at different levels of the hierarchy can also have different lifetimes, thereby changing the observed PDF; for example, if the clumps have mass-independent surface densities and survive for a time proportional to the free-fall time, the observed value of $\aclump$ is larger than the intrinsic value by 1/4 \citep{Fall10a}. This theory of the clump mass function is thus not very precise, but it is broadly consistent with observation.

\paragraph{From clumps to clusters}

As noted above, the ICMF is related to the mass distribution of the natal molecular clumps by the star formation efficiency (SFE), $\epss(M)$, so that $dN(M_*)\propto\epss(M)dN_{\rm clump}(M)$ approximately, where it is possible that more than one cluster forms in  a cloud. Photoionization and direct radiation pressure are the most broadly effective feedback mechanisms (\autoref{ssec:feedback}). Photoionization removes a mass
$\Delta M_{\rm ion}\propto S^{4/7}t^{9/7}/\rho^{1/7}$ over a time $t$, where $S$ is the rate of production of ionizing photons \citep{Whitworth79a}. If this is the dominant feedback mechanism, star formation will continue until $\Delta M_{\rm ion}\sim M$. In terms of $\epsilon_*$, this will occur at the star formation time $t_{\rm sf}=\epss\tff/\epsff$, so that
\beq
\epss\propto \epsff^{9/13}\Sigma^{33/52}M^{1/52}.
\eeq
Bound clumps ($\avir\sim 1$) have surface densities that depend only on the turbulence parameter, $C=\sigma/R^{1/2}$ (\autoref{eq:avir}), which is observed to be approximately constant in a galaxy \citep[e.g.,][]{Sun18}. For constant $\epsff$ (\autoref{ssec:epsff}), this implies $\epsilon_*$ is nearly independent of $M$, so the ICMF has the same slope as the clump mass function. If on the other hand stars form in a fixed number of free fall times so that $\epss\propto\epsff$, then $\epss\propto \Sigma^{33/16}M^{1/16}$, very close the results of \citet{Kim18a} for small $\epsilon_*$. 

Direct radiation pressure injects momentum at a rate proportional to the stellar mass. Stars will form until the injected momentum $\propto \epss M t_{\rm sf}$ reaches $Mv_{\rm esc}$, so that
\beq
\epss\propto (\epsff \Sigma)^{1/2},
\eeq
with no explicit dependence on $M$. If $t_{\rm sf} \propto t_{\rm ff}$, so that $\epss\propto\epsff$, then $\epss\propto\Sigma$, as found by \citet{Fall10a}. In this  case, or if $\epsff$ is about constant, the SFE is independent of mass for bound clumps, which have similar surface densities in a given galaxy as noted above. As a result, the slope of the ICMF is the same as that for the clump mass function, $\aclump\approx \alpha_M \approx -2$.

\subsubsection{Bound clusters}

With our broad definition of cluster (\autoref{ssec:working}), essentially all stars are born in clusters \citep[c.f.~][]{Lada03a}. However, most clusters are unbound by gas dispersal and thus disperse rapidly, so that $\Gamma\lesssim 0.1$ for $T\gtrsim 10$ Myr. In this section we investigate the origin of the low value of $\Gamma$ after gas dispersal. As discussed in \autoref{ssec:giant}, most GMCs appear to be bound (i.e., $\avir\lesssim 2$), as are the dense clumps within them \citep{Urquhart18a}, so most stars are born in gravitationally bound gas. Even in GMCs that are moderately unbound, $\avir \approx 5$, efficient star formation results in most stars being confined to gravitationally bound structures \citep{Clark05a}. Thus the low value of $\Gamma$ must be related to the star formation efficiency, $\epsilon_*$, and to how the gas not converted to stars is ejected.

The first author to estimate $\Gamma$ considered a spherical cloud in virial equilibrium that turns a fraction $\epsilon_*$ of its mass into stars and ejects the rest \citep{Hills80a}. If gas removal occurs over many dynamical times, the process is adiabatic, the cluster remains bound for any $\epsilon_*$, and its final and initial radii are related by $R_f/R_0=1/\epsilon_*$. If gas removal is rapid, ejection of more than half the mass ($\epsilon_*<0.5$) unbinds the stars; for $\epsilon_* > 0.5$ the radius expands, $R_f/R_0=\epsilon_*/(2\epsilon_*-1)$. Even for $\epsilon_* < 0.5$, however, the low-velocity part of the cluster can remain bound \citep{Lada84a}. In simulations, \citet{Kroupa01a} find bound remnants remain for $\epss$ as small as 0.3.

More realistic starting conditions can modify these conclusions. If the stars are initially subvirial, with $Q_0=2T_*/|W_*|<1$, where $T_*$ and $W_*$ are the initial kinetic and potential energies of the stars, respectively, then one can show that the minimum $\epsilon_*$ for the entire star cluster to be bound is $\epsilon_{*,\rm min}=Q_0/2$. If a cluster has substructure, then the additional binding energy can help it survive \citep{Allison09a}; whether the cluster remains bound depends unpredictably on the exact spatial distributions of gas and stars and on the timing of gas removal \citep{Smith11a, Smith13a}. The minimum value of $\epss$ for a bound cluster is also reduced if the stars are more concentrated than the gas, so that the local value of $\epss$ increases inward. \citet{Adams00a} showed that part of a cluster embedded in a gas distribution approximating an isothermal sphere could survive sudden gas expulsion down to very low values of $\epss$. A cluster that forms in a centrally concentrated gas cloud with a constant value of $\epsff$ has a star formation rate $\dot\rho_*\propto\rho^{3/2}$ \citep{Parmentier13a}; in a particular implementation of this model, \citet{Shukirgaliyev17a} found a minimum global star formation efficiency $\epsilon_{*,\rm min}=0.15$ for some stars to remain bound. Analysis of a simulation of the formation of a cluster led \citet{Kruijssen12c} to conclude that most cluster stars form in gas-poor subclusters that have locally high $\epsilon_*$ and are therefore relatively immune to gas expulsion; however, this work did not include feedback or magnetic fields, and thus its $\epsilon_*$ values are unrealistically high.

Real star-forming regions are hierarchically-structured, containing both dense parts for which mass removal is slow compared to the local dynamical time, and diffuse parts for which it is fast; we see this directly in the Orion star forming complex (\autoref{ssec:tsf}). \citet{Elmegreen06a, Elmegreen08a} was the first to point out that this configuration naturally gives rise to a gravitationally-bound central region and an unbound periphery. \citet{Kruijssen12a} provided a quantitative model for $\Gamma$ in such an environment, based on three main premises: first, the density PDF is lognormal, as expected in supersonically-turbulent flows; second, $\epsff \approx 0.01$ independent of density, as observations suggestion (\autoref{ssec:epsff}); third, gas at all densities is removed rapidly after some fixed timescale $t_{\rm sf}$ set by stellar feedback. Given these ingredients, he calculated a density-dependent star formation efficiency $\epsilon_*(\rho) = \epsff [t_{\rm sf}/\tff(\rho)]$, and integrated over the density PDF to derive the fraction of mass for which $\epsilon_* \gtrsim 0.5$, which he associated with the bound fraction $\Gamma$. A generic prediction of this model is that $\Gamma$ increases with mean ISM density or pressure, although such a prediction is not unique to this model, since many of the feedback mechanisms discussed in \autoref{ssec:feedback} become less effective at higher gas surface density. 

In this view, the low value of $\Gamma$ is not so much ``infant mortality'' as ``infant weight loss'': stars form in hierarchical structures where only a small fraction remains bound after gas dispersal, so rather than $\Gamma = 0.1$ meaning that 90\% of clusters disperse and 10\% remain bound, it means that 90\% of the stellar mass is in the unbound part of the structure that escapes immediately after gas dispersal. Of course in reality both mortality and weight loss may operate: hierarchical structures may lose much of their stellar mass in low-density regions, and then even the denser
regions that survive may disperse soon thereafter if they are insufficiently concentrated to survive stellar evolution and tidal shocking (\autoref{sec:life_and_death}).

Although the case in which stars form in a pre-existing density structure (either uniform or hierarchically-structured) has garnered the most attention in the literature, we have argued that it is more likely star formation and cloud assembly are contemporaneous. One can generalize the star formation model in \autoref{ssec:epsff} to allow for gas accretion while the stars form (see ``Conveyor-Belt Model for Cluster Formation" in the Supplementary Materials). However, provided the time over which the cloud accretes is $\la t_{\rm sf}$ (\autoref{eq:definitions}) the results are similar to those above. If the accretion stops due to feedback, then it is likely that the subsequent mass loss would be fast, since the same processes stopping the accretion would eject the gas in the clump. On the other hand, if the accretion stops because the supply of new gas runs out, then stars would continue to form and the mass loss would be slow.

One final consideration is that clusters form in tidal gravitational fields (\autoref{ssec:cluster_struct}), which sets a minimum density for the bound part of the cluster, $\rho_{\rm ti}$. After mass loss, the final mean density of a cluster with initial radius $R_0$ and density $\rho_0$ is
\citep{Mathieu83a}
\beq
\frac{\rho_{*f}}{\rho_0}=f_b\epss\left(\frac{R_0}{R_f}\right)^3\rightarrow \left\{
\begin{array}{l} \displaystyle\frac{(2\epss-1)^3}{\epsilon_*^2}\mbox{~~~fast mass loss,}\\
\displaystyle\epsilon_*^4\mbox{ ~~~~~~~~~~~~ slow mass loss,}
\end{array} \right.
\eeq
where 
$f_b$ is the fraction of the stellar mass that is bound, and we have assumed $f_b=Q_0=1$ in the expressions for the limiting cases.\footnote{The expression for the fast case is valid only for $\epss\gtrsim 0.6$ \citep{Lada84a}.} Clusters are generally centrally concentrated, and stars in the outer parts of the cluster in which the final stellar density, $\rho_{*f}$, is less than the tidal limit will be lost. For a simple spherically symmetric model in which (1) the gas density varies as $r^{-k}$ with $k\simeq 2$ \citep{Schneider15a}, (2) $\epsilon_* \propto 1/t_{\rm ff}$ (as in the \citealt{Kruijssen12a} model), so $\epsilon_* \propto \rho^{1/2}\propto 1/r$, and (3) $\epsilon_*$ near the center approaches unity (see ``Cluster-Forming Clouds and Clumps" in the Supplementary Materials), the bound fraction is large, $f_b\gtrsim 0.5$, for values of the cloud-averaged $\epsilon_*\gtrsim 0.01$; in this model, the mean value of $\epsilon_*$ for the stars that remain bound is $\epsilon_{*b} \gtrsim 0.3$.\footnote{There are several distinct bound fractions: $\Gamma$ is the fraction of all stars that are bound, $f_b$ is the fraction of stars that are bound in a particular cloud, and $\epsilon_{*b}$ is the fraction of gas that went into stars in the bound part of a cluster.} 

Altogether, theoretical estimates of the bound fraction are the product of several uncertain factors. First, after the accretion of gas onto the cluster-forming clump has ceased, what fraction of the clumps are bound?
Second, what fraction of the stars remain bound after the residual gas is ejected? Observations suggest that star formation is coincident with clump formation, at least in the early stages, but it is not clear whether this process is halted by feedback, leading to rapid mass loss and requiring a relatively high star formation efficiency, or by a lack of gas, which would be consistent with a lower efficiency. Tides due to the local environment can also strip stars from the outer regions of the cluster when mass is lost, a point to which we return in \autoref{sssec:tidal}.

\subsubsection{Numerical results}
\label{ssec:icmf_numeric}

A few authors have attempted to simulate the full range of processes -- formation of a mass spectrum of clumps, conversion of gas into stars, and gas removal by feedback -- that determine the ICMF and $\Gamma$. This is very challenging numerically, since capturing the formation of a statistically-meaningful sample of clusters requires simulating either an entire galaxy or a substantial portion thereof. Modern isolated galaxy or cosmological zoom-in simulations generally reach spatial resolutions no better than $\sim 1$ pc, and mass resolutions of $\sim 10^2 - 10^3$ $M_\odot$. Comparing these figures to the star cluster properties shown in \autoref{fig:mass_radius}, even for GCs this corresponds to $\lesssim 10^3$ mass resolution elements per cluster, and $\lesssim 3-10$ spatial resolution elements per half-mass radius. Thus simulations must rely on parameterized subgrid models to handle both star formation and feedback. 

Perhaps not surprisingly, this situation yields little consensus. \citet{Renaud15a} simulate an Antennae-like galaxy merger, and find that $\Gamma$ is a few percent pre-merger, but rises to $\approx 10\%$ during the main burst of star formation. In contrast, \citet{Li18a} find $\Gamma$ values from $\approx 1\%$ to $\approx 50\%$ in their cosmological simulations that run to $z\approx 1.5$. \citet{Renaud15a} and \citet{Maji17a} find lognormal ICMFs in simulations of both mergers and quiescent spiral galaxies (contrary to observations), while \citet{Dobbs17a}, \citet{Li17a}, and \citet{Li18a} find ICMFs that are well-described by power-laws or Schechter functions with $\alpha_M \approx -2$. \citet{Li18a} test a variety of star formation and feedback prescriptions, all tuned to reproduce the galaxy-scale Kennicutt relation and various other constraints, and find that both $\Gamma$ and the ICMF are extremely sensitive to the choice of subgrid model.

We conclude that present galaxy simulations have little predictive power when it comes to star clusters; instead, a more productive approach would be to use the observed properties of star clusters as an additional constraint to calibrate the subgrid models. There do, however, appear to be two qualitative results that persist across recipes. First, all simulations find that $\Gamma$ increases at higher surface densities of star formation, which occur in regions of higher pressure. Second, all authors find that the ICMF extends to higher masses during mergers. This is also consistent with the results of a number of other cosmological simulations showing that early, gas-rich galaxies are capable of producing clusters with initial masses $\gtrsim 10^6$ $M_\odot$ \citep{Kimm16a, Ricotti16a, Kim18b}.

\subsection{Elemental abundance distributions}
\label{ssec:chem_form}

There has been surprisingly little theoretical work on the physical origins of abundance homogeneity in star clusters. \citet{Murray1990} observed that narrow giant branches in GC CMDs implied they were highly uniform in [Fe/H], and suggested that turbulent diffusion could explain why clouds are highly uniform prior to the onset of star formation. This process would mix out abundance inhomogeneities at the scale of a cloud over a cloud crossing time, and smaller-scale inhomogeneities even more quickly. \citet{Feng2014} provide numerical support for this idea, finding in a simulation of star formation in a colliding flow of warm neutral gas that the mixing process washes out homogeneities in the initial colliding streams by a factor of 5 or more. This may be telling us that the small inhomogeneities detected in a few open clusters \citep[e.g.~M67;][]{Liu2016a} reflect metal anomalies (predicted to be $\sim 0.1$ dex -- \citealt{Krumholz2018}) in the birth cloud (\autoref{ssec:chemical_comp}). \citeauthor{Feng2014} also find that even unbound clusters can also be homogeneous, a prediction that has yet to be tested by observations. \citet{Armillotta2018} revisit the \citet{Feng2014} simulations using more realistic initial conditions, a cloud extracted from a galaxy simulation, and zoom in to follow it collapse into individual stars at a resolution of $\approx 10^{-3}$ pc. They find that the abundance scatter between formed stars decreases on shorter length scales, indicating that turbulent mixing becomes more efficient on smaller and smaller scales ($\lesssim 5$ pc), in agreement with the discussion above. Interestingly, they also find that some stellar clusters are separable from others within the collapsing clouds based on abundances, but not all. 

A few caveats are in order. First, the simulations of \citet{Feng2014} and \citet{Armillotta2018} assume that metals are well-coupled to the gas, but this need not be the case for metals in the form of dust grains. \citet{Padoan2006} first raised the prospect that small-scale turbulence could generate low-level variations in the dust to gas ratio. More recently, \citet{Hopkins2016} and \citet{Lee2017} find far more dramatic variations in their simulations and even suggested that some stars could form in ``totally metal'' regions \citep{Hopkins14b}. But this work has been refuted by \citet{Tricco2017} who find no evidence for significant variations; the use of tracer particles in supersonic flows lead to numerical artifacts that exaggerate density contrasts. Second, the cluster formation simulations to date have not considered the effects of magnetic fields, which \citet{Sur2014} suggest can suppress the mixing over a wide range in Mach number. Here, the same random stretching and folding that is responsible for turbulent diffusion also amplifies the magnetic energy density by the dynamo mechanism, which in turn can suppress the mixing on small scales.

A third caveat is that, while galaxy-scale simulations are starting to consider injection of new metals by stellar evolution \citep[e.g.,][]{Emerick19a}, simulations that resolve cluster formation have yet to do so. \citet{BlandHawthorn2010} point out that, even if turbulence homogenizes a cloud, it will only remain homogeneous if star formation ceases before the first SN explodes, since even a single SN will produce measurable abundance variations. For a gas cloud with mass $M$, surface density $\Sigma$, and virial ratio $\alpha_{\rm vir}$, the crossing time is $\tcr \approx (\alpha_{\rm vir} G)^{-1/2} M^{1/4} \Sigma^{-3/4}$ (c.f.~\autoref{ssec:giant}). If we adopt $\alpha_{\rm vir}=1.5$, and the cluster forms over $\tau_{\rm sf} = t_{\rm sf}/t_{\rm cr}$ crossing times, then
\begin{equation}
t_{\rm sf} \approx 3.0 \left(\frac{\tau_{\rm sf}}{4}\right) \left(\frac{\epsilon_*}{0.2}\right)^{-1/4} \left(\frac{M_*}{10^4\,\msun}\right)^{1/4} \left(\frac{\Sigma}{0.3\mbox{ g cm}^{-2}}\right)^{-3/4}\mbox{ Myr},
\end{equation}
where $\epsilon_*=M_*/M$ is the star formation efficiency. Thus $t_{\rm sf} < t_{\rm SN}$, which is conservative given that most SN progenitors have longer lifetimes, holds for OCs ($\Sigma\approx 0.3\,\msun$ pc$^{-2}$) up to $10^5\msun$, and for GCs ($\Sigma\approx 3\,\msun$ pc$^{-2}$) up to $10^7\msun$. This last estimate is interesting in light of the fact that globulars can show high levels of Fe homogeneity; as we have seen, for other elements, the story is more complicated. How there can be differential scatter across elements is not understood, unless it reflects the state in the collapsing cloud (see below). In \autoref{fig:mass_radius}, there is evidently a lot of scatter but our point is to emphasize the difference between OCs and denser GCs.

Both this calculation and the simulations of \citet{Armillotta2018} raise the question of how to apply estimates of chemical homogeneity or inhomogeneity in the context of hierarchical star-forming systems that do not have clear edges. This topic is currently under investigation using a combination of \Gaia\ data and ground-based, million-star surveys at high spectroscopic resolution \citep[e.g.~GALAH;][]{DeSilva2015}. The astrometric information can be used to identify every aggregate within dozens of nearby stellar hierarchies \citep{Gouliermis18a}, while the spectra provide abundances for all stars cooler than $\approx 7000$ K. This work will provide the first clues for how to map stars between abundance and physical space, i.e., the possibility of determining how closely together in space two stars formed based on the level of similarity or difference in their elemental composition, with major implications for the future of chemical tagging in Galactic archaeology \citep{Freeman2002,BlandHawthorn2004}. It is also not clear how to apply these models in the context of multiple generations and detectable abundance anomalies in globular clusters (\autoref{ssec:chemical_comp}). 

\section{LIFE AND DEATH}
\label{sec:life_and_death}

Since the slope of the CAF $\alpha_T < 0$ even for ages older than $\approx 10$ Myr (\autoref{ssec:age}), cluster disruption cannot depend solely on gas expulsion immediately after cluster formation. Instead, there must be processes that destroy star clusters during the gas-free phase of their evolution. It is to these processes that we now turn. We begin with a brief review of star cluster structure and tidal fields in \autoref{ssec:cluster_struct}, then examine disruption processes driven primarily by mechanisms internal to the cluster (\autoref{ssec:internal}) and those driven primarily by the external environment (\autoref{ssec:external}), before concluding with attempts to synthesize both into a coherent explanation for the demographics of star clusters \autoref{ssec:population_models}. Since our focus is on the demographics and evolution of the star cluster population as a whole, we will not discuss processes that alter the internal structure of star clusters while leaving their bulk properties unchanged, e.g., dynamical alteration of the binary fraction or stellar collisions. We refer readers to \citet{Portegies-Zwart10a} and \citet{Renaud18a} for a discussion of these processes.

\subsection{Star cluster structure and tides}
\label{ssec:cluster_struct}

A crossing time after gas is removed, star clusters that remain bound relax to a nearly-spherical, virialized state. From this point up to ages of several hundred Myr, their density profiles are well described by the \citet[EFF hereafter]{Elson87a} distribution, which has surface density as a function of radius $\Sigma(r) = \Sigma_0 \left(1 + r^2/a^2\right)^{-\gamma/2}$. Here $\Sigma_0$ is the central surface density, $a$ defines the radial scale, and $\gamma$ describes the fall off in surface density in the outer, power-law region; observed values of $\gamma$ are generally in the range $\approx 2.2 - 3.2$ \citep{Elson87a,Mackey03b,Mackey03a}. The core radius for an EFF profile, defined to match that of a King profile (see below) is $r_c \approx a \sqrt{2^{2/\gamma}-1}$.

The EFF profile is not in energetic equilibrium, and thus clusters observed to have this distribution cannot have reached either internal relaxation or equilibrium with the tidal field of their host galaxy. At ages beyond a few hundred Myr, clusters relax and become well-fit by \citet{King62a, King66a} models where stellar energies are distributed as a ``lowered Maxwellian'' of the form $f(E)\propto (2\pi \sigma^2)^{-3/2} \left[\exp\left(-E/\sigma^2\right)-1\right]$ for some constant velocity dispersion parameter $\sigma$. Physically, the $-1$ in square brackets reflects the fact that the velocity distribution cannot be a pure Maxwellian, because the cluster has a finite escape speed, and thus any star that is too far out on the tail of the distribution will leave. Combining this assumed energy distribution with the Poisson equation for self-gravity yields an equation for the radial density distribution and potential. Solutions to this equation form a one-parameter family, with the parameter conventionally specified by setting the dimensionless potential well depth at the cluster center $W_0 = (v_{\rm esc}/\sigma)^2$, where $v_{\rm esc}$ is the escape speed from the cluster center. Most observed OCs and GCs are reasonably well-fit by $W_0 \approx 3-7$, though a small number of core-collapsed globulars and some of the most concentrated YMCs have $W_0 \gtrsim 10$.

For a King model, one can define the core radius $r_c = \sqrt{9\sigma^2/4\pi G \rho_0}$, where $\rho_0$ is the central density. The core radius is distinct from the half-mass radius $r_h$ (\autoref{ssec:size}), and $r_h/r_c$ is a function of $W_0$, but for $W_0\approx 3-7$, $r_h/r_c$ is within a factor of a few of unity. A second useful quantity is the truncation radius $r_{\rm tr}$, the radius at which the density falls to zero. This can be combined with the core radius to yield the concentration $c = \log (r_{\rm tr}/r_c)$. This is a monotonically increasing function of $W_0$, so clusters with deeper potential wells are more centrally concentrated. For this reason, clusters are often parametrized in terms of values of $c$ instead of values of $W_0$. The observed range $W_0 \approx 3-7$ corresponds to $c \approx 0.6 - 1.5$. Note that the EFF model does not have a truncation radius.

\begin{figure}
\begin{minipage}{0.5\textwidth}
\includegraphics[width=\textwidth]{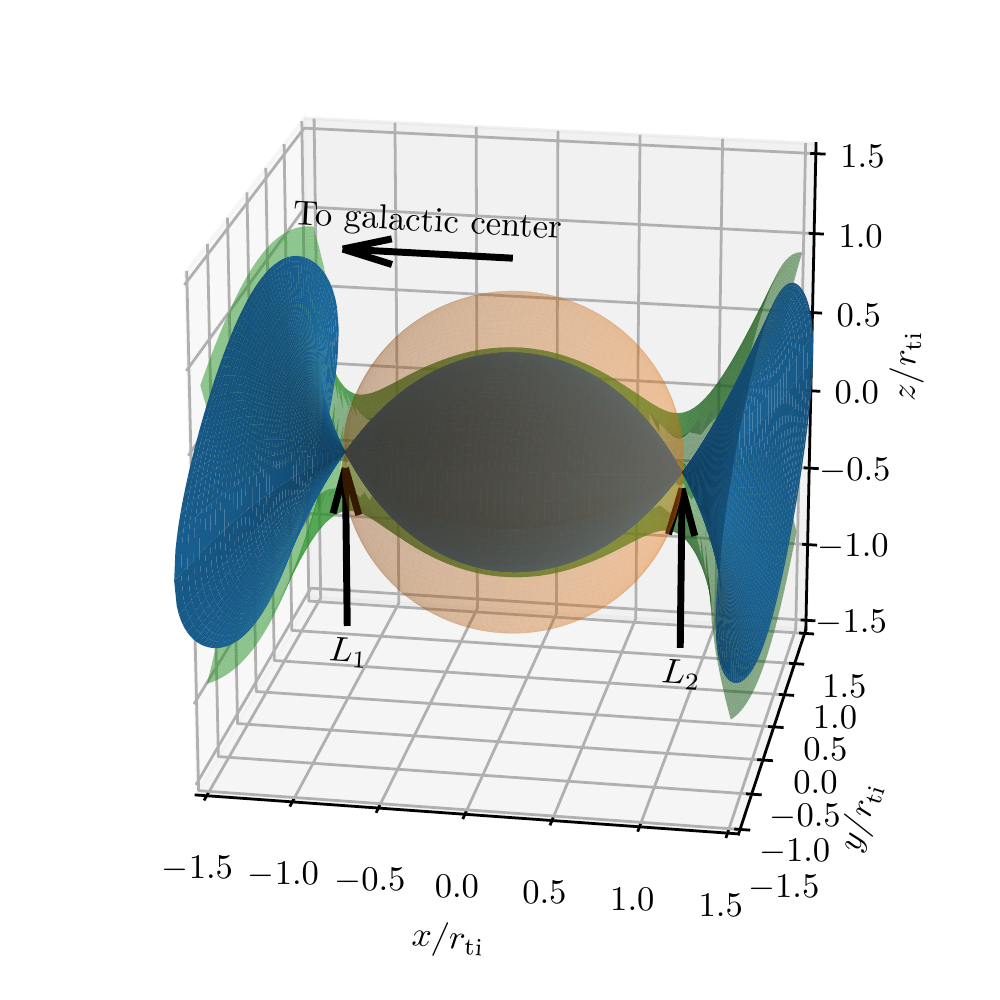}
\end{minipage}\begin{minipage}{0.45\textwidth}
\includegraphics[width=\textwidth]{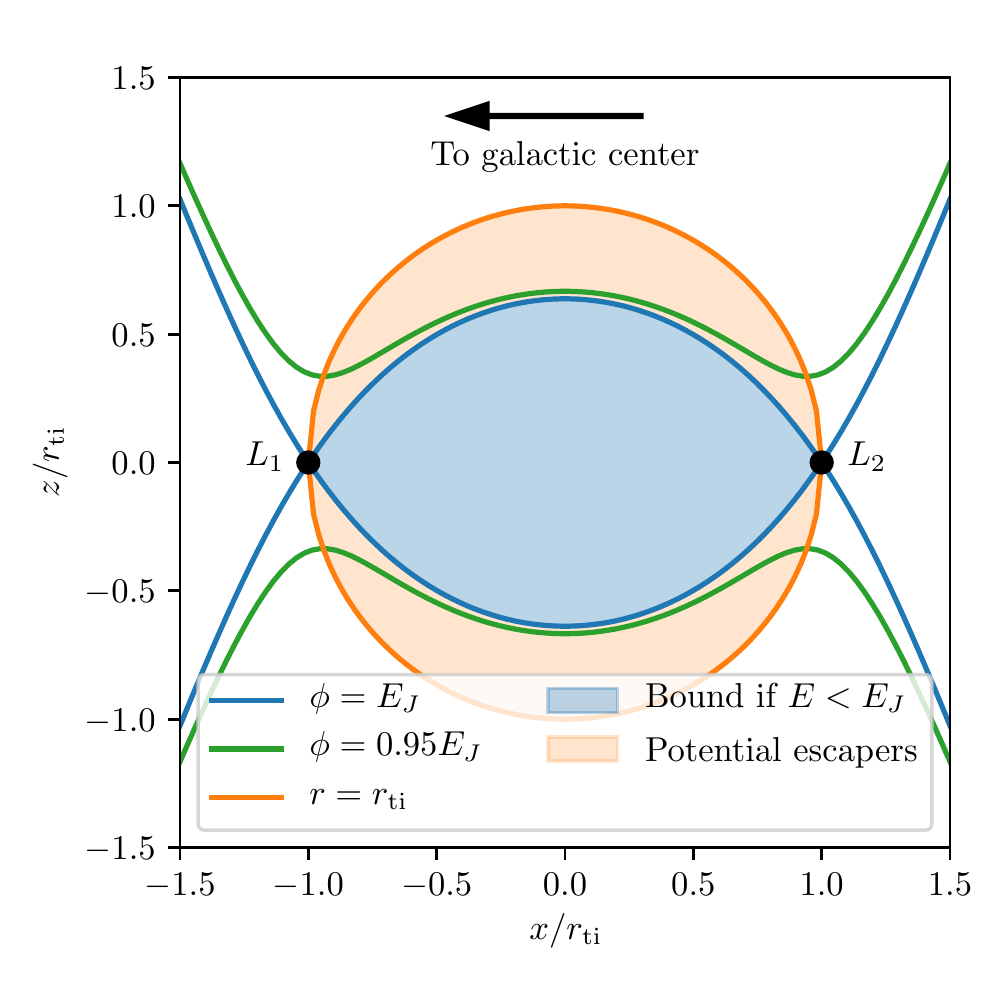}
\end{minipage}
\caption{
\label{fig:tidal}
Isosurfaces of the potential $\phi(x,y,z)$ (\autoref{eq:potential}) for a point-like star cluster embedded in a galaxy, seen in 3d (\textit{left}) and in a cut through the $xz$ plane (\textit{right}). The cluster center is located at the origin, while the galaxy center is located at $(-R_g,0,0)$ for some distance $R_g$ much larger than the physical size of the cluster. In the example shown, $\lambda_y = 0$ and $\lambda_z/\lambda_x = -1/3$. Blue shows the Jacobi surface, defined by $\phi = E_J$, while green shows a potential $\phi = 0.95 E_J$, i.e., with energy 5\% greater than the Jacobi energy. Orange shows a sphere of radius $r = r_{\rm ti}$ (\autoref{eq:rti}). The Lagrange points $L_1$ and $L_2$ are as indicated. Stars inside the Jacobi surface, shaded blue in the right panel, are bound cluster members if their total energy $E < E_J$. Stars outside the Jacobi surface but at $r < r_{\rm ti}$ (shaded orange), or those inside the Jacobi surface with $E\geq E_J$, are potentially unbound from the cluster, and can escape through the ``windows'' around $L_1$ and $L_2$. 
}
\end{figure}

King models in principle apply to an isolated star cluster, but it is common to assume that the truncation at $r_{\rm tr}$ is due to the tidal field of the cluster's host galaxy.\footnote{We emphasize that this assumption is not physically required. While a cluster in equilibrium cannot be larger than the size imposed by galactic tidal truncation, it could be smaller, e.g., if the cluster were truncated by gas removal unbinding its outer regions (c.f.~\autoref{ssec:emergence}). Conversely, clusters can remain out of equilibrium and larger than their tidal radii for hundreds of Myr, which is why EFF profiles usually provide better fits than King profiles for younger clusters.} One can derive an effective potential in the frame co-moving with a cluster orbiting with a galaxy by Taylor-expanding the galactic potential and adding this to the centrifugal acceleration in the rotating reference frame. If one approximates the cluster as point-like, the resulting total potential is \citep[e.g.,][]{Renaud11a}
\begin{equation}
\label{eq:potential}
\phi(x,y,z) = -\frac{GM}{\sqrt{x^2+y^2+z^2}} - \frac{\lambda_x}{2}\left(x^2 + \frac{\lambda_y}{\lambda_x} y^2 + \frac{\lambda_z}{\lambda_x} z^2\right),
\end{equation}
where $M$ is the cluster mass, $\lambda_x$, $\lambda_y$, and $\lambda_z$ are the three eigenvalues of the tidal plus centrifugal tensor,\footnote{Because $\lambda_{x,y,z}$ are eigenvalues that arise from derivatives of the gravitational plus centrifugal tensor, \citet{Renaud11a} refer to them as the effective eigenvalues, to distinguish them from the eigenvalues of the tidal tensor by itself, computed without including the centrifugal potential.} and the coordinate system is oriented so that the center of the galaxy lies at on the $x$ axis at $x<0$, and the tidal plus centrifugal tensor has zero off-diagonal components. For a star cluster on a circular orbit at the midplane of an azimuthally-symmetric galactic potential $\phi_G$, adopting a coordinate system where the $y$ direction is spinward, the $z$ direction is out of the galactic plane, and the coordinate system is right-handed, the eigenvalues of the tensor are $\lambda_x = -\partial^2 \phi_G/\partial x^2+\partial^2 \phi_G/\partial z^2$, $\lambda_y = 0$, and $\lambda_z = -\partial^2 \phi_G/\partial z^2$.

The potential $\phi$ has five extrema or saddle points, known as the Lagrange points; of these, by convention the points $L_1$ and $L_2$ lie at positions $L_{1,2} = (\mp r_{\rm ti}, 0, 0)$, where
\begin{equation}
\label{eq:rti}
r_{\rm ti} = \left(\frac{GM}{\lambda_x}\right)^{1/3}
\end{equation}
is known as the tidal radius. At these points, the acceleration due to the gravity of the cluster is equal to that due to the combination of galactic gravity and the centrifugal force. The value of the potential at $L_1$ and $L_2$ is the Jacobi energy $E_J = -(3/2) GM_c/r_{\rm ti}$, and the locus where $\phi(x,y,z) = E_J$ is called the Jacobi surface (\autoref{fig:tidal}).

Stars with total energy $E < E_J$ are unable to cross the Jacobi surface, so any such stars located within it are bound to the cluster. Since the volume within this surface defines the volume of the star cluster, tidal truncation therefore defines a characteristic density $\rho_{\rm ti} \approx 3 M / 4\pi r_{\rm ti}^3 = 3\lambda_x/4\pi G$, which depends only on the tidal tensor and thus on the cluster's orbit and environment, and not on the cluster mass. Stars with energies $E\geq E_J$ are not bound to the cluster, but if they are located inside the Jacobi surface then they can only escape through a ``window'' around $L_1$ and $L_2$, with the size of the window depending on $E/E_J$. We show an example escape window for $E/E_J = 0.95$ (i.e., for a star with energy 5\% greater than the Jacobi energy) in \autoref{fig:tidal}. With the exception of those ejected during violent tidal shocks, most stars escaping a cluster have energies only slightly larger than $E_J$, so escape tends to be primarily near the Lagrange points.

\subsection{Internal evolutionary processes}
\label{ssec:internal}

With these preliminaries understood, we can now ask what mechanisms can alter clusters' properties during the gas-free phase of their evolution.

\subsubsection{Stellar evolution}
\label{sssec:stellar_evolution}

At ages from $\approx 3-40$ Myr, stellar populations eject $\approx 20\%$ of their mass in SNe, which produce high-speed gas that almost certainly escapes the parent cluster; moreover, the remnant neutron stars or black holes left behind by SNe receive kicks of several hundred km s$^{-1}$ \citep[e.g.,][]{Faucher-Giguere06a}, so most escape the star cluster as well. After $\approx 40$ Myr, SNe cease and mass loss becomes dominated by asymptotic giant branch (AGB) stars shedding their envelopes. This process is more gradual, so that the total mass loss reaches $\approx 40\%$ only after $\approx 1$ Gyr, and does not reach 50\% until ages of order a Hubble time. Moreover, while AGB wind velocities are high enough (tens of km s$^{-1}$) that little of the wind mass is likely to be retained, the white dwarfs left behind receive much smaller kicks, and thus are likely to remain in the cluster \citep[e.g.,][]{Kruijssen09a}.

By itself a factor of two decrease in mass would be a small effect. However, the effects of mass loss can be greatly amplified by a tidal field. Adiabatic mass loss makes the cluster radius expand while the tidal radius shrinks; stars that find themselves outside $r_{\rm ti}$ as a result are lost. For clusters that begin their evolution filling their tidal radii, this effect roughly doubles the mass loss rate due to stellar evolution alone \citep{Takahashi00a, Baumgardt03a, Lamers10a}. Moreover, in the presence of a tidal field, for any specified cluster density profile there is a maximum ratio of half-mass radius to tidal radius, $r_h/r_{\rm ti}$, for which virial equilibrium is possible \citep{Chernoff90a, Fukushige95a}. If mass loss drives $r_h/r_{\rm ti}$ above this maximum, the stars cannot find a new virial equilibrium, and the entire cluster disrupts. Both the additional mass loss and the propensity to catastrophic disruption are enhanced in mass-segregated clusters, where the stars losing mass carry a disproportionate share of the potential energy \citep{Vesperini09a, Gieles10a}. The importance of stellar evolution depends on a star cluster's initial concentration and ratio of truncation radius to tidal radius $r_{\rm tr}/r_{\rm ti}$. Clusters with $r_{\rm tr} / r_{\rm ti} \ll 1$ or $W_0 \gtrsim 5$ ($c \gtrsim 1$) are dense enough to behave qualitatively like clusters that are not in a tidal field. Clusters with low concentration that already fill their tidal radii at the start of gas-free evolution are much more vulnerable.

Unfortunately there have been only limited observational studies of $c$ and $r_{\rm tr}/r_{\rm ti}$ for young clusters. \citet{Ryon15a, Ryon17a} survey massive ($> 5000$ $M_\odot$) clusters in M83, NGC 628, and NGC 1313, and find that about half have $r_h/r_{\rm ti} \lesssim 0.15$ independent of cluster mass, small enough to render mass loss unimportant. The fraction with larger values of $r_h/r_{\rm ti}$ increases with age, as previously observed by \citet{Elson89a} in the LMC, but too late for stellar mass loss to be effective. However, we caution that accurate measurements of $r_h$ require accurate photometry in clusters' extended, low surface brightness halos, which is very difficult to obtain for galaxies at the distances of \citeauthor{Ryon17a}'s sample. The result that the concentration does not correlate with mass is more robust, and appears to hold for OCs in the Milky Way (based on our examination of \citealt{Kharchenko13a}'s sample) and the Magellanic Clouds as well \citep[using the measurements of \citealt{McLaughlin05a}]{Fall12a}, where measurements are easier. For Milky Way GCs, \citet{McLaughlin00a} reports a weak correlation between concentration and mass, but no such correlation is apparent in the more recent fits of \citet{Baumgardt18a}. Thus while there are hints that clusters might be born too concentrated for mass loss to be a significant effect, the observational case is far from settled. Moreover, we caution that all observations are potentially subject to both selection bias in favor of more compact clusters and survival bias -- a substantial fraction of gas-free clusters at an age of $\approx 10$ Myr might be only weakly bound and have $W_0 \approx 2-3$, for example, but these would be dramatically under-represented in OC catalogs because almost none would survive to reach ages more than a few tens of Myr. 

\subsubsection{Relaxation-driven mass loss}
\label{sssec:relaxation}

Stellar evolution has its greatest effects in the first $\sim 10 - 100$ Myr of a cluster's lifetime. On much longer time scales, the dominant internal process is mass loss driven by relaxation, sometimes also referred to as dissolution or evaporation. The underlying mechanism is simple: because stars with too much energy escape from clusters, the high-energy tail of a cluster's velocity distribution is underpopulated compared to a Maxwellian. As stars randomly exchange energy via gravitational interactions, the velocity distribution attempts to relax to a Maxwellian, repopulating the tail. Once stars scatter to high enough energy by this process, however, they escape through the windows around $L_1$ and $L_2$, leaving the tail again underpopulated and continuing the cycle. Throughout the Galactic halo, we see direct evidence of this process through the disruption of GCs, e.g., Pal 5 \citep{Odenkirchen97a,Odenkirchen02a}, NGC 5466 \citep{Grillmair06a}, and Pal 14 \citep{Sollima11a}. Stars escaping through $L_1$ are at smaller galactocentric radii and thus move faster in the galaxy than the cluster which gives rise to a leading stream; stars escaping through $L_2$ are slower than the cluster and therefore trail. The escaping stars give rise to spectacular, cold stellar streams extending up to 60$^\circ$ across the sky \citep{Odenkirchen02a,Grillmair06a}. Limited evidence for the tidal disruption of OCs has been claimed but is rather less convincing \citep[e.g., Berkeley 17;][]{Bhattacharya17a}. 

Since this mass loss process is driven by repopulation of the Maxwellian tail as the velocity distribution relaxes, the loss rate scales approximately as $\dot{M}_{\rm rlx} \propto M/t_{\rm rlx}$, where $t_{\rm rlx} \approx 0.1 (N/\ln N) (r_h/\sigma)$ is the relaxation time at the half-mass radius, $M$ is the cluster mass, $N$ is the number of stars in the cluster, and $\sigma$ is the velocity dispersion. \citet{Henon61a, Henon65a, Henon65b} carried out some of the earliest modeling of this process, finding that $\dot{M}_{\rm rlx} = f_{\rm rlx} M/t_{\rm rlx}$ with $f_{\rm rlx} \approx 0.05$, i.e., clusters lose about 5\% of their mass per relaxation time. If we set $N = M/\overline{m}$, where $\overline{m}$ is the mean stellar mass, and $\sigma \approx \sqrt{G M/r_h}$, the mass loss rate scales as $\dot{M}_{\rm rlx} \propto \sqrt{\rho}$, where $\rho \propto M/r_h^3$ and we have dropped the weak mass dependence in $\ln N$. If clusters have density independent of mass (requiring $r_h$ vs.~$M$ only a slightly steeper than the relation in \autoref{fig:mass_radius}), or if they are tidally limited and thus have $\rho\approx \rho_{\rm ti}$ independent of $M$ (\autoref{ssec:cluster_struct}), then $\dot{M}_{\rm rlx}$ is independent of $M$ as well. This process is the justification for the ``evolved Schechter'' function form introduced in \autoref{ssec:mass} to fit the globular CMF, and explains how a Schechter-like ICMF could evolve into such a distribution \citep{Fall01a, McLaughlin08a}.

While $\dot{M}_{\rm rlx} \propto M/t_{\rm rlx}$ is a reasonable first approximation, $N$-body simulations suggest a slightly more complex dependence. \citet{Fukushige00a}, \citet{Baumgardt01a} and \citet{Baumgardt03a} show that, due to the finite time required for potential escapers to find the $L_1$ or $L_2$ windows and pass through them, the mass loss timescale depends on the crossing time $t_{\rm cr} = r_h/\sigma$ as well as the relaxation time, so $\dot{M}_{\rm rlx} \approx f_{\rm rlx} M / (t_{\rm rlx}^p t_{\rm cr}^{1-p})$, where $p\approx 0.7 - 0.8$ depending on $W_0$. The corresponding cluster dissolution timescale is
\begin{equation}
\label{eq:t_dis_relax}
t_{\rm dis} = \frac{M}{\dot{M}_{\rm rlx}} \approx \left\{
\begin{array}{ll}
1.6 \rho_2^{-1/2} M_3\,\mathrm{Gyr}, & p = 1 \\
0.6 \rho_2^{-1/2} M_3^{0.7}\,\mathrm{Gyr}, & p = 0.7
\end{array}
\right.,
\end{equation}
where $\rho_2 = \rho/(100$ $M_\odot$ pc$^{-3}$), $M_3 = M/(10^3$ $M_\odot$), and for the numerical evaluation we have used $f_{\rm rlx} = 0.05$, $\overline{m} = 0.5$ $M_\odot$, and $\ln N = 7.6$.
Subsequent studies have shown that, in addition to $W_0$, $p$ depends on the extent to which clusters initially fill their tidal radii \citep{Gieles08a, Madrid17a}, the ratios of the eigenvalues of the tidal tensor \citep{Tanikawa10a, Madrid17a}, and the mass spectrum and degree of mass segregation in the cluster \citep{Hong13a}. \citet{Goudfrooij16a} argue that $p$ should also depend on cluster mass, and that the observed GC luminosity function implies $p \approx 1$ for globulars; however, this analysis does not include possible mass-to-light ratio effects due to correlations between GC metallicity and formation history.

A number of authors have proposed analytic models for cluster evolution that attempt to capture the findings of the $N$-body simulations, though none thus far have been able to capture the full range of numerical results over all parameter space. \citet{Lamers05a} and \citet{Lamers10a} propose a model for cluster evolution incorporating the dependence of mass loss on the crossing time and the initial tidal radius, which gives $\dot{M}_{\rm rlx} \approx M^{1-\gamma}/t_0$, with $\gamma \approx 0.65$ and $t_0 \propto 1/\Omega$, where $\Omega$ is the angular velocity of the cluster's orbit through its host galaxy. \citet{Gieles11b} propose a pair of ``unified evolution equations'' that capture the process of expansion and subsequent contraction that occurs when a cluster born with $r_{\rm ti} < r_{\rm tr}$ expands to its tidal radius and then shrinks as mass loss reduces that radius:
\begin{equation}
\xi \equiv -{\frac{\dot{N}}{N} t_{\rm rlx}} = \frac{3}{5}\zeta \frac{t_{\rm cr}}{t_{\rm cr,1}}
\qquad\qquad
\chi \equiv \frac{\dot{t}_{\rm cr}}{t_{\rm cr}} t_{\rm rlx} = 
\frac{3}{2}\zeta \left(1 - \frac{t_{\rm cr}}{t_{\rm cr,1}}\right),
\end{equation}
where $\xi$ and $\chi$ are dimensionless rates of mass loss and cluster expansion and $t_{\rm cr,1}$ is the crossing time of the cluster at the point where it has expanded to its tidal radius. The quantity $\zeta$ is the fraction of the total energy conducted per $t_{\rm rlx}$, and has a complex dependence on the stellar mass spectrum and on the cluster core properties that must be calibrated by $N$-body simulations. The idea that clusters are born with $r_{\rm tr} < r_{\rm ti}$ and expand during their first $\approx 200$ Myr of life is consistent with the tentative findings of \citet{Ryon15a, Ryon17a}, and with the observation that a higher proportion of the older GC population, though not all, fill their tidal radii \citep{Innanen83a, Baumgardt10a}.

\subsection{External and environmental processes}
\label{ssec:external}

Our discussion of internal processes implicitly assumed a constant tidal field. However, as clusters orbit the tidal field they experience may change due to cluster encounters with massive objects, or because the cluster's orbit is not a simple circle in the galactic plane. Such changes in the tidal potential cause the Jacobi surface to move, and sudden perturbations can also alter stellar energies. Both effects can move stars across the Jacobi surface, thereby unbinding part of a cluster. We now explore processes of this type.

\subsubsection{Tidal perturbations}
\label{sssec:tidal}

For star clusters with near-circular orbits in a galactic plane, the primary source of tidal perturbations is encounters with GMCs \citep{Spitzer58a}. 
If an encounter takes place over a time shorter than the crossing time of a cluster, one can approximate it as impulsive \citep[e.g.,][]{Binney08a}. Formally, this requires that the encounter velocity obey $v \gtrsim b \sqrt{GM/r_h^3} = 2.1 b_0 r_{h,0}^{-3/2} M_3^{1/2}$ km s$^{-1}$, where $b$ is the impact parameter, $b_0 = b/1$ pc, $M$ is the cluster mass, $M_3=M/10^3$ $M_\odot$, and $r_h$ is the cluster half-mass radius, $r_{h,0}=r_h/1$ pc. Since the typical encounter velocity in the ISM will be of order the ISM velocity dispersion ($\approx 10$ km s$^{-1}$ in local galaxies), encounters between GMCs and all but the most massive clusters tend to be impulsive. Impulsive encounters can be further classified into catastrophic ones that completely disrupt the cluster, and diffusive ones that do not. Catastrophic disruption occurs only for perturbers that are denser than the cluster, while most star clusters are at least somewhat denser than Milky Way GMCs, $n\approx 100$ cm$^{-3}$ (c.f.~\autoref{fig:mass_radius}), so the diffusive regime is more common.



The specific energy added per tidal shock in the diffusive regime, and the resulting mass loss rate, has been discussed by many authors \citep[e.g.,][]{Gnedin99a, Gieles07a, Prieto08a,Binney08a}. For the case of shocking by GMCs of finite size, \citet{Gieles06a} obtain the approximate relations
\begin{equation}
\label{eq:de_tidal}
\frac{\Delta E}{E} \approx f(\tilde{b}) \left(\frac{M_{\rm GMC}}{M}\right)^2 \left(\frac{\sigma}{v_{\rm max}}\right)^2
\qquad\qquad
\frac{\Delta M}{M} \approx f_M \frac{\Delta E}{E},
\end{equation}
where $E$ and $M$ are the cluster total energy and mass, $\Delta E$ and $\Delta M$ are the changes in these quantities due to the encounter (with a sign convention whereby $\Delta M > 0$ corresponds to mass loss), $f_M \approx 0.25$ is the ratio of fractional mass loss to fractional energy gain (less than unity since stars escape with finite energy), $\sigma = \sqrt{G M/r_h}$ is the stellar velocity dispersion in a cluster, $v_{\rm max}$ is the maximum velocity during the encounter, and $f(\tilde{b})$ is a numerical factor that depends on the dimensionless impact parameter $\tilde{b} \equiv b/r_h$ as $f(\tilde{b})\rightarrow\mbox{const}$ for $\tilde{b}\ll 1$ and $f(\tilde{b})\rightarrow \tilde{b}^{-4}$ for $\tilde{b}\gg 1$. The time to disrupt a cluster, $t_{\rm dis}=M/\dot{M}$, can be found by integrating over encounters with a population of GMCs. Second-order energy injection \citep{Kundic95a} reduces the \citet{Gieles06a} disruption time by a factor $C_{\rm sh}\simeq 0.3$ \citep{Kruijssen11a} so that the disruption time for a cluster of mass $M$ becomes
\begin{equation}
\label{eq:dmdt_tidal}
t_{\rm dis}
\approx
\frac{C_{\rm sh}}{\sqrt{\pi}} f_{\rm dis} \frac{\sigma_{\rm ISM}}{G \Sigma_{\rm GMC}} \left(\frac{\rho}{\rho_{\rm GMC}}\right) \approx 1.6 \sigma_1 \Sigma_2^{-1} \left(\frac{\rho/\rho_{\rm GMC}}{1000}\right)\mbox{ Gyr}
\end{equation}
where $\sigma_{\rm ISM}$ is the velocity dispersion of the ISM (assumed equal to the velocity dispersion of the clusters), $\Sigma_{\rm GMC}$ is the surface density of an individual GMC, $\rho_{\rm GMC}$ is the mass density of GMCs in the ISM (not the GMC internal density), $\rho = 3 M/8\pi r_h^3$ is the cluster density, and $f_{\rm dis}$ is a factor of order unity that depends on gravitational focusing and cluster structure; for \citet{Gieles06a}'s favored parameters, $f_{\rm dis} \approx 0.4$. We have scaled to Milky Way values (c.f.~\autoref{ssec:giant}) for the other parameters: $\sigma_1 = \sigma_{\rm ISM}/10\mbox{ km s}^{-1}\approx 1$, $\Sigma_2 = \Sigma_{\rm GMC}/100\mbox{ M}_\odot\mbox{ pc}^{-2}\approx 1$, and $\rho/\rho_{\rm GMC} \approx 1000$. The 1.6 Gyr disruption timescale is substantially longer than earlier estimates \citep[e.g.,][]{Binney08a}, largely because \citet{Gieles06a} find $f_M \approx 0.25$, while some earlier analytic models adopted $f_M \approx 1$.

One factor that could increase the disruption time is that tidal shocks tend to be self-limiting: a first tidal shock strips off the most weakly bound stars, but the timescale for the cluster to repopulate the high energy orbits these stars occupied is dictated by two-body relaxation (\autoref{sssec:relaxation}), so that subsequent tidal shocks separated in time by less than $t_{\rm rlx}$ remove much less mass \citep{Kruijssen15a, Gieles16a}. On the other hand, the numerical evaluation in \autoref{eq:dmdt_tidal} could underestimate the importance of tidal heating because it is based on mean conditions in the ISM. However, star clusters are formed in regions of higher than average gas density and structure, i.e., larger $\rho_{\rm GMC}$ and $\Sigma_{\rm GMC}$. Star cluster locations remain correlated out to ages $\approx 100$ Myr \citep{Grasha17a}, and during this time they likely experience enhanced tidal perturbations, a phenomenon that \citet{Kruijssen11a} dubbed the ``cruel cradle effect''. Several authors have suggested this effect may explain why cluster number counts decline from $10-100$ Myr, when no other mechanism except stellar mass loss (\autoref{sssec:stellar_evolution}) is expected to operate \citep{Elmegreen10a, Kruijssen12a, Miholics17a}. It is likely to be particularly important in gas-rich environments such as high-redshift galaxies and local mergers, and may explain why globular clusters have a lognormal rather than a power-law mass function: since $t_{\rm dis} \propto \rho$, if density increases with mass then low-mass clusters will be preferentially destroyed by tides, while massive clusters will survive \citep{Elmegreen10b, Kruijssen15a}. However, the importance of this effect depends on the assumed mass-radius relation. Most published models assume mass-independent cluster radii, so $\rho\propto M$, which tends to strengthen the preferential destruction of low-mass clusters compared to the mass-radius relation shown in \autoref{ssec:size}.

This situation is somewhat different for clusters in disk-crossing or bulge-crossing orbits, which experience changing tides as they pass through a disk or bulge. During each passage the galaxy exerts a pinch compression on the cluster that acts normal to the disk. The formalism for modelling the effects of a disk passage is largely the same as that for GMC encounters, simply with a different frequency and shock strength, and with some subtle modifications for the fact that GMC tidal shocks are 2D, while disk tidal shocks are 1D \citep{Gnedin97a}. Starting with \citet{Ostriker72a}, there is an extensive literature on transiting clusters, with an in-depth review provided by \citet{Binney08a}. The timescale for complete destruction of globulars through most of the Milky Way disk is much longer than a Hubble time, but erosion of the outer stars requires only $\approx 10$ Gyr. Thus, the stellar escape feeding the cold streams observed in Pal 5 \citep{Odenkirchen97a,Odenkirchen02a}, for example, may have been exacerbated by repeated crossings. Globulars that transit the bulge of a galaxy are subjected to much stronger tides, and may lose substantial fractions of their mass in well under a Hubble time \citep{Madrid17a}.

For open clusters, which are less massive and dense, disk crossing is more perilous. \citet[see also \citealt{Webb14a}]{Martinez-Medina17a} find that the destruction rate is nearly independent of vertical velocity ($\propto v_z^{-0.2}$) in conflict with the $v_z^{-2}$ dependence found by \citet{Ostriker72a}; \citeauthor{Martinez-Medina17a}~find the number of disk crossings is more important than the velocity. Clusters far from the plane transit the disk only rarely, whereas clusters close to the plane experience a weaker gradient when they do transit. Consequently, clusters a few hundred parsecs from the plane are the most vulnerable to destruction by tidal shocking. This may explain the clear dichotomy of OCs close to the plane and GCs far from the plane, with very little crossover in phase space. Interestingly, the vast majority of reported OCs more than a few hundred parsecs off the plane are not confirmed in the latest \textit{Gaia} DR2 analysis \citep{CantatGaudin2018}.

\subsubsection{Migration and environment}
\label{sssec:migrate}

Several mechanisms alter the orbits of star clusters as they evolve, potentially shaping their demographics. Here we discuss processes that move clusters within a single galaxy, deferring discussion of galaxy-galaxy interaction to \autoref{sssec:non_disk}. One potentially important mechanism is dynamical friction, which tends to move clusters toward galactic centers by causing them to lose angular momentum. For a cluster of mass $M$ moving with velocity $\mathbf{v}_c$ through an inertial stellar field described by a distribution function $f(v)$ for stars with mass $m$ and velocity $v$, \citet{Chandrasekhar43a} showed
\begin{equation}
\frac{d\mathbf{v}_c}{dt} = -16\pi^2 G \ln\Lambda\: \frac{m(m+M)}{v_c^3} \int_0^{v_c} v^2 f(v)\: dv\: \mathbf{v}_c
\end{equation}
for which $\ln\Lambda$ is the Coulomb logarithm; this expression can be simplified considerably if $f(v)$ is a Maxwellian. Massive clusters decelerate more rapidly than lower mass ones ($d\mathbf{v}_c/dt\propto M$ for $M\gg m$), because they create larger wakes. This is one of the few mechanisms that encourages the destruction of the most massive clusters: \citet{Bekki10a} finds that dynamical friction causes clusters with masses $\gtrsim 2\times 10^5$ M$_\odot$ in the disks of sub-$L_*$ galaxies to sink to the galactic center in less than a Hubble time, where stronger tides are likely to destroy them \citep{Gerhard01a}. The effect is smaller in larger galaxies like the Milky Way. 

Radial migration driven by transient spiral density waves, first discussed by \citet{Sellwood02a} and demonstrated numerically by \citet{Roskar08a}, can move stars inward or outward. The same mechanism works for clusters, and is non-destructive to them since clusters have internal crossing times much shorter than the time required for them to transit a spiral density wave \citep{Gieles07a}. Migration occurs because a spiral perturbation with pattern speed $\Omega_P$ can modify an object's energy $E$ and angular momentum $J$ while conserving its Jacobi integral $I_J$; in the ($E, J$) plane, objects move along lines of constant $I_J=E-\Omega_P J$. A single spiral wave near co-rotation can perturb the angular momentum by 20\% without significant heating, moving objects from one circular orbit to another, inwards or outwards. Multiple transient spirals lead to a random walk in the ($E, J$) plane at constant $I_J$ \citep{Dehnen00a}, so that substantial variations in the angular momentum of an object can occur over Gyr timescales. By contrast, long-term spiral arms produce no net effect, because perturbations as an object enters an arm are cancelled by those as it exits \citep{Lynden-Bell72a}. Objects that oscillate far from the plane or are on highly radial orbits are much less influenced by the migration mechanism. \citet{Minchev10a} and \citet{Minchev11a} propose a related mechanism that does not rely on transience: quasi-chaotic interference between resonances from multiple rotating patterns, e.g., the bar and the spiral arms, with different pattern speeds (c.f.~\citealt{Brunetti11a}). \citet{Jilkova12a} and \citet{Quillen18a} investigate this resonance overlap and find that outward migration is possible but relatively inefficient. External influences can also drive radial migration, in particular, radial in-plane orbiting galaxies that come close enough to strongly perturb the disk \citep{Quillen09a}. 

\citet{Minchev10a} provide a comprehensive discussion of the evidence for strong migration of individual stars over billions of years, in particular, the $\sim$1 dex scatter in [Fe/H] at all radii. The high local fraction of metal-rich stars may also be evidence of migration \citep{Kordopatis15a}, since the local ISM has somewhat sub-solar metallicity \citep{Nieva12a}. The case for migrating OCs is less clear. There are at least half a dozen supersolar metallicity OCs within $\sim 1$ kpc of the Solar Circle \citep{Quillen18a}. Moreover, \citet{Friel95a} finds that there is a large scatter in the metallicity and age of OCs at all radii.

However, given the lower statistics, it is unclear whether this provides compelling evidence for cluster migration within the disk, particularly since there is an alternative possibility: OCs that seem out of place with respect to the average properties of their environment may have been accreted from infalling dwarfs. \citet{Carraro09a} argue that the OCs Berkeley 29 and Saurer 1 appear to be associated with the disrupting Sgr dwarf galaxy, and the case for some Milky Way GCs having arrived with Sgr is compelling: \textit{Gaia} has shown that four GCs (Arp 2, Terzan 7, Terzan 8, M54) confined to the Galaxy's central core are co-moving \citep[q.v.~][]{Vasiliev2018}. An OC that is arguably an even stronger candidate for having arrived by infall rather than migration is NGC 6791, an old ($\sim 8.3$ Gyr), massive ($\sim 4000$ $M_\odot$), extremely metal-rich ($[\mathrm{Fe}/\mathrm{H}]\approx +0.3$) OC along the Solar Circle \citep{Boesgaard15a}, 4 kpc from the Sun and 0.8 kpc above the plane \citep{Carraro06a, Origlia06a}.  Its large vertical distance argues against radial migration from the bulge region. While a likely host for infall has not been identified to date, \Gaia\ DR2 data has begun to provide evidence that massive accreted systems like Sgr are dispersed throughout the Galaxy (e.g., Gaia-Enceladus; \citealt{Helmi18a}). We anticipate that the more comprehensive \Gaia\ data release (DR3) in 2020 will significantly clarify whether infall can account for all the anomalous OCs, or whether some must be the result of migration.

\subsection{Models of the full cluster population}
\label{ssec:population_models}

Having reviewed the individual processes affecting gas-free clusters, we now discuss attempts to combine them with cluster formation into models for the complete cluster population. We first consider clusters confined to galactic disks, and second examine clusters outside of disks.

\subsubsection{Disk clusters}

\def\ami		{{\alpha_{Mi}}}
\def\muev		{\mu_{\rm ev}}

Comparison of the observed distribution of cluster masses and ages with theoretical models sheds light on the rate at which clusters are born and on the processes that destroy them.  If the mass of a cluster is determined by its initial mass, $M_i$, and its age, $T$, then the number of surviving clusters in a given age range is the same as that when they were born, so that \citep{Fall01a,Lamers05a,Fall09a}
\beq
\frac{\partial^2 N}{\partial M \, \partial T}=C(T)M_i^\ami\left(\frac{\partial M_i}{\partial M}\right)_T,
\eeq
where $C(T)$ is proportional to the birthrate of clusters with a current age $T$ and where we have assumed that the ICMF is a power law in the range of masses considered; as discussed in \autoref{ssec:mass}, $\ami\approx -2$. Insofar as the effects of stellar evolution occur prior to relaxation (a good approximation) or tidal effects (a good approximation after the cluster has left its natal complex), one can define $M_0=\muev(T) M_i$ as the mass at the onset of relaxation and tidal effects \citep{Lamers05a}, where $\muev(T)$ drops from unity to $\sim 0.25-0.5$ during the first few $10^7$ years of the cluster's life.

Two classes of models for cluster evolution have been considered, mass-independent destruction (MID; \citealp{Fall05a,Whitmore07a,Fall09a}) and mass-dependent destruction (MDD; \citealp{Boutloukos03a,Lamers05a,Lamers10a}). For the MID case, the cluster distribution function breaks up into the product of a mass-dependent term and an age-dependent term, which are taken as power laws, so that
\beq
\frac{d^2N}{dM \,dT} \propto M^{\ami}T^{\alpha_T}.
\label{eq:d2n1}
\eeq  
Since the effects of stellar evolution are approximately independent of mass as long as initial concentration $c$ does not depend on mass, its effects are naturally included in this formulation. This model is phenomenological, since the rate of cluster destruction is not calculated.

For the MDD case, the destruction time is generally taken to be a power-law function of mass \citep{Boutloukos03a,Lamers05a}, $dM/dt=-M/(t_0 M^\gamma)$, where $M$ is measured in solar masses. If $t_0$ is constant, then \citep{Lamers05a}
\beq
\frac{\partial^2N}{\partial M \, \partial T}=\frac{C(T) \muev(T) M^\ami}{\left[1+\gamma T/(t_0 M^\gamma)\right]^{(\gamma+1)/\gamma}},
\label{eq:d2n2}
\eeq
Values of $\gamma$ estimated from observation (e.g., \citealp{Boutloukos03a,Gieles09a}) and $N$-body simulations \citep{Lamers05a,Lamers10a} are usually in the range $0.6-0.7$ (see \autoref{sssec:relaxation}).
At late times or low masses, the second term in the denominator dominates, so that the distribution function becomes proportional
to $M^{\ami+\gamma}T^{-(\gamma+1)/\gamma}$. However, because clusters fade with time, it is sometimes the case that in order to be complete, cluster samples must be constructed so that the minimum mass is large enough that the second term is not dominant. It is not clear that any extragalactic sample of clusters has sufficient dynamic range to clearly show the break in the slope of the mass function.

There are two complications with the MDD model. First, $t_0$ may not be constant: clusters move from a dense environment where they are born to the normal ISM over a period that can extend to 100 Myr (and much longer in galaxy mergers), and over this time $t_0$ is likely to increase \citep{Elmegreen10a,Kruijssen11a}. Second, the simulations used to calibrate this model \citep[e.g.,][]{Baumgardt03a,Lamers10a} have included only relaxation, not tidal losses due to encounters with GMCs, yet in the Galaxy the latter process is likely dominant. The characteristic destruction time for clusters due to GMCs is a few hundred Myr (\autoref{sssec:tidal}), while for a $10^4$ $M_\odot$ cluster the disruption time due to relaxation (confirmed by simulations -- \citealt{Lamers05a}) is multiple Gyr (\autoref{eq:t_dis_relax}). The time scale for destruction by encounters with GMCs depends only on the density of the cluster, so agreement between the observed and simulated values of $\gamma$ occurs only if $\rho \propto M^{0.6-0.7}$ -- possible given the uncertainties in \autoref{fig:mass_radius}, but by no means certain.

Observational evidence has been marshalled on both sides of this debate, which is closely tied to the issue of inclusive versus exclusive catalog construction discussed in \autoref{ssec:unresolved_pop} and \autoref{ssec:age}. \autoref{eq:d2n1} predicts a cluster age distribution with constant $\alpha_T$, and thus is ``universal'', while \autoref{eq:d2n2} predicts a value of $\alpha_T$ that smoothly varies from 0, for $\gamma T/(t_0 M^\gamma) \ll 1$, to $-(\gamma+1)/\gamma$, for the opposite limit. Thus authors who construct exclusive catalogs, which tend to produce $\alpha_T$ close to zero at early ages and then lower values at older ages, generally favor MDD models. Authors who construct inclusive catalogs, which generally have $\alpha_T$ close to $-1$ at all ages, tend to favor MID models. 

In the latter category, \citet{Whitmore07a} and \citet{Fall09a} show that the data on clusters in the Antennae are well described by a power law of the form in \autoref{eq:d2n1} and suggest that this is a ``quasi-universal law'' of cluster destruction. \citet{Fall12a} extended this analysis to clusters in several other galaxies, including the SMC, the LMC, and M83; in all cases, $dN/dM$ and $dN/dT$ exhibit power law behavior with indexes $\alpha_M\approx -2$ and $\alpha_T\approx -0.8$. However, for Milky Way OCs, $\alpha_T\simeq -0.6$ to $-0.5$ over the age range $10-300$ Myr (\autoref{fig:caf}), inconsistent with universality. \citealt{Fall12a} also search for deviations from power-law behavior in the LMC as predicted by the MDD model, by finding clusters with both $\gamma T>t_0 M^\gamma$ and $\gamma T<t_0 M^\gamma$. They find that simple MDD models fail for the LMC. However, they did not include stellar mass loss, which is often included in MDD models (e.g., \citealp{Lamers10a}), allow for a time-dependent destruction time (e.g., \citealp{Elmegreen10a}), or consider the possibility that some of the youngest clusters may be unbound and therefore dissipate rapidly.

Supporting the MDD picture, \citet{Lamers05a} show that the cluster population within 600 pc of the Sun \citep{Kharchenko05a} is consistent with $dN/dT$ calculated from \autoref{eq:d2n2} with $\gamma=0.62$ and a 1.0 Gyr destruction time for $10^4\,M_\odot$ clusters; the data cannot be fit by a power law. \cite{Bastian12a} generalize the cluster evolution model to include both a mass-dependent and mass-independent destruction term, and find that MID and MDD models give equally good fits to two fields in M83. \citet{Adamo17a} find evidence for mass-dependent destruction of clusters with $M<10^4\,M_\odot$ in NGC 628, and \citet{Messa18a} report a similar result for M51. In both cases, however, the evidence comes from the lowest mass bin, so deeper observations or statistical methods for making better use of partially-complete data \citep[e.g.,][]{Krumholz18a} would be beneficial.

Theoretically, the key difference between the two models lies in the mass dependence of the cluster disruption time, $t_{\rm dis}$. As discussed in \autoref{sssec:tidal}, the dominant destruction process for disk clusters is likely tidal shocking by GMCs, yielding disruption time proportional to the cluster density, $t_{\rm dis}\propto\rho$. If $\rho$ is approximately independent of mass, then so is $t_{\rm dis}$ and one gets the MID model. Mass-independent densities are in fact expected for tidally limited clusters \citep{Fall12a}, but the fact that observed young clusters have EFF rather than King profiles (\autoref{ssec:cluster_struct}) and (at least potentially) underfill their tidal radii \citep{Ryon15a,Ryon17a} suggests they have not yet reached equilibrium. On the other hand, advocates of the MDD model (e.g., \citealp{Lamers05a}) assume $\rho\propto M^{0.6-0.7}$, so that the mass dependence of tidal disruption is similar to that for 2-body relaxation. Their models can therefore implicitly include tidal disruption by GMCs, although not with the correct rate. Unfortunately, cluster densities are difficult to determine (\autoref{ssec:size}), so to date it has not been possible to use measurements of $\rho$ vs.~$M$ to distinguish the two models.

\citet{Miholics17a} carry out simulations of isolated galaxies using a subgrid model for the formation and evolution of star clusters \citep{Kruijssen11a,Kruijssen12b}; the model includes both relaxation and encounters with GMCs. They demonstrate that, in the MDD model, the median age of clusters in a sample is proportional to $t_0$, and the median age depends on the surface density and velocity dispersion of the gas. Comparing with observations of M31 (outer galaxy), M51 (three mass ranges), and M83 (inner and outer regions), they find the best fit for the mass dependence of the median age is $t_{\rm med}\propto M^{0.54}$. In all their models, $dN/dT$ decreases with age even though the star formation rates are roughly constant, and they are able to fit the observed median ages for the six cluster samples to within 0.2 dex, ranging from $10^{7.3}$ yr for clusters in M51 with masses greater than $10^3\,M_\odot$ to $10^{8.3}$ yr for clusters in the outer parts of M31 and M83. This more complete model for cluster evolution is more complex than the simple MID and MDD models discussed above, and it demonstrates that cluster destruction is environmentally dependent.

\subsubsection{Non-disk clusters}
\label{sssec:non_disk}

All star clusters, including the progenitors of GCs, likely began their lives in the disk of some galaxy (though see \citealt{Mandelker18a} for a model where this is not true). Thus clusters found outside galaxy disks necessarily have complex histories. Studies of these histories therefore focus on determining possible origin sites for the clusters we see today, on the processes by which those clusters reached their present locations. A great deal of this work has been devoted to modeling the cosmological assembly history of the Milky Way's GC system, and to explaining observed features such as its color bimodality and color-magnitude relationships (the ``blue tilt''; \citealt{Brodie06a}). These topics are beyond the scope of our review, which we limit to processes affecting star clusters in general, rather than the specific histories of GCs.

Since the transfer of clusters out of disks involves strong tides, simulations of the origin of non-disk clusters must model tidal effects. Since galactic and cosmological simulations cannot resolve cluster interiors (\autoref{ssec:icmf_numeric}), they cannot do so directly, and instead the usual approach is to treat each star cluster as a point mass that is either formed self-consistently in the simulation or inserted following some prescribed initial distribution. As the particle moves along its orbit, one records the tidal potential to which it is subject, and uses this to calculate cluster evolution from an analytic model \citep[e.g.,][]{Prieto08a, Kruijssen11a, Brockamp14a, Rossi15b, Carlberg18a, Pfeffer18a} or via a direct $N$-body simulation \citep[e.g.,][]{Renaud13a, Rieder13a, Rossi15a, Mamikonyan17a}. The latter approach is more accurate, but due to its expense one can only simulate a few examples, and for less than a full Hubble time.

One obvious potential site for the formation of non-disk clusters is in major galaxy mergers such as the Antennae. \citet{Renaud13a} and \citet{Kruijssen12b} both simulate Antennae-like mergers, but reach opposite conclusions: \citeauthor{Renaud13a} find that the merger has little effect on the cluster population, while \citeauthor{Kruijssen12b}~find that it transforms the CMF from Schechter-like to lognormal like. The \citeauthor{Kruijssen12b}~results are likely more reliable. They include gas and form clusters self-consistently during the merger, while \citeauthor{Renaud13a} carry out pure $N$-body simulations and study only pre-existing clusters; both differences produce weaker tides in \citeauthor{Renaud13a}'s simulations. \citeauthor{Renaud13a} also initialize their clusters with much smaller radii, making them more resistant to tidal disruption.

Dwarf galaxies represent another potential formation site for clusters that end up in the halos of larger galaxies such as the Milky Way: clusters form in the dwarf (and are therefore metal-poor, as observed), and then become part of the halo when the larger galaxy accretes and tidally strips the dwarf. As discussed in \autoref{sssec:migrate}, there are multiple Milky Way clusters that may have originated in this way. Simulations suggest that the accretion process is gentle, so accreted clusters do not experience major tidal losses, and quickly adjust to their new environments and become indistinguishable, except in their abundances, from the pre-existing cluster population \citep{Miholics14a, Miholics16a, Bianchini15a}.

Exploring the full GC population requires cosmological simulations. Unfortunately, it is not presently possible to use the same simulations to study both the formation of GCs and their subsequent evolution (see \citealt{Forbes18a} for more discussion). Simulations that can resolve cluster formation \citep[e.g.,][]{Kimm16a, Ricotti16a, Li17a, Li18a, Kim18b} are too expensive to run to redshift zero, so studies that reach the present use potential GC progenitors inserted by hand. Different authors make different assumptions about how this should be done. \citet{Prieto08a} and \citet{Carlberg18a} use dark matter-only simulations; the former initialize cluster particles based on an analytic model of the baryonic mass distribution, while the latter adopt a randomly-oriented disk. They also assume different redshift distributions. \citet{Renaud17a} and \citet{Pfeffer18a} use dark matter plus baryon cosmological simulations, with \citeauthor{Renaud17a} assuming that clusters follow the distribution of all newly-formed stars and \citeauthor{Pfeffer18a}~using the analytic models of \citet{Kruijssen12a} and \citet{Reina-Campos17a} to vary $\Gamma$ and the ICMF depending on local ISM conditions.

The conclusions are as diverse as the sets of initial conditions: \citet{Prieto08a} and \citet{Pfeffer18a} find that tides and two-body relaxation are able to transform an initially-power-law distribution of cluster masses into a lognormal, consistent with the observed globular CMF. \citet{Renaud17a} reach the opposite conclusion, finding that their test clusters experience tides too weak to produce such a radical change. The difference is likely because their recipe for cluster formation -- clusters simply follow star formation, in contrast to \citeauthor{Prieto08a}'s prescribed distribution or \citeauthor{Pfeffer18a}'s variable $\Gamma$ and ICMF -- places clusters on orbits where they experience weaker tides. Similarly, \citet{Carlberg18a} finds that the effectiveness of tides is sensitive to the assumed distribution of redshifts for cluster formation. As for the ICMF (\autoref{ssec:icmf_numeric}), at present simulation outcomes appear to depend primarily on subgrid recipes, and thus have little predictive power. Fortunately, we are likely to gain far better observational constraints once \textit{JWST} launches, since it should narrow down the range of plausible models considerably.

\section{AFTERLIFE}
\label{sec:afterlife}

The clusters we observe today are the surviving remnant of a much larger population that have now dissolved. However, within the Milky Way it is possible to reconstruct some of these clusters via a variety of techniques. Doing so provides important clues about the cluster formation and dispersal process, the Galactic potential and, for clusters in the halo, the nature of dark matter. We therefore focus this final section on cluster reconstruction using kinematic (\autoref{sec:kin}), action-angle (\autoref{sec:action}), and chemical tagging methods (\autoref{sec:chem}). This field is in its infancy, so the discussion is necessarily more speculative than that up to this point.

\subsection{Kinematics}
\label{sec:kin}

The traditional method for cluster reconstruction is to find ``moving groups'': clusters of stars in velocity space.\footnote{This space is usually described by the three velocity components $(U,V,W)$, where $U$ is toward the Galactic Center, $V$ is the velocity in the Galactic plane along the direction of the Sun's orbit, and $W$ is the velocity out of the plane, oriented so the coordinate system is right-handed. Velocities are measured relative to the Local Standard of Rest \citep{BlandHawthorn2016a}.} Clustering in velocity space is useful because unbound stars disperse in position on a timescale of a crossing time (a few Myr), but only spread out in velocity over timescales of a substantial fraction of their orbital period in the Galaxy (tens of Myr). This technique has a long and checkered history \citep{Griffin1998} with some early identified groups being questioned in later analysis \citep[q.v.~][]{Taylor2000}, reminiscent of the ``faux clusters'' identified in \Gaia\ DR2 \citep{CantatGaudin2018}. Nonetheless, it remains a powerful method for finding disrupted clusters.

Within the disk, \citet{Dehnen1998} first used the {\it Hipparcos} survey to identify kinematic substructure in the $(U,V)$ plane for a very local sample ($D\lesssim 100$ pc); he identified known star clusters, but also discovered the previously-unknown Hercules stream. \citet{Bensby2007} showed that this structure is inhomogeneous in its abundances, and therefore more likely to be stars trapped in a resonance \citep[e.g.,][]{Quillen2018} than a disrupting cluster. Subsequent wavelet analysis of the {\it Hipparcos} survey combined with the RAVE survey identified many more local clumps in the $(U,V)$ plane, although they have proven difficult to interpret \citep{Antoja2012}. While resonant streams are interesting, they are ``false positives'' from the standpoint of cluster reconstruction. One major challenge for kinematic searches is that the expected rate of such false positives is unknown. The 2D simulations of \citet{DeSimone2004} suggest it is high, but we suspect that it will be lower in a 3D simulation due to the extra degree of freedom. For objects that are real disrupted clusters, one can use dynamical traceback methods to determine the age and other properties of the original cluster \citep{deZeeuw1999,Riedel2017}.

The halo has a great deal of substructure, but most known structures are associated with the disrupting Sgr dwarf \citep{Helmi2008}. Exceptions include cold halo streams like those extending from the GC Pal 5 (\autoref{sssec:relaxation}). There are similar linear structures such as GD-1 \citep{Grillmair06a} and the ``Jet'' stream \citep{Jethwa2018} not associated with a surviving progenitor; these are likely clusters that have fully disrupted \citep[e.g.,][]{Koposov2010}. Some halo substructures of comparable mass have only been identified kinematically \citep{Kepley2007} or, for an assumed Galactic potential, by converting from kinematic coordinates to energy-angular momentum \citep{Helmi2017} or angular momentum-eccentricity space \citep{Helmi2006}. Since \Gaia\ DR2, new studies \citep[q.v.][]{Antoja2018} reveal many more substructures in the $(U,V)$ plane, but follow-up work to identify them has only just begun.

\begin{textbox}[ht!]
\section{DISRUPTED STAR CLUSTERS AS PROBES}
The coldest streams from disrupting star clusters are proving to be useful probes of the shape and enclosed mass of the Galaxy \citep{Bowden2015,Bovy2016}. Recent observations confirm that GD-1 has high-contrast gaps along its length \citep{PriceWhelan2018}. Several authors have argued that these are useful probes of the halo's dark matter substructure, since substructures would cause diffusion of stellar orbits that would blur out sharp features over time \citep{Carlberg2016,Bovy2017}.
\end{textbox}

\subsection{Action-angle space}
\label{sec:action}

Kinematic reconstruction fails at ages $\gtrsim 100$ Myr because differential acceleration along stars' orbits makes them disperse in velocity space. However, while velocities change as stars orbit, there are conserved quantities that do not, at least to the extent that we can approximate the Galactic potential as smooth and time-independent. In principle these quantities act as labels that are fixed when stars form, allowing reconstruction of clusters long after they have dispersed in velocity. This insight motivates the idea of cluster reconstruction in action-angle space (${\mathbf J,\vectheta}$), a set of canonical conjugate coordinates defined by \citep{Binney08a}
\begin{equation}
2\pi J_i = \oint \dot x_i\,d x_i
 = \frac{1}{\Omega_i}\int_0^{2\pi}(\dot x)^2\,d\theta_i
\end{equation}
where $x_i$ are Cartesian coordinates and $\Omega_i$ are the orbital frequencies. For each star, the equations of motion with respect to the Hamiltonian ${\mathbf H}$ are
\begin{equation}
\dot{\vectheta} = \frac{\partial\mathbf{H}}{\partial\mathbf{J}} = \Omega(\mathbf{J}) 
\qquad\qquad
\dot{\mathbf{J}} = -\frac{\partial\mathbf{H}}{\partial\vectheta} = 0
\end{equation}
An orbit can be traced with the solution $\theta_i(t)=\Omega_i(t)+\Omega_{i,0}$. For the inertial Galactic system, $(R,\phi,z)$ coordinates\footnote{In this coordinate system, $R$ is the distance from the Galactic Center, $z$ is distance out of the Galactic plane, and the line from the Sun to Sgr A$^*$ defines $\phi=0$.} are most often used (even for halo studies) such that $\vectheta=(\theta_R,\theta_\phi,\theta_z)$ and ${\mathbf J}=(J_R,J_\phi,J_z)$ where $J_\phi\equiv L_z$, i.e., $J_\phi$ is the angular momentum with respect to the spin axis. A stellar orbit is described by the three actions $\mathbf{J}$. Orbits with $J_z = 0$ lie in the Galactic plane, orbits with $J_R = 0$ are circular, and we choose units so that a circular orbit at the Solar Circle has $J_\phi = 1$; circular orbits with $J_\phi < 1$ lie inside the Solar Circle. The orbits of all stars in the Galaxy are described by the distribution function $f({\mathbf J})$.

To find clusters in action-angle space one must convert the observed positions and velocities of stars into actions. This requires knowledge of the Galactic potential $\Phi$, which enters the Hamiltonian $\mathbf{H}$. We will not fully review efforts to measure the Galactic potential and the stellar density distribution, which must be determined simultaneously. Arguably, the best efforts to date are from the RAVE survey, which provides an explicit fit for the vertical stellar density distribution \citep{Piffl2014} as opposed to the parameterized potentials used in earlier work \citep[e.g.][]{McMillan2011}. Moreover, $\Phi$ is at least weakly time-dependent. In addition to the outer warp, there is clear evidence near and far of large-scale corrugation waves propagating throughout the disk \citep[e.g.][]{Xu2015,Antoja2018}. The spiral arms and bar provide additional challenges.

Even with the uncertainties, the action-angle approach reveals rich structure throughout the halo and Galactic disk. \citet{Sellwood2010} mapped {\it Hipparcos} moving groups to action space and found that this improved the delineation of the Hyades stream, but that other structures were barely evident. With \Gaia\ DR2, the situation has improved and over a much larger volume than {\it Hipparcos}. \citet{Trick2018} show that the highest density peaks in the $(U,V)$ plane map to the highest density peaks in the action plane, specifically ($\sqrt{J_R},J_\phi=L_z$), but the mapping is not simple. They trace structure out to 1.5 kpc from the Sun and find much of it has low vertical action, $J_z$, 
suggestive of disk resonances that operate most efficiently in the plane \citep{Katz2018}. The Helmi streams were discovered with a dozen or fewer stars using action analysis \citep{Helmi2008}, structures that were invisible against the background in kinematic space. Arguably the most impressive action-based discovery to date is Gaia-Enceladus, the remnants of a massive dwarf spread throughout the inner Galaxy that was accreted long before Sgr \citep{Helmi2018}. To date, no OCs or GCs (or their streams) have been discovered using actions, but these are early days. There are many clumps in this space that have yet to be followed up. Stars bunched in phase space are typically spread across the sky requiring extended observing programs over $1-2$ years.

\subsection{Chemical tagging}
\label{sec:chem}

Dynamical ``invariants'' like actions are useful labels for reconstruction, but they are not strictly conserved over a star's lifetime, e.g., when transiting the spiral arm or bar. These time-dependent features of the Galactic potential induce perturbations in the actions of the stars they affect; indeed, non-conservation of $J_\phi$ due to arms and bars is responsible for stellar migration (\autoref{sssec:migrate}), and the same structures also likely cause non-conservation of $J_R$ \citep{Solway2012}. This ultimately limits how far back in time action-angle reconstructions for clusters in the disk can go. Longer-term reconstructions therefore require another invariant, for which elemental abundances are a natural choice. Stars conserve their abundances over almost their entire lives, and stars in the same cluster have nearly-identical abundances (\autoref{ssec:chemical_comp}). This led \citet{Freeman2002} to suggest that, with abundance data of sufficient quality and enough elements, clusters that have dispersed even in action space could be detected via clustering in ${\cal C}$-space, defined so each star is a point in a high dimensional space, i.e., ${\cal C}$([Fe/H], [$\alpha$/Fe], [X$_1$/Fe], [X$_2$/Fe], $\ldots$).

Weak chemical tagging, defined as identifying distinct components separated by a surface in $\mathcal{C}$-space, has been used to good effect. For example, stars in the Solar neighborhood show a clear bimodality in ${\cal C}$([Fe/H], [$\alpha$/Fe]) \citep[e.g.,][]{Bensby2005} and ${\cal C}$([$\alpha$/Fe], [(C+N)/Fe], [Al/Fe], [Mg/Mn]) \citep{Hawkins2015}. The Sgr dwarf and stream separate cleanly from all components of the Milky Way in the high-dimensional space ${\cal C}$([(C+N)/Fe], [O/Fe], [Mg/Fe], [Al/Fe], [Mn/Fe], [Ni/Fe]) \citep{Hasselquist2017}. Reconstructing star clusters, however, requires strong chemical tagging, defined as identifying a cluster in $\mathcal{C}$-space against the background (q.v.~\citealp{Ting2012,Ting2015}). In searching for dissolved clusters, the choice of elements depends on the target. OCs are highly homogeneous so many elements are prospective candidates, while GCs are inhomogeneous in some or all elements, but have distinctive anti-correlations, e.g., O/Mg vs.~Na/Al. \citet{Martell2016} use the latter signature to identify stars in the Galactic halo from long-dissolved GCs. \citet{Ting2015} search for clustering of abundances in the APOGEE survey of the $\alpha$-rich disk, but do not identify any previously-unknown clusters. Related techniques applied to the $\alpha$-poor disk using the GALAH survey have met with more success \citep{Quillen2015,Kos2018a}. Even when clusters are found, however, there may not be a simple one-to-one mapping between $\mathcal{C}$-space clusters and OCs. For example, \citet{DeSilva2015} identify a pair of OCs that have identical radial velocities and overlapping stars in ${\cal C}$-space. These were likely formed in close proximity, so abundance analysis does correctly identify real structures, but they are not necessarily the same clusters one would identify in physical space. 

As for kinematics, strong chemical tagging suffers from a poorly-constrained false positive rate. \citet{Ness2018} identify pairs of field stars with identical abundance patterns that are unlikely to have been born in the same cloud, but the overall frequency of such chance pairs remains uncertain. The problem should be less severe at low [Fe/H], where there are fewer stars and thus a reduced background of interlopers. This led \citet{BlandHawthorn2010b} to propose applying chemical tagging to the Milky Way's dwarf satellites rather than in its disk. Prima facie, this appears to work \citep[e.g.,][]{Karlsson2012,Webster2016}, and if confirmed the patterns identified thus far are evidence of star clusters significantly more metal-poor than the lowest metallicity GCs, which have $\mbox{[Fe/H]}\approx -2.4$.

The challenges of strong chemical tagging mean that, thus far, it has been most effective when used to characterize groups found by other methods, and to test for potential membership. For example, most stellar streams identified to date appear to be inhomogeneous, implying that they are not disrupting star clusters, and are instead either stars trapped by disc resonances \citep[e.g.,][]{Bensby2007} or the remnants of much more massive ($> 10^8\msun$) systems \citep[e.g.,][]{Hasselquist2017}. However, a few streams have proven to be homogeneous, e.g., the 2-Gyr old moving group HR 1614 \citep[][]{Feltzing2000,DeSilva2007}, first identified by \citet[][]{Eggen1978} and confirmed with {\it Hipparcos} \citep[][]{Eggen1998}. For these systems, elemental homogeneity provides strong evidence that the progenitor is a disrupted star cluster. In the future, it seems likely that the most powerful applications of chemical tagging will continue to be in conjunction with other methods \citep[e.g.,][]{Quillen2015} rather than by itself. \citet{Kos2018a} provide an impressive example of how such combinations can work, by combining chemical tagging and \Gaia\ proper motions to identify Pleiades members several degrees from the parent cluster. To date, other identified streams for which chemical tagging may prove useful are faint, so follow-up work awaits multi-object high resolution spectrographs on the ELTs (e.g., MANIFEST on GMT).

\section{CONCLUSIONS AND FUTURE PROSPECTS}
\label{sec:conclusions}

Star clusters stand at a crossroads of scales. Traditionally star cluster research has focused on clusters as discrete entities whose formation, evolution, and eventual dissolution can be viewed in isolation, at most treating the background galaxy as a source of a passive tidal field. In this paradigm, groups of stars can be neatly classified into open clusters, globular clusters, associations, and field stars, each with distinct properties and formation histories. Such isolation is no longer viable. Observations now reveal that the objects that go on to form bound star clusters are merely the innermost parts of a hierarchy that extends to the scales of galaxies. Once clusters form, the tidal perturbations to which they are subject are affected first by the immediate star-forming environment, and over longer times by the full cosmological history of galaxy assembly. Going to high redshift brings us to a point where the distinction between open clusters and globular clusters dissolves. Consequently any modern understanding of star clusters must bridge from sub-pc to cosmological scales. While many questions remain, the outlines of a model are starting to become clear:

\begin{summary}[SUMMARY POINTS]
\begin{enumerate}
\item Stars form in giant molecular clouds (GMCs) over a few free-fall times, at a rate per free-fall time $\epsff \sim 1\%$, with a spread of $\sim 0.5$ dex. At formation, the stars are hierarchically-structured and cannot be separated neatly into clusters. Distinct star clusters emerge only after gas is cleared, which occurs in less than a few million years. After about 30 Myr in Milky Way-like galaxies, $\sim 10\%$ of the stars are in clusters. There are good theoretical arguments that this fraction should vary with galactic environment, but the evidence either for or against variation is thus far unconvincing.
\item The portions of clouds that form bound clusters have at most mildly elevated $\epsilon_{\rm ff}$ values, but form stars over multiple free-fall times, allowing them to reach elevated total star formation efficiencies. Extended star formation renders these regions dynamically relaxed and well-mixed in abundance. In contrast, the stars that are not part of the bound regions never dynamically relax, and may or may not be well-mixed. They only gradually drift apart; they are not part of a coherent flow either collapsing towards or expanding away from the dense regions that become bound. 
\item Star formation is ultimately terminated by feedback, but which type of feedback is dominant depends on the local environment. Outflows dominate in clouds too small to contain massive stars, and radiation pressure dominates in the most massive and dense systems, but for the bulk of star clusters photoionization is likely to be most important. Feedback sets the timescale for gas clearing, and thus determines the density at which star formation transitions from taking place over one to many dynamical times, marking the boundary between the bound and unbound parts of a hierarchical star-forming cloud. The initial cluster mass function (ICMF) of both bound and unbound clusters is well described by a power-law $dN/dM\propto M^{\alpha_M}$ with $\alpha_M\approx -2$, likely with an environmentally-dependent high-mass cutoff. Evidence for any stronger variation in cluster formation with environment is weak.
\item In the first $\approx 10-100$ Myr after a gas-free cluster population emerges, it is subject to mass loss via stellar evolution-driven expansion and tidal shocking by gas in its immediate environment. The rate at which the cluster age function (CAF) declines from $\approx 10 - 100$ Myr, as parameterized by its slope $\alpha_T$ ($dN/dT\propto T^{\alpha_T}$), provides a powerful constraint on the importance of these processes. In the Solar neighborhood $\alpha_T \approx -0.5$, implying moderately strong cluster destruction, but the CAF slope outside the Milky Way is uncertain because present samples are compromised by uncertainties in cluster selection and age assignment. The strength of early mass loss mechanisms depends strongly on the distributions of cluster concentration and ratio of size to tidal radius at the end of gas clearing, which are poorly constrained both theoretically and observationally, though there are observational hints that clusters in Milky Way-like galaxies are born fairly concentrated and under-filling their tidal radii. There is tension between these hints and the Milky Way cluster age distribution, which requires moderately strong early cluster disruption.
\item Over timescales of Gyr, the dominant cluster destruction mechanisms are relaxation and tidal shocking (by GMCs for disk clusters, by disk traversal for halo clusters). Relaxation preferentially destroy low-mass clusters; tidal shocking likely does as well, though this depends on how strongly density increases with mass. These mechanisms thus provide a plausible mechanism for transforming the power-law mass distribution ubiquitously observed for young clusters into the peaked distribution observed for globular clusters. However, enhanced tidal shocking during clusters' first $\approx 100$ Myr of life also preferentially destroys low-mass clusters, and potentially provides an alternate explanation. The relative importance of the various processes, and the timescale over which the CMF shape changes, likely depends on the galactic environment.
\item Even after clusters come apart in physical space, the stars remain coherent in kinematic, action-angle, and chemical space for timescales of tens, hundreds, and thousands of Myr, respectively. In principle it should be possible to reconstruct clusters in these spaces. However, efforts to do so are still in their infancy, and most present techniques still suffer from high rates of false positives.
\end{enumerate}
\end{summary}

\begin{issues}[FUTURE ISSUES]
\begin{enumerate}
\item A recurring theme of this review is the uncertainty in inferences of various quantities -- cutoffs in the CMF, cluster ages, etc. -- based on integrated light. When one cross-checks these results against more accurate methods based on resolved stellar populations and young stellar objects (YSOs), the level of agreement is often worse than a naive interpretation of the stated error bars would suggest. However, integrated light observations will remain a vital tool for the foreseeable future, since even in the era of \textit{JWST} we will not be able to resolve stellar populations in clusters at even $\sim 100$ Mpc distances, let alone at high redshift. Consequently, there is an urgent need for a program of checking integrated light techniques against resolved stellar populations, and developing new methods that return realistic error bars.
\item There is a pressing need to extend extragalactic cluster catalogs to lower mass and larger age. At present the Milky Way and extragalactic samples are almost completely non-overlapping in mass, and the extragalactic samples have very limited dynamic range in mass, making it hard to draw conclusions. The best prospects for improving this situation are in pushing to lower mass clusters in galaxies at distances of a few Mpc. This will become possible with extreme AO-fed instrumentation on the next generation of extremely large telescopes. The same instruments will also make it possible to probe star formation in cosmologically denser regimes compared to the low overdensities in the Local Volume.
\item Both analytic models and simulations of star cluster formation that begin from isolated clouds, ignoring the galactic context, are likely reaching the end of their utility. They consistently fail to reproduce observed star formation histories and stellar kinematic, and are likely incapable of correctly predicting the cluster formation efficiency overall. The future is in simulations that start self-consistently from the galactic or cosmological scale, including all relevant physics, and then zoom in to individual clusters, or that do form and evolve clusters using semi-analytic models calibrated to zoom simulations.
\item The largest theoretical uncertainties in cluster demographics now lie at the transition between gas-dominated and gas-free evolution. Pure $N$-body simulations of the gas-free phase have advanced to the point where it is now possible to simulate a cluster of $10^6$ stars directly, but a continuing effort to push these simulations to ever-larger numbers of stars and longer integrations may have hit the point of diminishing returns because the largest uncertainty is the choice of initial conditions, which cannot be addressed without modeling the gas. 
\item Observations of 6D phase space from ESA \textit{Gaia} and ESO Gravity may address the issue of the poorly known initial conditions, since they can simultaneously supply initial conditions to the $N$-body modelers and provide a benchmark against which gas simulators can test their work. This project will, however, require new statistical tools. Statistics like the CMF are problematic for unrelaxed populations that cannot be uniquely decomposed into clusters, while structure-agnostic statistics such as two point correlation functions are less informative than statistics like the CMF when applied to populations of relaxed, discrete clusters. Observed star-forming complexes such as Orion, however, appear to transition smoothly between smooth and relaxed at high density and fractal and unrelaxed at low density. We need statistical tools that can characterize this transition and work across it.
\item The most powerful future methods for reconstructing disrupted clusters are likely to combine elemental abundance and kinematic data. Each method suffers from significant false positives by itself, but the false positives to which each are subject are quite different, and thus the two methods can be used to cross-check one another. However, this will require the availability of abundance measurements going to fainter magnitudes than is now possible. Ground-layer, AO-corrected wide-field spectrographs planned for the next generation of telescopes will allow detailed abundance analysis on stars to fainter limits ($V\approx 20$) than possible today, greatly aiding this effort.
\end{enumerate}
\end{issues}

\section*{DISCLOSURE STATEMENT}
The authors are not aware of any affiliations, memberships, funding, or financial holdings that
might be perceived as affecting the objectivity of this review. 

\section*{ACKNOWLEDGMENTS}
We thank A.~Adamo, N.~Bastian, R.~Chandar, E.~van Dishoeck, B.~Elmegreen, E. Falgarone, M.~Fall, D.~Kruijssen, E.~Ostriker, and A.~Sternberg for helpful comments on manuscript. MRK acknowledges support from the Australian Research Council's (ARC) \textit{Discovery Project} and \textit{Future Fellowship} funding schemes, awards DP160100695 and FT180100375.  CFM acknowledges support by NASA through NASA ATP grant NNX13AB84G. JBH acknowledges support through the ARC  \textit{Discovery Project} and \textit{Laureate Fellowship} funding schemes, awards DP150104667 and FL140100278.  JBH acknowledges the hospitality of UC Berkeley and thanks the Miller Foundation for their generous support of the 2018 Visiting Professorship. MRK and JBH acknowledge support from an ARC Centre of Excellence for All Sky Astrophysics in 3 Dimensions (ASTRO 3D), through project number CE170100013.

\bibliographystyle{ar-style2}
\bibliography{refs,joss}

\end{document}